\documentclass[iop]{emulateapj}

\usepackage{amsmath}
\usepackage{rotating}
\usepackage{float}
\usepackage{verbatim}
\usepackage[caption = false]{subfig}

\shorttitle{The Ages of A-Stars I}
\shortauthors{Jones, J. et al.}

\begin{document}
\renewcommand{\arraystretch}{1.5}

\title{The Ages of A-Stars I: Interferometric Observations and Age Estimates for Stars in the Ursa Major Moving Group}

\author{
	Jeremy Jones\altaffilmark{1}\altaffilmark{*}, 
	R. J. White\altaffilmark{1}, 
	T. Boyajian\altaffilmark{2}, 
	G. Schaefer\altaffilmark{1}, 
	E. Baines\altaffilmark{3}, 
	M. Ireland\altaffilmark{4}, 
	J. Patience\altaffilmark{5}, 
	T. ten Brummelaar\altaffilmark{1}, 
	H. McAlister\altaffilmark{1}, 
	S. T. Ridgway\altaffilmark{6}, 
	J. Sturmann\altaffilmark{1}, 
	L. Sturmann\altaffilmark{1}, 
	N. Turner\altaffilmark{1}, 
	C. Farrington\altaffilmark{1},
	P. J. Goldfinger\altaffilmark{1}
	}
\altaffiltext{*}{Correspondence to: jones@astro.gsu.edu}
\altaffiltext{1}{Center for High Angular Resolution Astronomy and Department of Physics and Astronomy, Georgia State University, 25 Park Place, Suite 605, Atlanta, GA 30303, USA}
\altaffiltext{2}{Department of Astronomy, Yale University, New Haven, CT 06511}
\altaffiltext{3}{Remote Sensing Division, Naval Research Laboratory, 4555 Overlook Avenue SW, Washington, DC 20375}
\altaffiltext{4}{Research School of Astronomy \& Astrophysics, Australian National University, Canberra ACT 2611, Australia}
\altaffiltext{5}{School of Earth and Space Exploration, Arizona State University, PO Box 871404, Tempe, AZ 85287}
\altaffiltext{6}{National Optical Astronomy Observatory, 950 North Cherry Avenue, Tucson, Arizona 85719, USA}

\begin{abstract}
	We have observed and spatially resolved a set of seven A-type stars in the nearby Ursa Major moving group with the Classic, CLIMB, and PAVO beam combiners on the CHARA Array.  
	At least four of these stars have large rotational velocities ($v \sin i$ $\gtrsim$ 170 $\mathrm{km~s^{-1}}$) and are expected to be oblate. 
	These interferometric measurements, the stars' observed photometric energy distributions, and $v \sin i$ values are used to computationally construct model oblate stars from which stellar properties (inclination, 
		rotational velocity, and the radius and effective temperature as a function of latitude, etc.) are determined. 
	The results are compared with MESA stellar evolution models \citep{mesa1,mesa2} to determine masses and ages. 
	The value of this new technique is that it enables the estimation of the fundamental properties of rapidly rotating stars without the need to fully image the star.
	It can thus be applied to stars with sizes comparable to the interferometric resolution limit as opposed to those that are several times larger than the limit.
	Under the assumption of coevality, the spread in ages can be used as a test of both the prescription presented here and the MESA evolutionary code for rapidly rotating stars. 
	With our validated technique, we combine these age estimates and determine the age of the moving group to be 414 $\pm$ 23 Myr, which is consistent with, but much more precise than previous estimates. 
	
\end{abstract}

\section{Introduction and Motivation}
	\setcounter{footnote}{0}
	Population I stars with spectral type A have masses that range from 1.5 - 2.5 M$_{\sun}$, based on dynamical measurements of spectroscopic binaries \citep[e.g.,][]{torres_2010}, and corresponding main sequence 
		lifetimes of 3.6 - 1.0 Gyr (assuming $\tau_{\mathrm{MS}}$ $\propto$ M$^{-2.5}$, \citealt{sse_book}).
	However, measuring the age and mass of any single A-type star is often severely complicated by their peculiar characteristics, including chemically anomalous photospheres (e.g., Am stars, Ap stars, $\lambda$ Boo stars), 
		radial and non-radial pulsations (e.g., $\gamma$ Doradus-type, $\delta$ Scuti-type), and severely distorted shapes from rapid rotation (e.g., Regulus - \citealt{chara1}, Altair - \citealt{monnier_2007}, Vega - 
		\citealt{aufdenberg_2006,monnier_2012}). 
	Despite these challenges, interest in determining precise ages and masses for A-type stars has been heightened with the discoveries of directly imaged candidate planets orbiting several of them, including 
		HR8799 \citep{marois_2008}; Fomalhaut \citep{kalas_2008}; etc.
	In all cases the estimated masses of the planetary companions depend critically on the ages assigned to the host stars (e.g., \citealt{moya_2010} versus \citealt{marois_2010}), which in most cases are poorly determined.

\begin{figure*}
	\subfloat[]{\includegraphics[height =2.5in]{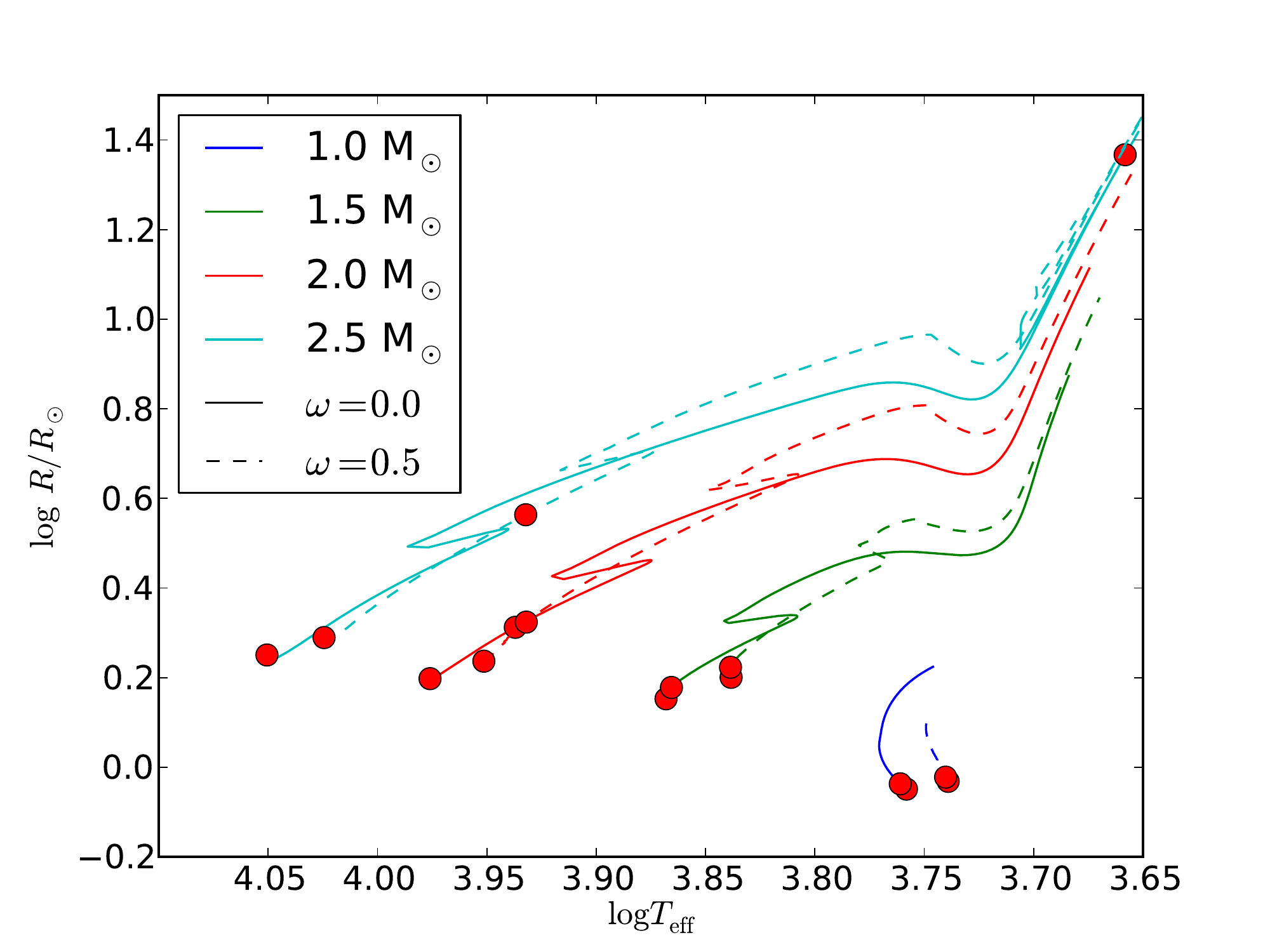}}
	\subfloat[]{\includegraphics[height =2.5in]{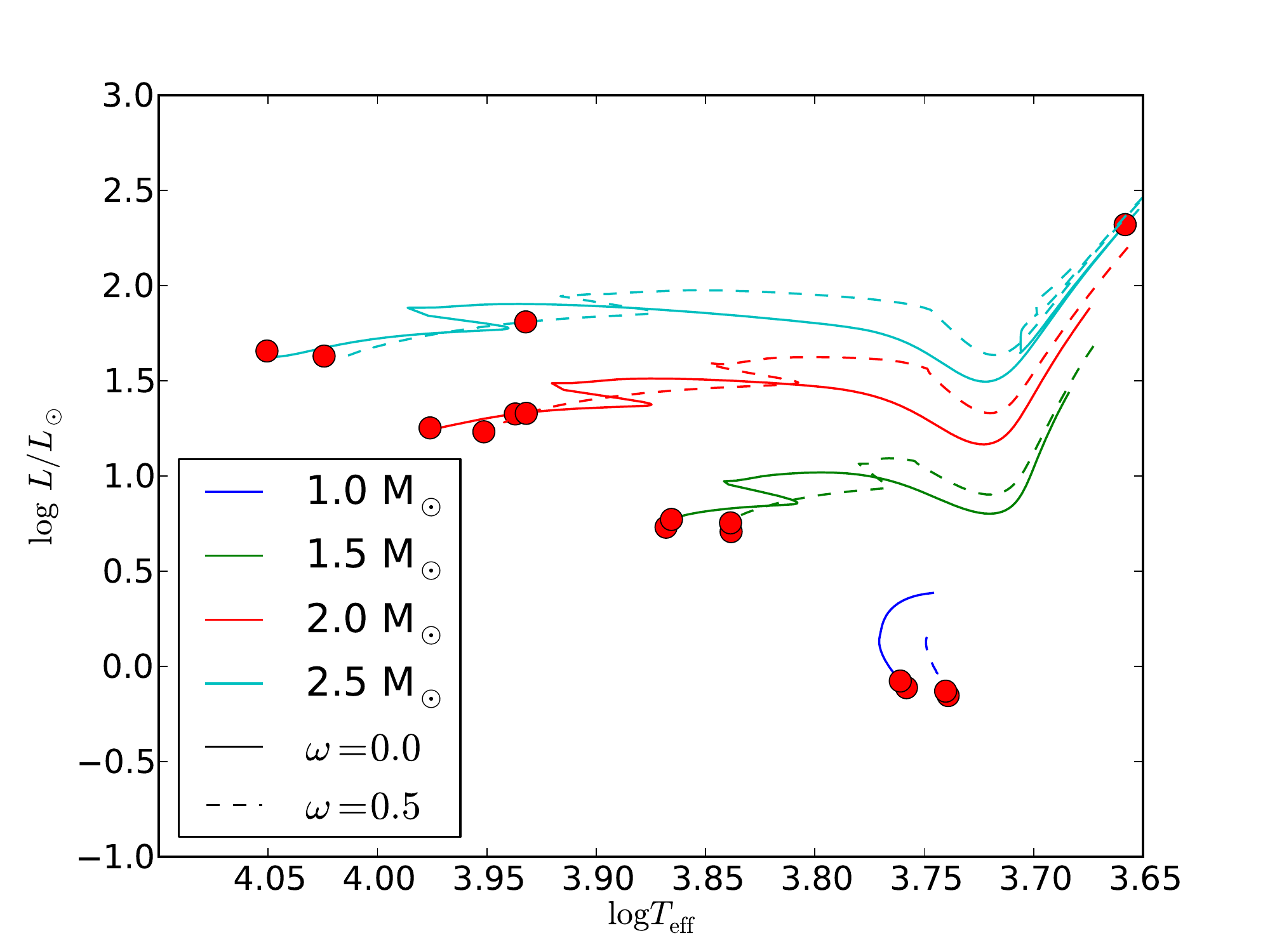}} \\
	\caption{Plot of temperature versus radius (left) and temperature versus luminosity (right) of the evolution tracks of eight stars with masses ranging from 1.0 to 2.5 M$_{\sun}$ and an angular rotation rate of either 
		0\% or 50\% that of the break-up velocity. 
	The red circles represent the properties of each star while on the zero age main sequence (at 41, 22, 9.5, and 5.7 Myr for the 1.0, 1.5, 2.0 and 2.5 M$_{\sun}$ stars, respectively for $\omega=0.0$ and 49, 26, 11, and 5.9 Myr 
		for $\omega=0.5$) and 500 Myr after that point.}
	\label{fig:HRD}
\end{figure*}

	Although a variety of independent techniques exist for estimating the ages of Sun-like stars, such as photospheric lithium depletion or chromospheric activity \citep[e.g.,][]{mamajek_2008}, these are ineffective for 
		A-type stars with predominantly radiative atmospheres. 
	For these stars, a more effective method is to compare observable stellar properties (e.g., radius, luminosity, and temperature) to the predictions of stellar evolutionary models \citep[e.g.,][]{david_2015,brandt_2015}. 
	This ``isochronal fitting" technique has the potential to work well for A-type stars since their radii, temperatures, and luminosities evolve much more substantially than Sun-like stars do during the first $\sim$Gyr of their
		main sequence lifetime. 
	For example, the MESA evolutionary models \citep{mesa1,mesa2} predict that the radius, luminosity, and temperature of a 2 M$_{\sun}$ star change by +32\%, +20\%, and $-$10\%, respectively, in just 500 Myr after 
		the zero-age main sequence.\footnote{The zero-age main sequence is defined for each star to be the point at which the contribution to the luminosity of the star due to gravitational contraction is $\sim$1\% that of 
		core fusion as predicted by the MESA evolution code.}
	This can be compared to a 1 M$_{\sun}$ star that, in the same time frame sees its radius, luminosity, and temperature change by only +3.0\%, +8.8\%, and +0.6\%, respectively. 
	Figure \ref{fig:HRD} illustrates these differences in evolutionary rates with four stars with masses between 1 and 2.5 M$_{\sun}$.
	
	Unfortunately, the peculiar properties of A-stars make these relatively straight-forward comparisons with models difficult in practice. 
	Pulsation and rapid rotation result in observable stellar properties that are both time and orientation dependent. 
	While the photometric variations due to pulsations are typically less than a few percent \citep{henry_2007}, rapid rotation distends the star so that its size is no longer defined by a single radius. 
	The resulting gravity darkening that occurs creates a temperature gradient on the star's surface \citep{von_zeipel_1924a,von_zeipel_1924b}, causing the star to no longer be defined by a single temperature. 
	The net effect is that the observed flux depends on the star's inclination, making the total luminosity challenging to determine since inclination is unknown for most stars. 
	In addition to the challenges in observationally determining an A-star's stellar properties, their peculiar characteristics must be accounted for in the adopted stellar evolution models. 
	Rapid rotation has been shown to dramatically affect the way the star evolves \citep{meynet_2000,maeder_2010}.
	For example, the MESA evolutionary code predicts that solar- to intermediate-mass stars rotating at 50\% of break-up velocity have average surface temperatures that are significantly cooler relative 
		to non-rotating stars, and evolve more slowly off the main sequence (Figure \ref{fig:HRD}).
	Finally, the anomalous surface abundances of many A-stars can complicate the choice of evolutionary model metallicity, which is usually scaled relative to solar.
	
	Fortunately, with the high angular resolution that optical/infrared interferometers provide, it is now possible to use interferometric imaging, often referred to as aperture synthesis \citep{baron_2010}, to directly determine 
		fundamental properties of rapidly rotating early-type stars \citep{vanbelle_2012,che_2011,monnier_2012}. 
	In these cases, the oblateness and gravity darkening can be observed directly, which enables more accurate determination of the star's luminosity and comparisons with evolutionary models. 
	However, there are only a handful of stars that are large enough and bright enough for this technique to work effectively with current facilities.
	
	We present a technique that allows for the correction of the effects of rotational distortion without having to fully image the star. 
	Fundamental parameters are determined by tuning a model of an observed rapidly rotating star such that the model-calculated interferometric visibilities match the observed visibilities obtained at multiple baseline orientations; 
		the model is further constrained by the star's photometric energy distribution and projected rotational velocity ($v \sin i$).
	The advantage of this technique is that it enables the determination of fundamental properties of rapidly rotating stars that are too small and/or too faint to be observed with imaging interferometric beam combiners.
	Of the 112 A-type stars within 50 parsecs that are observable with the CHARA Array (i.e., with $\delta$ $>$ $-10^{\circ}$ and with no known companions within 2$^{\prime\prime}$ and with $\Delta M_V$ $<$ 5 mag), only 13 have 
		estimated angular diameters large enough ($\theta$ $\gtrsim$ 1 milliarcsecond) to fully benefit from imaging.
	Another advantage of this technique is that it does not require the measurement of closure phases, so it is not necessary to use the many simultaneous baselines that are necessary for the aperture synthesis imaging technique.
	In this paper, we demonstrate the success of this technique by comparing the relative ages of rapidly and non-rapidly rotating stars in the Ursa Major moving group. 
	These ages are determined by comparing modeled luminosities and radii with the predictions of the MESA evolution model. 
	Indirectly, the results thus also provide a new age estimate for this moving group and tests of gravity darkening laws and stellar evolutionary models that include rotation.
\section{The Sample and CHARA Observations}
	\label{sec:sampobs}
	With a nucleus distance of 25 pc, the Ursa Major moving group is one of the closest and best-studied moving groups. 
	It consists of 15 nucleus stars and 47 likely stream members with an estimated age of 500 $\pm$ 100 Myr and a metallicity of Z=0.016 \citep{king_2003}.
	As summarized in Table \ref{tab:uma_ages}, previous studies have found an age for the moving group ranging from 200 to 1000 Myr. 
	The introduction of \cite{ammlervon_2009} provides an excellent history of the study of the UMa moving group.
	
	We define a sample of A-stars in the Ursa Major moving group for interferometric observations by selecting all stars with $B-V$ colors less than 0.31 from the ``UMa nucleus stars" list in \citet{king_2003}. 
	The hottest of these stars, has a $B-V$ color of $-$0.022 \citep{vanleeuwen_2007} and an assigned spectral type of A1 \citep{gray_2003}. 
	The resulting list consists of 7 stars of which 2 stars (Mizar A = HD 116656 and Mizar B = HD 116657) form a spectroscopic binary pair of comparable brightness ($\Delta M_V$ = 1.68 mag).
	Mizar A and B are consequently excluded from this sample because the close proximity ($\sim$4 milliarcseconds) and small $\Delta M_V$ of this pair would bias interferometric observations, making it difficult to 
		distinguish the physical properties of each star individually. 
	Another of these seven nucleus stars (Alioth = HD 112185) has a possible companion star. 
	\cite{roberts_2011} identify a companion to Alioth with a projected separation of $0.11^{\prime\prime}$ and a $\Delta M_I$ of 2.31 mag.
	A fourth of these seven stars (Alcor = HD 116842) has an observed stellar companion of spectral type M3-M4 and with a projected separation of $1.11^{\prime\prime}$ \citep{zimmerman_2010,mamajek_2010}. 
	However, with a $\Delta M_H$ of $\sim$6, the companion is too faint to contaminate the interferometric observations, so it is not excluded from the sample.	
	None of the other nucleus stars have known companions \citep{derosa_2014}.
	The four nucleus member stars that are included in this sample are Merak = HD 95418, Phecda = HD 103287, Megrez = HD 106591, and Alcor = HD 116842.
	
	There are 6 additional A-stars that are likely stream members of the moving group (listed as ``Y" or ``Y?" in \cite{king_2003}).
	Two of these 6 (Menkalinan = HD 40183 and Alphecca = HD 139006) are spectroscopic binaries with $\Delta M_V$ values of $\sim$1 and $\sim$4, respectively (\citealt{pourbaix_2000}; \citealt{tomkin_1986}) and so are not observed.
	Of the remaining four, one star (21 LMi = HD 87696) was not observed due to limited telescope time. 
	The remaining three (Chow = HD 141003, 16 Lyr = HD 177196, and 59 Dra = HD 180777) are included in the sample.
	One of these stream stars (59 Dra) has a candidate brown dwarf companion \citep{galland_2006}, but this is too faint to contaminate the interferometric observations. 
	
	In total, we obtained new interferometric observations for 6 Ursa Major A-type stars (3 nuclear members and 3 stream members).
	One additional star, Merak, was observed interferometrically by a previous study \citep{boyajian_2012}. 
	These seven stars have spectral types ranging from A0-A7. 
	Merak also has a peculiar metallicity \citep{royer_2014} and is an apparent slow rotator with a $v \sin i$ of 46 $\pm$ 2.3 $\mathrm{km~s^{-1}}$. 
	While it is possible that Merak is a rapidly rotating star oriented pole-on, there is some suggestion that the peculiar metallicity of Ap stars is due in part to their slow rotation \citep[and references therein]{abt_2009}.
	Another apparent slow rotator in the observed sample is 59 Dra with a $v \sin i$ of 70 $\pm$ 3.5 $\mathrm{km~s^{-1}}$.
	59 Dra shows a normal A-star metallicity suggesting that it may be a rapidly rotating star oriented pole-on.
	The four stars in this set that are nuclear members have distances within the very narrow range of 24.4 to 25.5 pc, while the three stream members are more spread out, having distances of 27.3, 37.4, and 47.6 pc.
	The properties of all seven stars in the set are summarized in Table \ref{tab:sample}, which includes spectral type, projected rotational velocity, \emph{Hipparcos} distance, photometry, and UMa membership as determined 
		by \cite{king_2003}.

\begin{figure*}
	\subfloat[\label{fig:HD103287_ELR_ellplot}]{\includegraphics[height =2.5in]{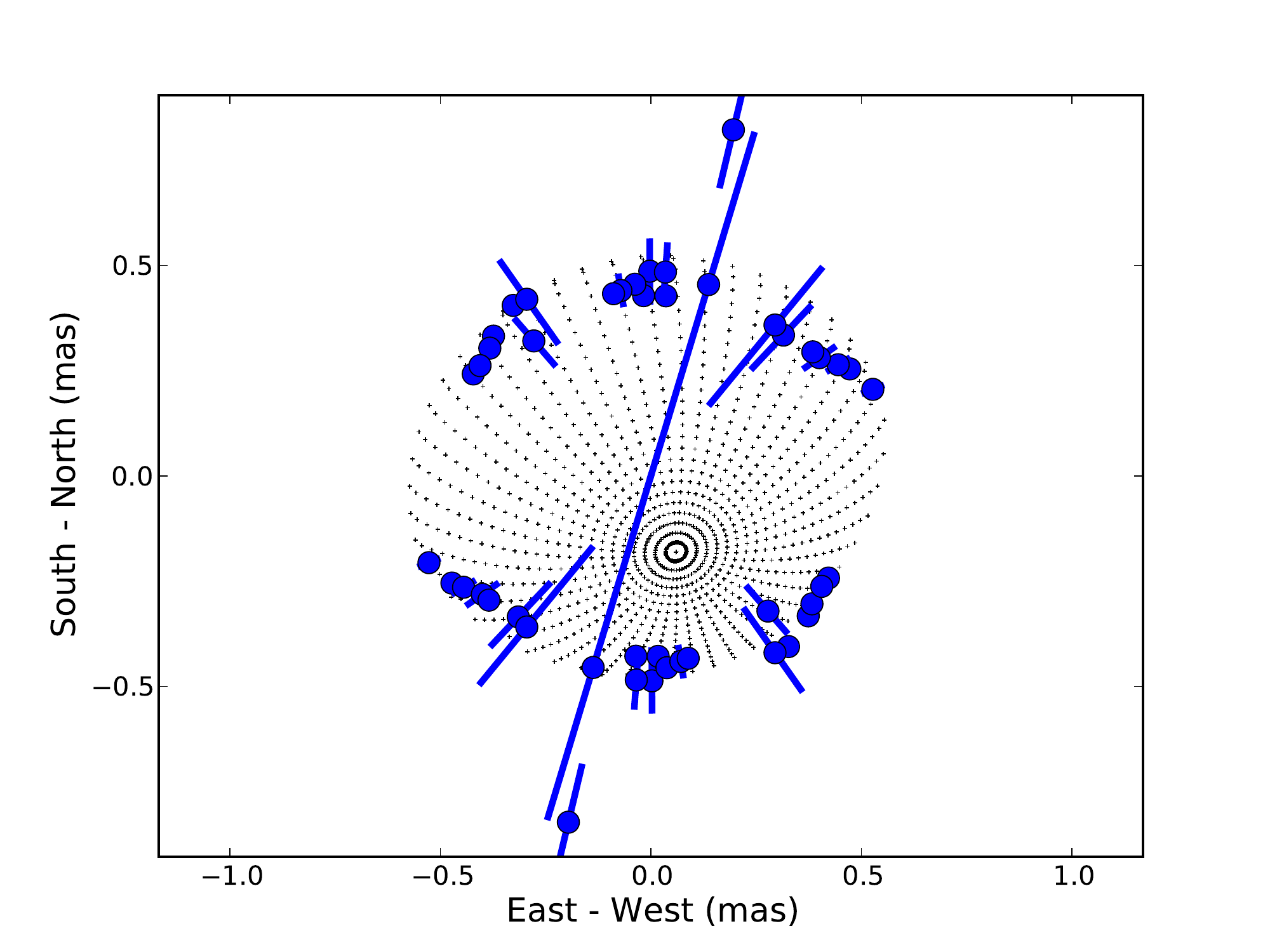}}
	\subfloat[\label{fig:HD106591_ELR_ellplot}]{\includegraphics[height =2.5in]{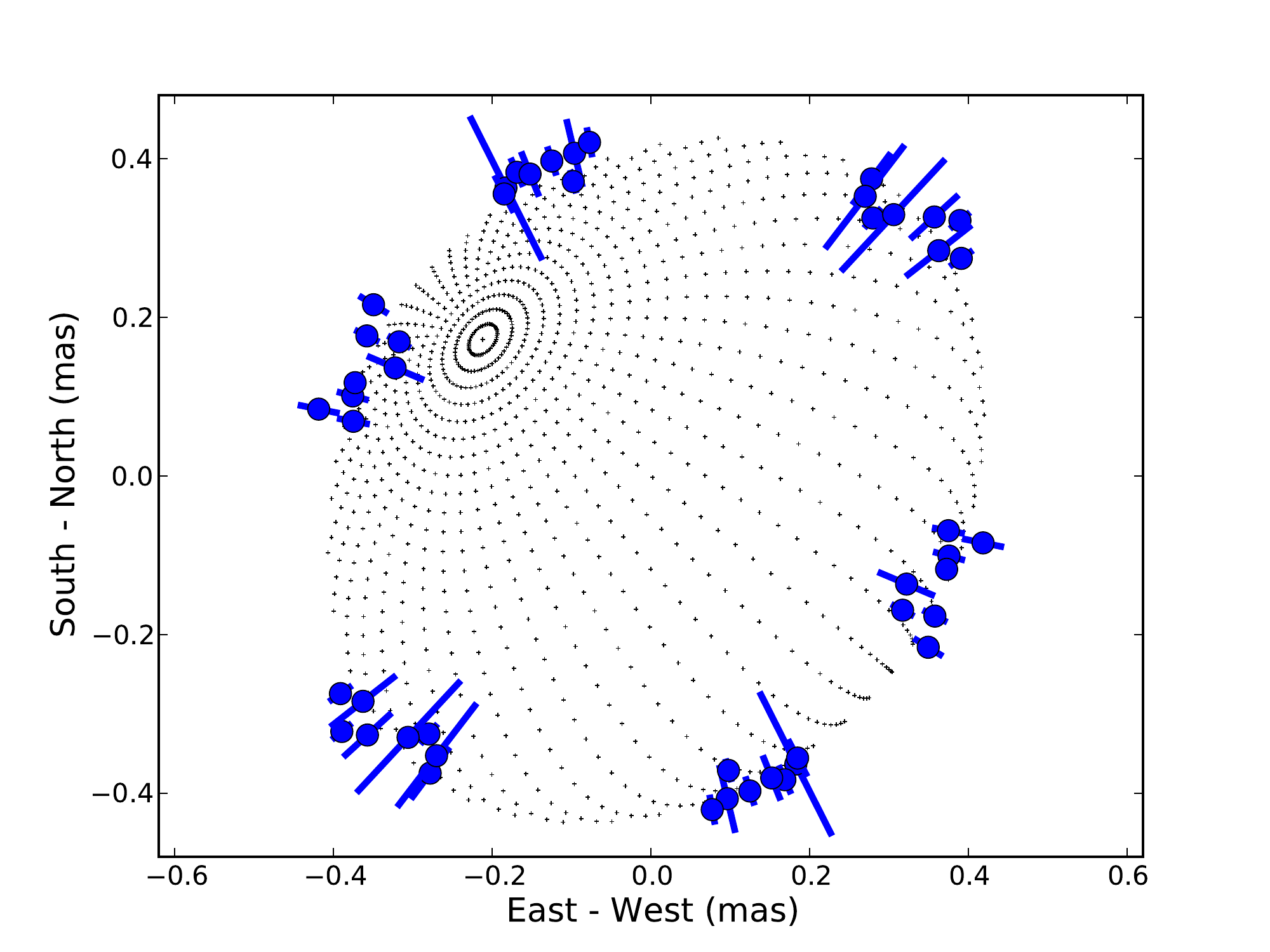}} \\
	\subfloat[\label{fig:HD116842_ELR_ellplot}]{\includegraphics[height =2.5in]{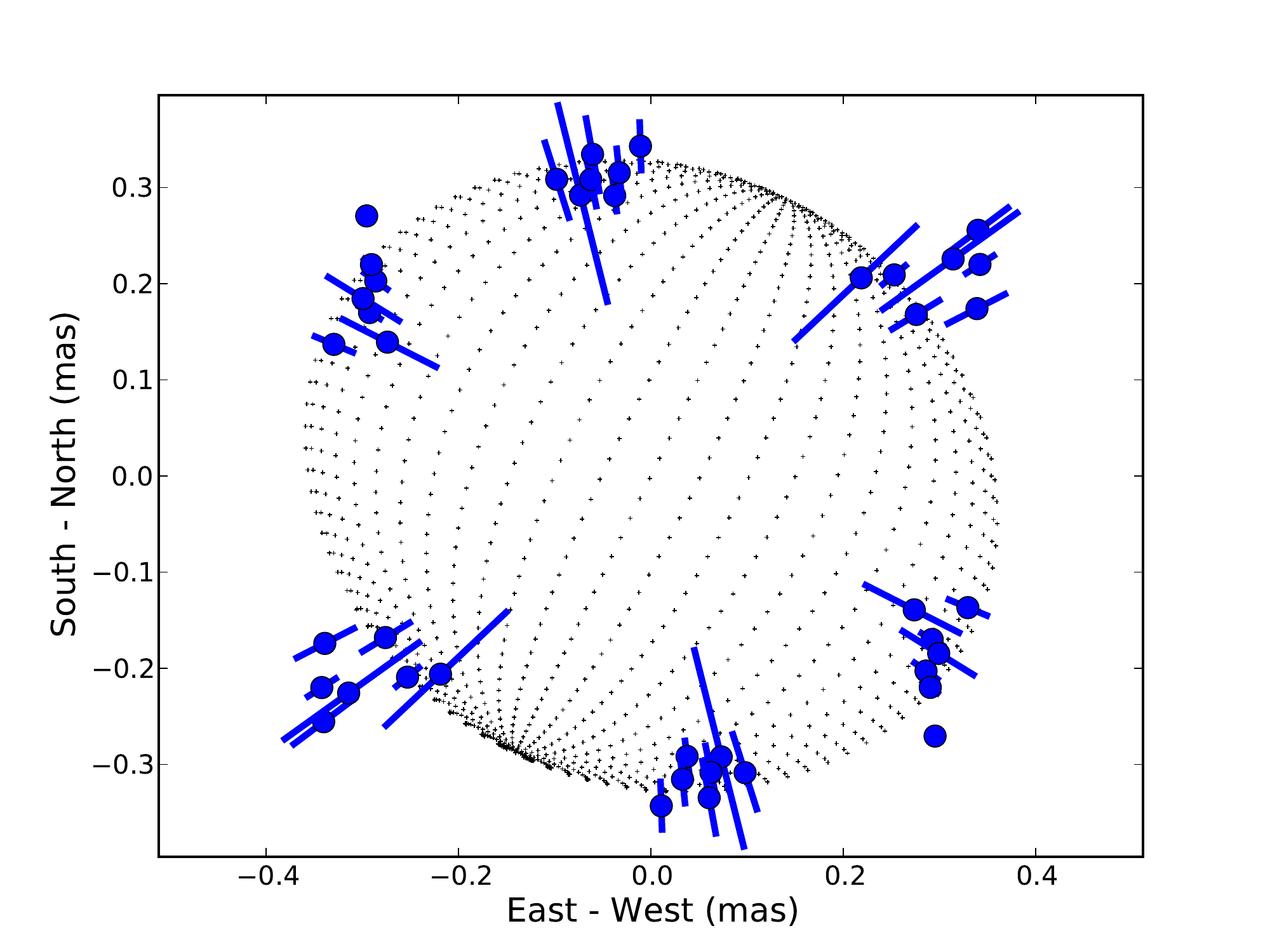}}
	\caption{The photospheres of the best fitting models for the three UMa nucleus stars modeled here: \ref{fig:HD103287_ELR_ellplot} - Phecda (HD 103287), \ref{fig:HD106591_ELR_ellplot} - Megrez (HD 106591), 
		\ref{fig:HD116842_ELR_ellplot} - Alcor (HD 116842).
		The black points represent a grid of colatitudes and longitudes on the near side of the model star.
		The blue circles represent the a uniform disk radius fitted to each individual visibility at the appropriate baseline orientation observed.
		The data are duplicated at 180$^\circ$ orientation.}
	\label{fig:ellplots1}
\end{figure*}

\begin{figure*}
	\subfloat[\label{fig:HD141003_ELR_ellplot}]{\includegraphics[height =2.5in]{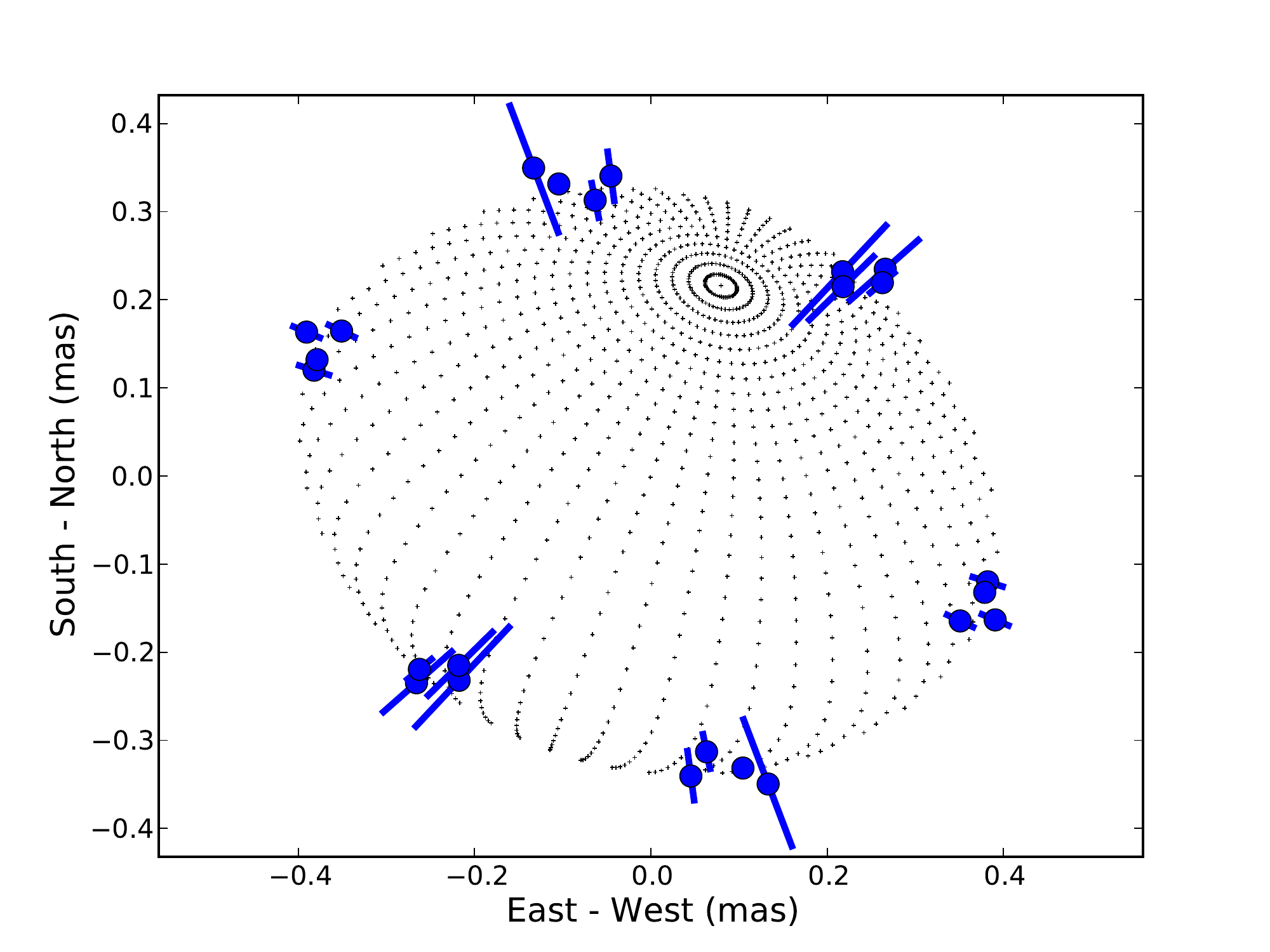}} 
	\subfloat[\label{fig:HD177196_ELR_i57_ellplot}]{\includegraphics[height =2.5in]{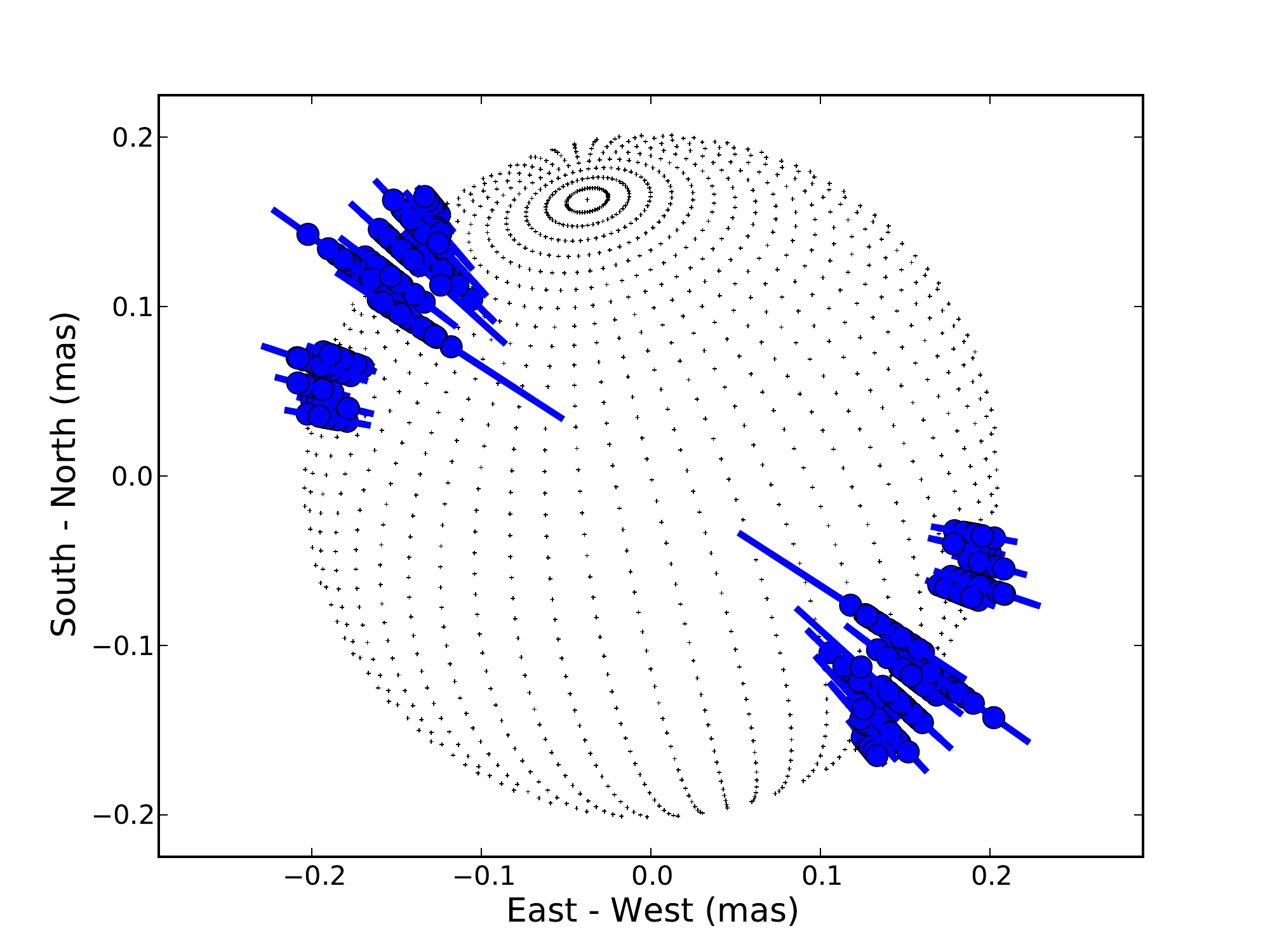}} \\
	\subfloat[\label{fig:HD180777_ELR_i28_ellplot}]{\includegraphics[height =2.5in]{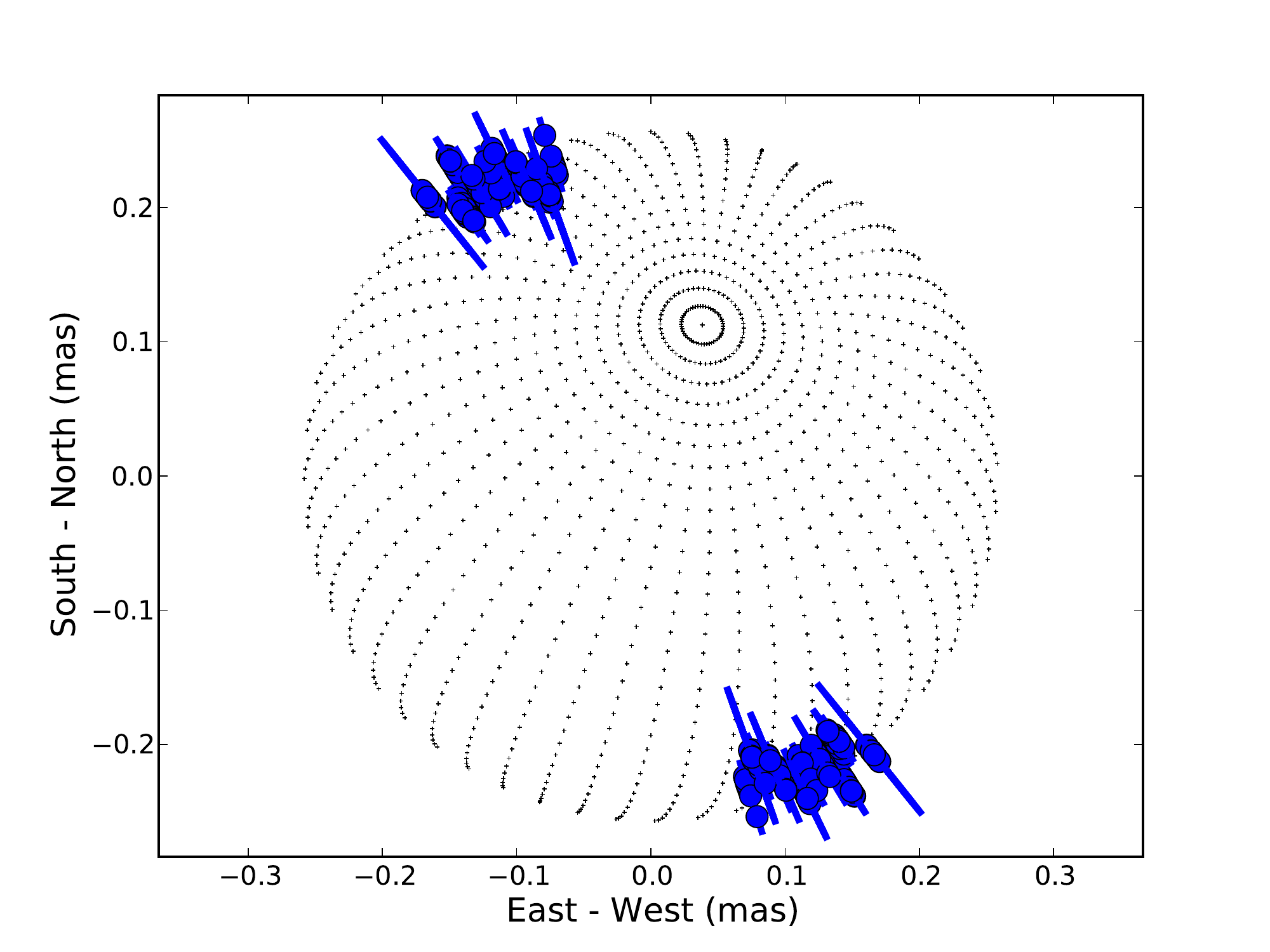}}
	\caption{Same as Figure \ref{fig:ellplots1}, but for the three UMa stream stars modeled: \ref{fig:HD141003_ELR_ellplot} - Chow (HD 141003), \ref{fig:HD177196_ELR_i57_ellplot} - 
		16 Lyr (HD 177196), \ref{fig:HD180777_ELR_i28_ellplot} - 59 Dra (HD 180777).
		The baseline orientations of 16 Lyr and 59 Dra are undersampled, making it difficult to measure their oblateness directly.}
	\label{fig:ellplots2}
\end{figure*}

	All observations were obtained using Georgia State University's Center for High Angular Resolution Astronomy (CHARA) Array. 
	The CHARA Array is a six telescope interferometer which operates at optical and near-infrared wavelengths \citep{chara2}. 
	The CHARA Array's six telescopes are arranged in a Y-shaped configuration with baselines ranging from 34-331 m. 
	The naming convention for these six telescopes consists of a letter representing one of three arms of the ``Y" (``S" for south, ``E" for east, and ``W" for west), and a number indicating the outer telescope (1) or the inner 
		telescope (2) of each arm.
	Data were obtained using three beam combiners: Classic, CLIMB, and PAVO. 
	All three beam combiners measure the contrast of the interference pattern produced by the light from each of the telescopes used. 
	This contrast is known as a visibility. 
	The two-telescope Classic beam combiner takes a single visibility measurement per observation in a broadband near-infrared filter ($K$-band for this work). 
	The three-telescope CLIMB beam combiner, which also operates in the near-infrared (in either the $H$- or $K$-band), takes three simultaneous visibility measurements for each broadband observation 
		(one for each combination of two telescopes).
	The PAVO beam combiner was used in its two-telescope mode and each observation yields 23 visibilities spectrally dispersed across a wavelengths ranging from 0.65-0.79 $\micron$.
	Because PAVO and Classic observations were taken using two telescopes at a time, only a narrow range of baseline orientations was used. 
	This is illustrated in Figures \ref{fig:ellplots1}-\ref{fig:ellplots2}.
	We note that for two stars (16 Lyr and 59 Dra), we do not have sufficient baseline orientations to measure oblateness. 
	A general observing strategy was adopted whereby calibrator stars (described in Section \ref{sec:calvis}) were observed both before and after each target star. 
	This set of observations is referred to as a visibility bracket. 
	Over 8 nights of observing, a total of 56 visibility brackets yielding 724 individual visibility measurements were obtained on 6 stars. 
	\cite{boyajian_2012} obtained 25 brackets on Merak with the two-telescope Classic beam combiner. 
	Table \ref{tab:log} lists the calibrators, beam combiners, baselines, and wavelengths used during each observation as well as how many brackets were obtained for each star.

\begin{table*}
	\begin{center}
	\caption{Age Estimates for the Ursa Major Moving Group. \label{tab:uma_ages}}
	\begin{tabular}{cc}
	\tableline\tableline
	Age 		& Reference \\
	(Myr) 		& \\
	\tableline
	$\sim$300 	& \cite{vonhoerner_1957} \\
	300$\pm$100 	& \cite{giannuzzi_1979} \\
	630-1000 	& \cite{eggen_1992} \\
	300-400 		& \cite{soderblom_1993} \\
	$\sim$500 	& \cite{asiain_1999} \\
	$\sim$200 	& \cite{konig_2002} \\
	500$\pm$100 	& \cite{king_2003} \\
	$\sim$600 	& \cite{king_2005} \\
	393\footnote{\cite{david_2015} do not report an age for the UMa moving group. The value listed here corresponds to the median of the ages they report for the 7 Ursa Majoris stars studied here (Table \ref{tab:sample}).}	& \cite{david_2015} \\
	530 $\pm$ 40 	& \cite{brandt_2015} \\
	414 $\pm$ 23 	& This work \\
	\tableline
	\end{tabular}
	\end{center} 
\end{table*}
	

\begin{table*}
	\begin{center}
	\caption{Presented Sample. \label{tab:sample}}
	\begin{tabular}{cccccccccc}
	\tableline\tableline
	Common 	& HD 		& HIP 		& Spectral 	& $v$sin$i^b$  		& $D^c$  		& $V_\mathrm{T}^d$ 	& $B-V^d$ 	& $K_S^e$ 	& UMa\\
	Name 	& Number 	& Number 	& Type$^a$ 	& (km/s) 			& (pc) 		& (mag) 			& (mag) 		& (mag) 		& Membership$^g$\\
	\tableline
	Merak 	& 95418 		& 53910 		& A1 IVps (SrII) 	& 46 $\pm$ 2.3 		& 24.4 $\pm$ 0.1 	& 2.35 			& 0.033 		& 2.285 		& Nuclear \\
	Phecda 	& 103287 	& 58001 		& A1 IV(n) 	& 178  $\pm$ 8.9 		& 25.5 $\pm$ 0.3 	& 2.43 			& 0.044 		& 2.429 		& Nuclear \\
	Megrez 	& 106591 	& 59774 		& A2 Vn 		& 233  $\pm$ 11.7 		& 24.7 $\pm$ 0.1 	& 3.34 			& 0.077 		& 3.104 		& Nuclear \\
	Alcor 	& 116842 	& 65477 		& A6 Vnn 	& 228  $\pm$ 11.4 		& 25.1 $\pm$ 0.1 	& 4.05 			& 0.169 		& 3.145 		& Nuclear \\
	Chow 	& 141003 	& 77233 		& A2 V 		& 207  $\pm$ 10.4 		& 47.6 $\pm$ 0.6 	& 3.68 			& 0.073 		& 3.546 		& Stream \\
	16 Lyr 	& 177196 	& 93408 		& A7: V 		& 124 $\pm$ 6.2 		& 37.4 $\pm$ 0.2 	& 5.07 			& 0.186 		& 4.505 		& Stream \\
	59 Dra 	& 180777 	& 94083 		& F0 Vs 		& 70 $\pm$ 3.5 $^f$ 	& 27.3 $\pm$ 0.1 	& 5.19 			& 0.308 		& 4.313 		& Stream \\
	\tableline
	\end{tabular}
	\end{center}
	\small
	Notes - (a) Nucleus Stars - \cite{gray_2003}, Stream Stars - \cite{levato_1978}; (b) \cite{royer_3}; (c) \cite{vanleeuwen_2007}; 
		(d) \cite{hipparcos}; (e) \cite{2MASS}; (f) \cite{glebocki_2005}; (g) \cite{king_2003}.
\end{table*}

\begin{table*}
	\begin{center}
	\caption{Observing Log. \label{tab:log}}
	\begin{tabular}{ccccccccc}
	\tableline\tableline
	Target Name/HD 	& Cal HD 		& Cal Diameter (mas) 	& Combiner 	& Baseline 	& Bandpass 	& \# brackets 	& \# visibilities 	& Date \\
	\tableline
	Phecda 		& 99913 		& 0.582 $\pm$ 0.058 	& Classic 		& E2-W2 		& $K$ 		& 2 		& 2 		& 4/23/2012 \\
	103287 		& 99913 		& 0.582 $\pm$ 0.058 	& CLIMB 		& S2-E2-W2 	& $K$ 		& 2 		& 6 		& 6/2/2012 \\
	 		& 105525 	& 0.392 $\pm$ 0.039 	& CLIMB 		& S1-E1-W1 	& $K$ 		& 2 		& 6 		& 5/11/2013 \\
	 		& 99913 		& 0.582 $\pm$ 0.058 	& CLIMB 		& S1-E1-W1 	& $K$ 		& 3 		& 9 		& 5/11/2013 \\
	\tableline
	Megrez 		& 108954 	& 0.451 $\pm$ 0.045 	& CLIMB 		& S1-E1-W1 	& $H$ 		& 4 		& 12 		& 4/20/2012 \\
	106591 		& 108845 	& 0.481 $\pm$ 0.048 	& CLIMB 		& S1-E1-W1 	& $H$ 		& 2 		& 6 		& 4/21/2012 \\
	 		& 108954 	& 0.451 $\pm$ 0.045 	& CLIMB 		& S1-E1-W1 	& $H$ 		& 2 		& 6 		& 4/21/2012 \\
	\tableline
	Alcor 		& 119024 	& 0.306 $\pm$ 0.031 	& CLIMB 		& S1-E1-W1 	& $H$ 		& 4 		& 12 		& 4/20/2012 \\
	116842 		& 108954 	& 0.451 $\pm$ 0.045 	& CLIMB 		& S1-E1-W1 	& $H$ 		& 1 		& 3 		& 4/21/2012 \\
	 		& 118232 	& 0.465 $\pm$ 0.047 	& CLIMB 		& S1-E1-W1 	& $H$ 		& 2 		& 6 		& 4/21/2012 \\
	\tableline
	Chow 		& 140160 	& 0.293 $\pm$ 0.029 	& CLIMB 		& S1-E1-W1 	& $H$ 		& 2 		& 6 		& 4/21/2012 \\
	141003 		& 137510 	& 0.525 $\pm$ 0.053 	& CLIMB 		& S1-E1-W1 	& $H$ 		& 2 		& 6 		& 4/21/2012 \\
	\tableline
	16 Lyr 		& 177003 	& 0.156 $\pm$ 0.016 	& PAVO 		& S2-E2 		& $R$ 		& 3 		& 69 		& 7/10/2012 \\
	177196 		& 172883 	& 0.181 $\pm$ 0.018 	& PAVO 		& S2-E2 		& $R$ 		& 2 		& 46 		& 7/10/2012 \\
	 		& 177003 	& 0.156 $\pm$ 0.016 	& PAVO 		& E2-W2 		& $R$ 		& 3 		& 69 		& 8/4/2013 \\
	 		& 185872 	& 0.256 $\pm$ 0.026 	& PAVO 		& E2-W2 		& $R$ 		& 3 		& 69 		& 8/4/2013 \\
	 		& 177003 	& 0.156 $\pm$ 0.016 	& PAVO 		& E1-W2 		& $R$ 		& 3 		& 69 		& 8/5/2013 \\
	 		& 185872 	& 0.256 $\pm$ 0.026 	& PAVO 		& E1-W1 		& $R$ 		& 2 		& 46 		& 8/5/2013 \\
	\tableline
	59 Dra 		& 184102 	& 0.263 $\pm$ 0.026 	& PAVO 		& S2-E2 		& $R$ 		& 3 		& 69 		& 7/10/2012 \\
	180777 		& 201908 	& 0.187 $\pm$ 0.019 	& PAVO 		& S2-E2 		& $R$ 		& 3 		& 69 		& 7/10/2012 \\
	 		& 184102 	& 0.263 $\pm$ 0.026 	& PAVO 		& E2-W2 		& $R$ 		& 3 		& 69 		& 8/4/2013 \\
	 		& 201908 	& 0.187 $\pm$ 0.019 	& PAVO 		& E2-W2 		& $R$ 		& 3 		& 69 		& 8/4/2013 \\
	 \tableline\tableline
	\end{tabular}
	\end{center}
\end{table*}

\section{Data Reduction and Calibrated Visibilities}
	\label{sec:calvis}
	Interferometric data from the Classic and CLIMB beam combiners were reduced using the \emph{redclassic} and \emph{redclimb} pipelines, respectively \citep{classic_climb_pipeline}, yielding reduced visibilities for each observation 
		made.
	The pipeline used to reduce the observations made with the PAVO beam combiner is described by \cite{pavo}.
	Many factors, both atmospheric and instrumental, serve to decrease the visibility measured by an interferometer. 
	This decrease depends in part on atmospheric turbulence at the time of observation and the airmass at which the star is observed \citep[e.g.,][]{boden_2007,roddier_1981}. 
	Correcting for these temporal effects on the visibility requires frequent observation of a star with a known angular diameter that is ideally smaller than the interferometric resolution ($\lambda / 2B$).
	Such a star is called a calibrator star. 
	When observed near the target star both in time ($\lesssim$ 30 minutes) and on the sky ($\lesssim$ 10$^\circ$), the target star's intrinsic visibility (V$\mathrm{^*_i}$) should be observed (V$\mathrm{^*_m}$) to be reduced by the same 
		amount as the calibrator's (intrinsic - V$\mathrm{^c_i}$, measured - V$\mathrm{^c_m}$):
	\begin{equation}
		\frac{V\mathrm{^*_i}}{V\mathrm{^c_i}}=\frac{V\mathrm{^*_m}}{V\mathrm{^c_m}}
		\label{cal}
	\end{equation}
	
	A common method for estimating a calibrator star's size (if it is not known from previous interferometric measurements) is by fitting a photometric energy distribution (PED) to measured photometry (see Appendix \ref{app:phot}).
	\cite{tabby_thesis} found an average difference between angular sizes determined by PED fitting and angular sizes measured by interferometry to be $\sim$10\%, so a 10\% error in the angular size is adopted for the calibrator stars 
		observed for this work.
	Small calibrator stars are used because the smaller a star is, the less its estimated intrinsic visibility is affected by inaccuracies in its size estimate. 
	For example, a small calibrator with a 10\% error (angular diameter, $\theta = 0.2 \pm 0.02$ mas) observed with the CHARA Array's longest baseline ($B=331$ m) in the $K$-band will have an estimated intrinsic 
		visibility of $0.974 \pm 0.005$ (a 0.5\% error due to the inaccuracy of an PED-determined size).
	A calibrator that is twice as large ($\theta = 0.4 \pm 0.04$ mas) and observed in the same way will have an estimated intrinsic visibility of $0.90 \pm 0.02$ (a 2.2\% error due to the inaccuracy of an PED-determined size).
	As a rule of thumb, good calibrators are ones that are smaller than approximately half the resolution of the observation to avoid significant errors in the calibrator's visibility \citep{vanbelle_2005}. 
	
	For this work, at least two calibrator stars were observed for each target star to help mitigate calibrator size errors. 
	Their angular diameters are estimated by fitting PHOENIX model PEDs \citep{phoenix} to photometry gathered from the literature.
	Three of the calibrators used here (HD 177003, HD 185872, and HD 201908) had temperatures greater than the PHOENIX model grid (which goes as high as 12000 K). 
	For these three calibrators, the PED fits were made using ATLAS9 model PEDs \citep{castelli_2004}.
	Calibrator angular diameters are listed in Table \ref{tab:log} and range from 0.156 to 0.582 mas.
	
\section{Fundamental Stellar Properties}
	\label{sec:mod}
	\subsection{Oblate Star Model}
		\label{sec:omod}
		The limb-darkened disk model traditionally used to analyze interferometric visibilities takes neither the distended shape of rapidly rotating stars nor the gravity darkening caused by this distended shape into account.  
		The model used here employs a Roche geometry and is based on the models used in \cite{vanbelle_2012}, \cite{aufdenberg_2006}, and \cite{monnier_2012}. 
		In order to determine the fundamental properties of rapid rotators, the observed visibilities (Section \ref{sec:calvis}) and broadband photometry are compared to model-predicted visibilities and photometry; 
			the adopted photometry for each star is assembled in Appendix \ref{app:phot}. 
		The eight input parameters for the model star are its equatorial radius ($R_\mathrm{e}$), its mass ($M_\mathrm{*}$), its equatorial rotational velocity ($V_\mathrm{e}$), the inclination of its polar axis relative to our line-of-sight ($i$), 
			the gravity darkening coefficient used in the model ($\beta$), the temperature at its pole ($T_\mathrm{p}$), the parallax of the observed star ($\pi_\mathrm{plx}$), and the position angle of its pole ($\psi$) with a 180$^{\circ}$ 
			ambiguity.
		Of these, the parallax is set by \emph{Hipparcos} measurements, the gravity darkening coefficient is set by one of two possible relations (see below), and the mass is estimated from evolution models (see below). 
		The remaining five parameters ($R_\mathrm{e}$, $V_\mathrm{e}$, $i$, $T_\mathrm{p}$, and $\psi$) are allowed to vary under the constraint that the equatorial velocity ($V_\mathrm{e}$) must yield a 
			model $v \sin i$ that is consistent with the observed $v \sin i$.
		
		Two gravity darkening laws are incorporated here. 
		With the canonical gravity darkening law \citep[hereafter, vZ]{von_zeipel_1924a,von_zeipel_1924b,claret_2000}, the stars modeled here are hot enough to have fully radiative envelopes, giving them 
			a gravity darkening coefficient, $\beta$, of 0.25. 
		However, a modern gravity darkening law, tested with results from interferometric observations of rapidly rotating stars \citep[hereafter ELR]{ELR_2011} shows that $\beta$ is dependent on 
			the angular rotation rate, $\omega$, and ranges from 0.25 for a non-rotating star ($\omega = 0$) to $\sim$0.09 for a star rotating at its breakup velocity ($\omega = 1$). 
		
		The oblateness of a star depends not only on its rotation, but also its mass.
		After the best fitting free parameters are determined, the age and mass are calculated using evolution models.
		The mass used in the oblate star model is then updated to match the mass determined by the evolution model.
		The oblate star model and evolutionary model are run iteratively until neither the mass nor the free parameters change by more than $\sim$0.1\% after a series of consecutive runs, corresponding to 
			an $R_\mathrm{e}$ of $\sim$0.002 R$_{\sun}$, a $V_\mathrm{e}$ of $\sim$0.2 km s$^{-1}$, an $i$ of $\sim$0.1$^{\circ}$, a $T_\mathrm{p}$ of $\sim$8 K, and a $\psi$ of $\sim$0.4$^{\circ}$.

		To determine the ages and masses of the rapidly rotating stars in this paper the star's average radius ($R\mathrm{_{avg}}$), total luminosity ($L\mathrm{_{tot}}$), and equatorial velocity ($V\mathrm{_e}$), 
			as determined by the oblate star model are compared to the predictions of MESA evolutionary models \citep{mesa1,mesa2}. 
		These three parameters ($R\mathrm{_{avg}}$, $L\mathrm{_{tot}}$, and $V\mathrm{_e}$) correspond to a star with a unique mass, age and angular velocity.
		The mass used by the oblate star model is set equal to the mass determined by this comparison in the iterative process described above. 
		For this project, MESA evolution tracks\footnote{See the supplemental material for examples of MESA inlists used in this project.} were computed for a grid of masses and angular velocities (with resolution of 
			0.1 M$_\sun$ and 10\% breakup velocity, respectively) at a metallicity of Z = 0.016 as measured by \cite{king_2003} for the UMa moving group.
		
		The stellar model is constructed by calculating the stellar intensity at each point on an oblate spheroidal grid, constructed of 51 points along the colatitudinal axis ($\vartheta$) and 
		51 points along the longitudinal axis ($\varphi$) for a total of 2601 points on the star. 
		Then, a radius ($R(\vartheta)$) and surface gravity ($g(\vartheta)$, with radial component, $g_\mathrm{r}(\vartheta)$ and polar component, $g_\mathrm{\vartheta}(\vartheta)$) are calculated for each point on the 
			grid \citep{vanbelle_2012}: 
		\begin{equation}
			R(\vartheta)=3 \frac{R_\mathrm{p}}{\omega \sin \vartheta} \cos [\frac{\pi + \arccos (\omega \sin \vartheta)}{3}]
		\end{equation}
		\begin{equation}
		\begin{split}
			g(\vartheta)=\sqrt{g_\mathrm{r}(\vartheta)^2+g_{\mathrm{\vartheta}}(\vartheta)^2} \mathrm{, where} \\ 
			g_\mathrm{r}(\vartheta)=\frac{-GM_\mathrm{*}}{R(\vartheta)^2}+R(\vartheta)(\Omega \sin \vartheta)^2 \\
			g_\mathrm{\vartheta}(\vartheta)=R(\vartheta)\Omega^2 \sin \vartheta \cos \vartheta.
		\end{split}
		\end{equation}
		In this prescription, $R_\mathrm{p}$ is the model star's polar radius: 
		\begin{equation}
			R_\mathrm{p}=[\frac{1}{R_\mathrm{e}} +  \frac{V_\mathrm{e}^2}{2GM_\mathrm{*}}]^{-1},
		\end{equation}
		$\omega$ is the angular velocity of the star relative to its critical velocity, $\Omega_\mathrm{crit}$: 
		\begin{equation}
		\begin{split}
			\omega=\sqrt{\frac{27}{4} w_0 (1-w_0)^2} \\
			w_0=\frac{V_\mathrm{e}^2 R_\mathrm{p}}{2GM_\mathrm{*}},
		\end{split}
		\end{equation}
		and $\Omega$ is the angular velocity of the star in radians per second: 
		\begin{equation}
			\Omega=\omega \Omega_\mathrm{crit}=\omega(\frac{8}{27}\frac{GM_\mathrm{*}}{R^3_\mathrm{p}})^{1/2}.
		\end{equation}
		This allows the gravity dependent surface temperature ($T(\vartheta)$) to be calculated at each point on the grid:
		\begin{equation}
			\label{eqn:temp}
			T=T_\mathrm{p} \left(\frac{g(\vartheta)}{g_\mathrm{p}}\right)^\beta
		\end{equation}
		where $g_\mathrm{p}$ is the surface gravity at the model star's pole: 
		\begin{equation}
			g_\mathrm{p}=\frac{GM_\mathrm{*}}{R_\mathrm{p}^2}.
		\end{equation}
		A grid\footnote{Grid step sizes are 0.5 in $\log{g}$ and 200 K in $T_{eff}$} of PHOENIX atmosphere models \citep{phoenix} are interpolated to determine the intensity spectrum at each point on the stellar model 
			surface grid based on the temperature and surface gravity of those points.
		
		Model photometry is calculated by integrating the 2601 intensity spectra that cover the star to compute the flux spectrum of the star, $F_\mathrm{\lambda}$: 
		\begin{equation}
			F_\mathrm{\lambda}=\int_{\vartheta=0}^{\pi} \! \int_{\varphi=0}^{2 \pi} \! I_\mathrm{\lambda}(\vartheta,\varphi) \theta_\mathrm{R}^2(\vartheta) \sin(\vartheta)\, \mu(\vartheta,\varphi) \, \mathrm{d}\varphi \, \mathrm{d}\vartheta
		\end{equation}
		$I_\mathrm{\lambda}(\vartheta,\varphi)$ is the intensity spectrum given by the PHOENIX model. $\theta_\mathrm{R}(\vartheta)$ is the angular radius of the model star as a function of colatitude.
		$\mu(\vartheta,\varphi)$ is the cosine of the angle between the observer and the normal of the star:
		\begin{equation}
		\begin{split}
			\mu(\vartheta,\varphi)=\frac{1}{g(\vartheta)}[-g_\mathrm{r}(\vartheta)(\sin(\vartheta) \sin(i) \cos(\varphi) + \cos(\vartheta) \cos(i))- \\
			g_\mathrm{\vartheta}(\vartheta)(\sin(i) \cos(\varphi) \cos(\vartheta)- \sin(\vartheta) \cos(i))].
		\end{split}
		\end{equation}
		Note that $I_\mathrm{\lambda}(\vartheta,\varphi) = 0$ for $\mu < 0$ (i.e., only light directed at the observer is included in the integration). 
		The resulting flux spectrum is convolved with the appropriate bandpass filter to compute the specific flux from which the photometry is calculated.
		
		The bolometric flux is simply $F_\mathrm{bol}=\int \! F_\mathrm{\lambda} \, \mathrm{d}\lambda$ and the apparent luminosity is then $L_\mathrm{app}=4 \pi F_\mathrm{bol} d^2$. 
		The total luminosity, $L_\mathrm{tot}$, is calculated by determining $J_\mathrm{\lambda}$, the specific irradiance on each point
		\begin{equation}
			J_\mathrm{\lambda}(T_\mathrm{eff},g)=\int_{\mu=0}^1I_\mathrm{\lambda}(T_\mathrm{eff}, g,\mu) \mu \: \mathrm{d}\mu
		\end{equation}
		integrating over all wavelengths:
		\begin{equation}
			J_\mathrm{bol}(\vartheta)=2 \pi \int_{\lambda} J_\mathrm{\lambda} (T_\mathrm{eff}, g) \: \mathrm{d}\lambda
		\end{equation}
		and integrating over the model star's surface: 
		\begin{equation}
			L_\mathrm{tot}=2 \pi \int_{\vartheta=0}^\pi J_\mathrm{bol}(\vartheta) R^2(\vartheta) sin(\vartheta) \: \mathrm{d}\vartheta
		\end{equation}
		
		Model visibilities are calculated by first creating an image of the model star in the bandpass of the observations. 
		For example, if the visibilities are observed in $H$-band, the intensity spectra at the different points in the image are convolved with an $H$-band filter. 
		A 2D fast Fourier transform (FFT) is taken of that synthetic image. 
		This image is 4900$\times$4900 pixels with $\sim$1000 of those pixels (in the center of the image) being made up of synthetic starlight.
		This distribution is designed to produce an image that is high enough resolution to detect the oblateness and for the FFT to extract accurate visibilities.
		The model squared visibility is the complex square of that transform at the observed $u$ and $v$ spatial frequencies and the model visibilities are the square root of that quantity.

		The above prescription yields visibilities and photometry based on a model star that can be tuned to match the observations.
		A random search algorithm is employed to find the set of free-parameters ($R_\mathrm{e}$, $V_\mathrm{e}$, $i$, $T_\mathrm{p}$, and $\psi$) that minimizes the difference between observed and model predictions. 
		For each set of input parameters, a reduced $\chi^2$ goodness-of-fit metric is calculated with five degrees of freedom for both the visibilities and the photometry. 
		The final $\chi^2$ (hereafter, $\chi^2_\mathrm{tot}$) is then calculated by adding the $\chi^2$ values of the visibility data and those of the photometry, assuming equal weight for the two. 
		The search algorithm randomly selects a set of parameters within a given window of parameter space. 
		The initial window size for the set of parameters, ($R_\mathrm{e}$, $V_\mathrm{e}$, $i$, $T_\mathrm{p}$, $\psi$)
			is ($\pm 0.5$ R$_{\sun}$, $\pm \frac{\sigma_{v\mathrm{sin}i}}{\mathrm{sin}(i)}$ $\mathrm{km~s^{-1}}$, $\pm 20^{\circ}$, $\pm 500$ K, $\pm 30^{\circ}$), and this 
			search area is decreased over multiple steps, eventually reaching ($\pm 0.01$ R$_{\sun}$, $\pm 1$ $\mathrm{km~s^{-1}}$, $\pm 1^{\circ}$, $\pm 1$ K, $\pm 1^{\circ}$).
		This window is initially centered on the initial guess parameters, but it is re-centered whenever a model with a smaller $\chi^2_\mathrm{tot}$ is calculated.
		The best fitting model is determined by minimizing the $\chi^2_\mathrm{tot}$ after multiple iterations.
		The error for each of the five free-parameters is found by first varying them independently until the $\chi^2_\mathrm{tot}$ increases by 1 after first scaling the $\chi^2_\mathrm{tot}$ such that the 
			minimum $\chi^2_\mathrm{tot}=1$.
		
		Due to the large scatter in the broad-band photometric measurements relative to their error, the best fitting model finds an unscaled $\chi^2_\mathrm{tot}$ of  $\gtrsim$ 100 (dominated by the photometric 
			$\chi^2$) when adopting the published errors for the photometry measurements, the mean and median of which are 0.016 and 0.011 mag, respectively.
		More importantly, few of the photometric measurements overlapped with the model PED which could indicate underestimates of the photometric error, inaccuracies of the synthetic spectral energy distribution, 
			incorrect filter profiles or zero-points, etc.
		To account for this, photometric errors of 0.03 mag were adopted for all photometric values which had an error less than 0.03 mag.
		With these adopted photometric errors, all of the best fitting models had an unscaled $\chi^2_\mathrm{tot}$ of $<$ 15. 

\begin{figure*}
	\subfloat[\label{fig:HD103287_ELR_vis}]{\includegraphics[height =2.5in]{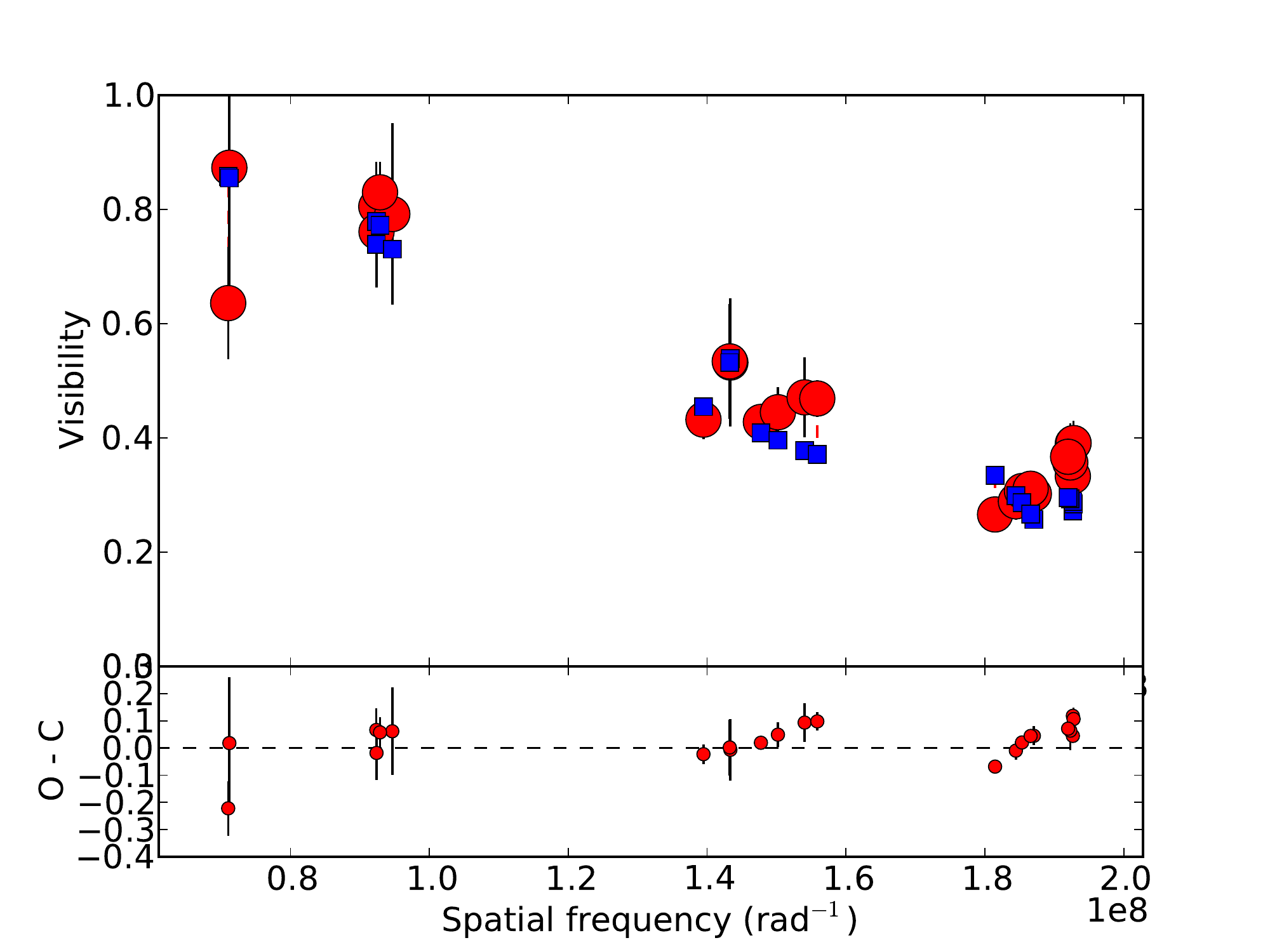}}
	\subfloat[\label{fig:HD103287_ELR_phot}]{\includegraphics[height =2.5in]{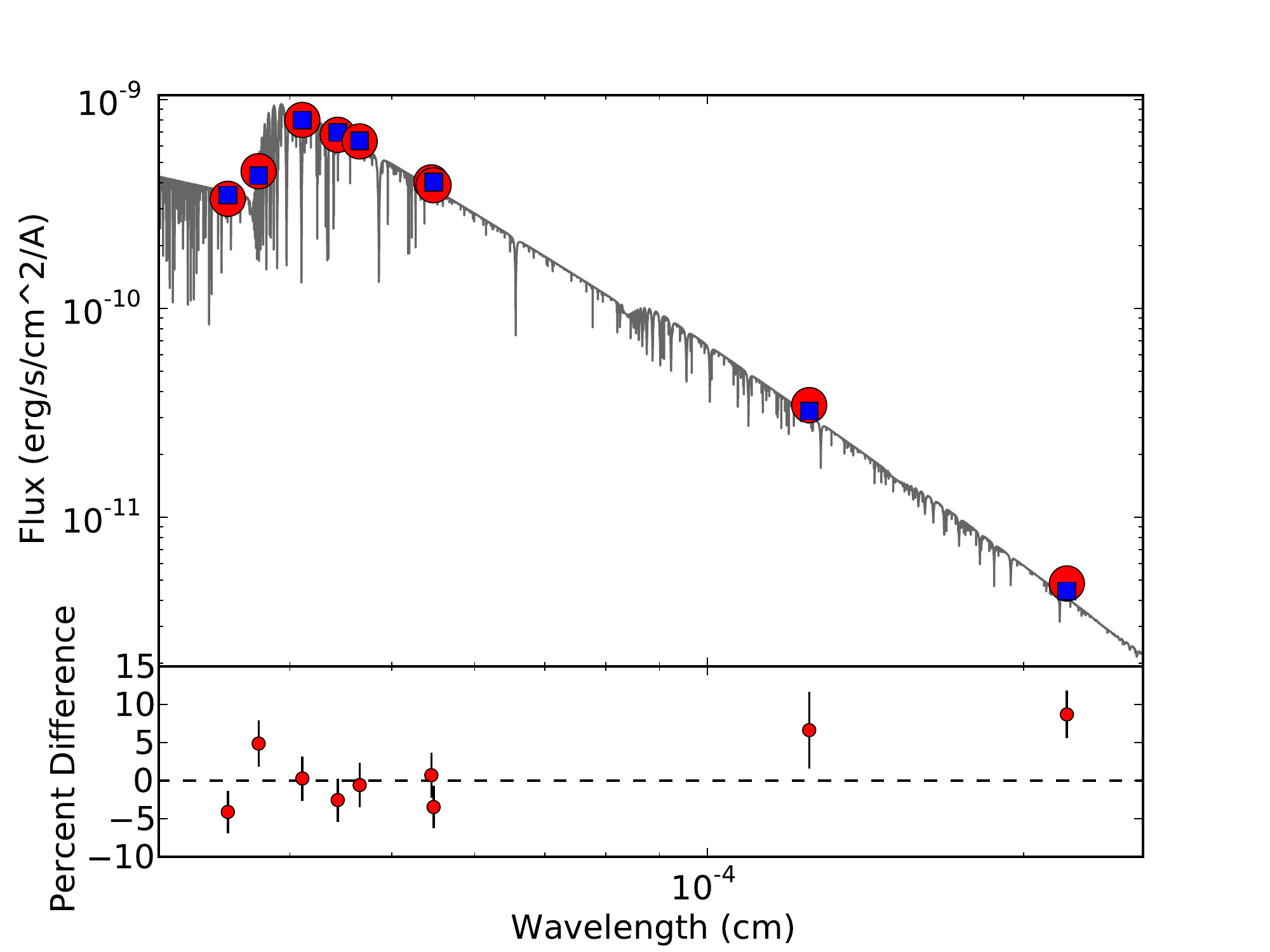}} \\
	\subfloat[\label{fig:HD103287_vZ_vis}]{\includegraphics[height =2.5in]{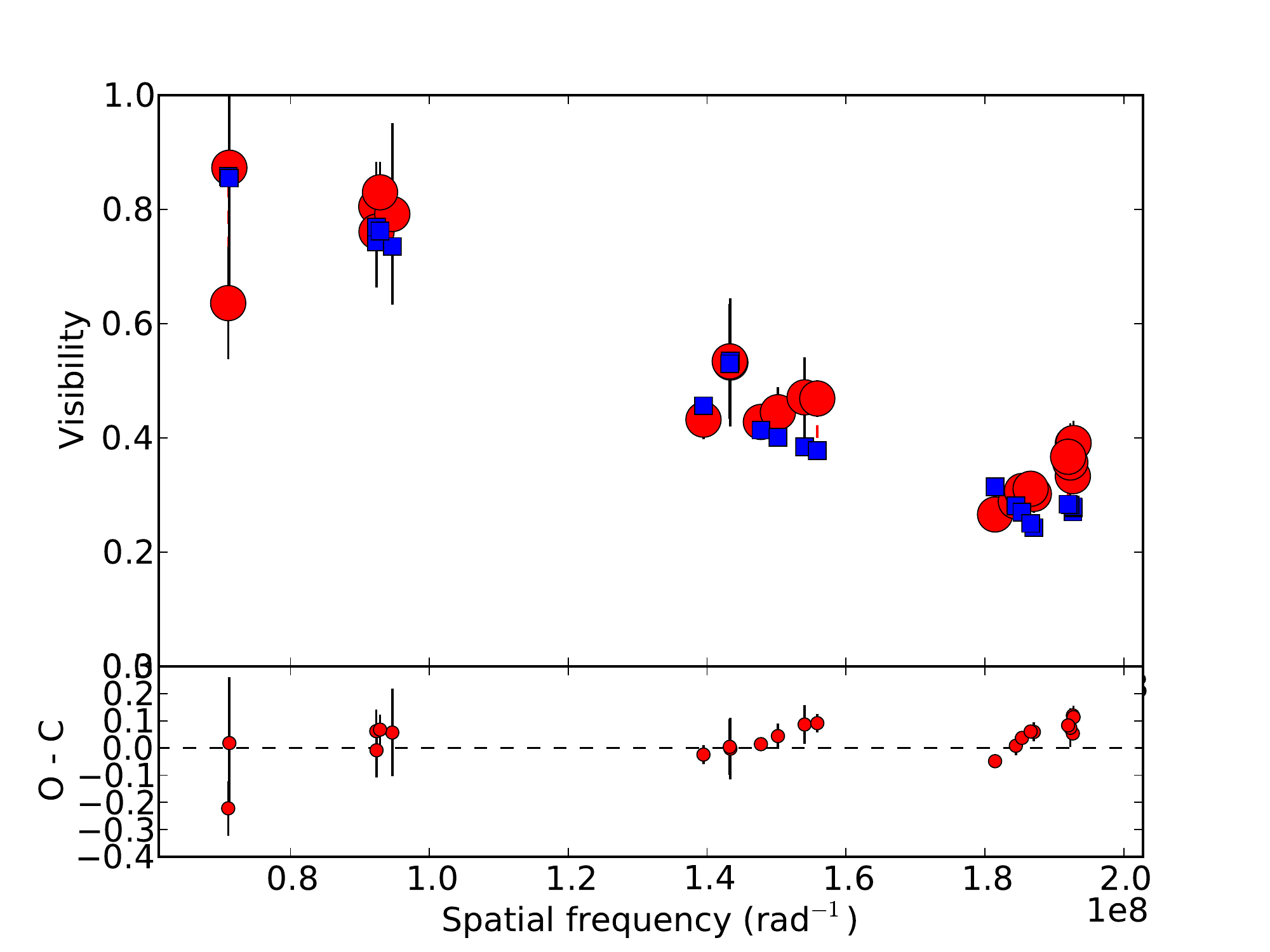}}
	\subfloat[\label{fig:HD103287_vZ_phot}]{\includegraphics[height =2.5in]{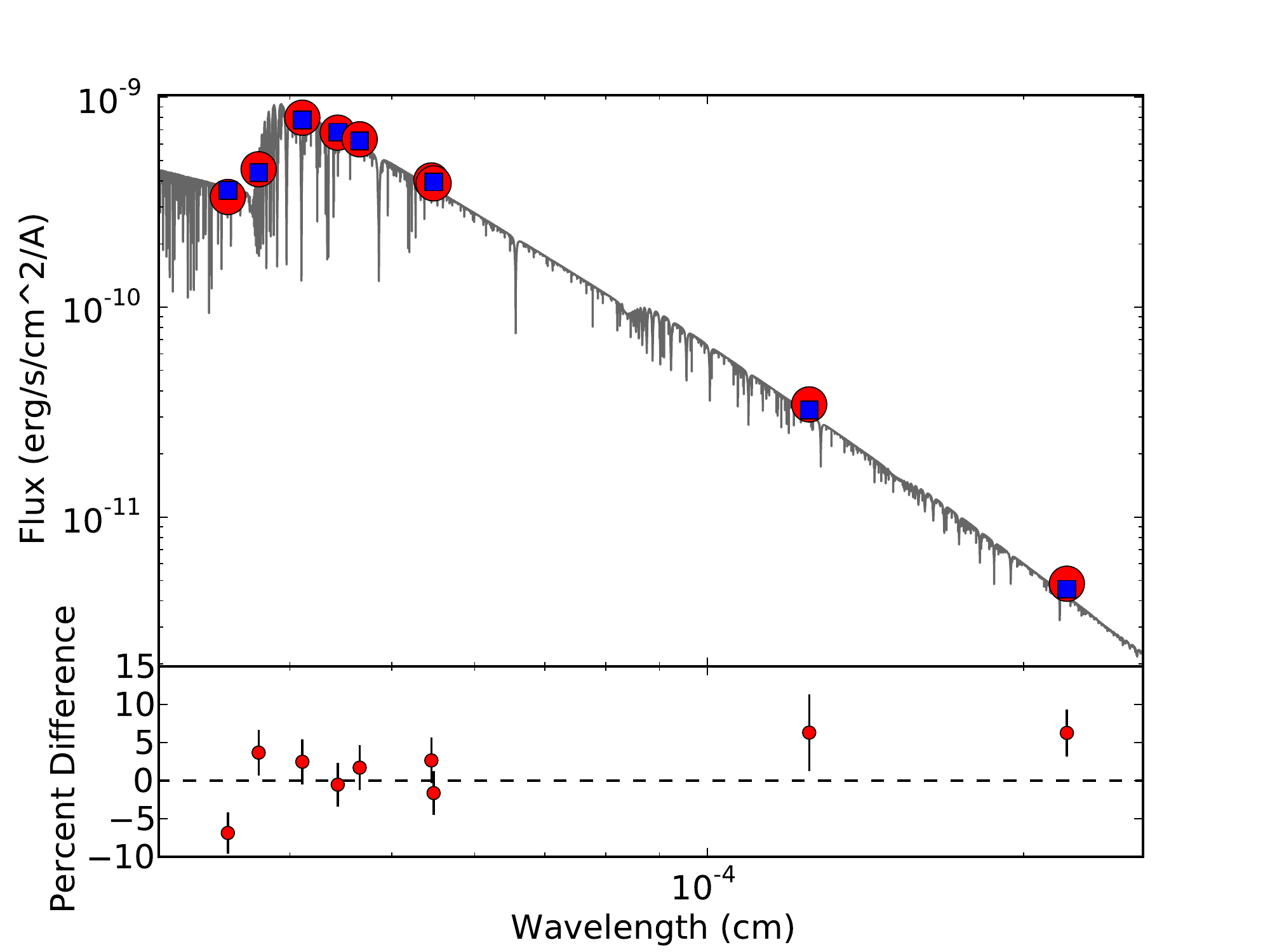}}
	\caption{Top Left - Visibility measurements (red circles) for Phecda (HD 103287) are compared to the best fit model visibilities (blue squares) assuming the ELR prescription for gravity darkening. 
			Dashed lines connect individual model and measured values and solid lines are the error bars.
		Top Right - Photometric measurements (red circles) for Phecda (HD 103287) are compared to the best fit model photometry (blue squares) assuming the ELR prescription for gravity darkening. 
			The spectral energy distribution from which the PED is calculated is plotted in grey for comparison.
		Bottom Left - Same as Top Left, but for the vZ gravity darkening law.
		Bottom Right - Same as Top Right, but for the vZ gravity darkening law.}
	\label{fig:HD103287_plots}
\end{figure*}

\begin{figure*}
	\subfloat[\label{fig:HD106591_ELR_vis}]{\includegraphics[height =2.5in]{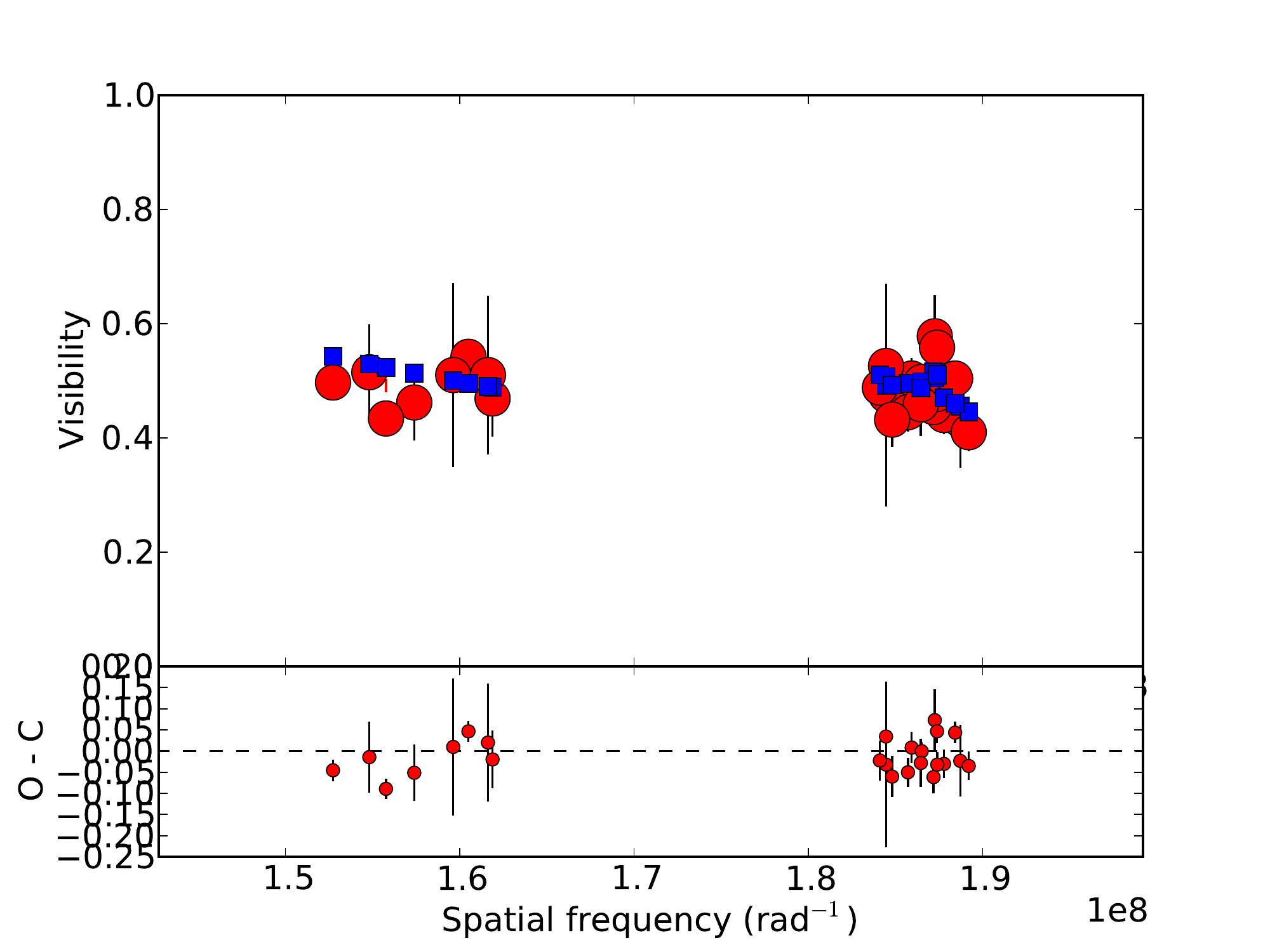}}
	\subfloat[\label{fig:HD106591_ELR_phot}]{\includegraphics[height =2.5in]{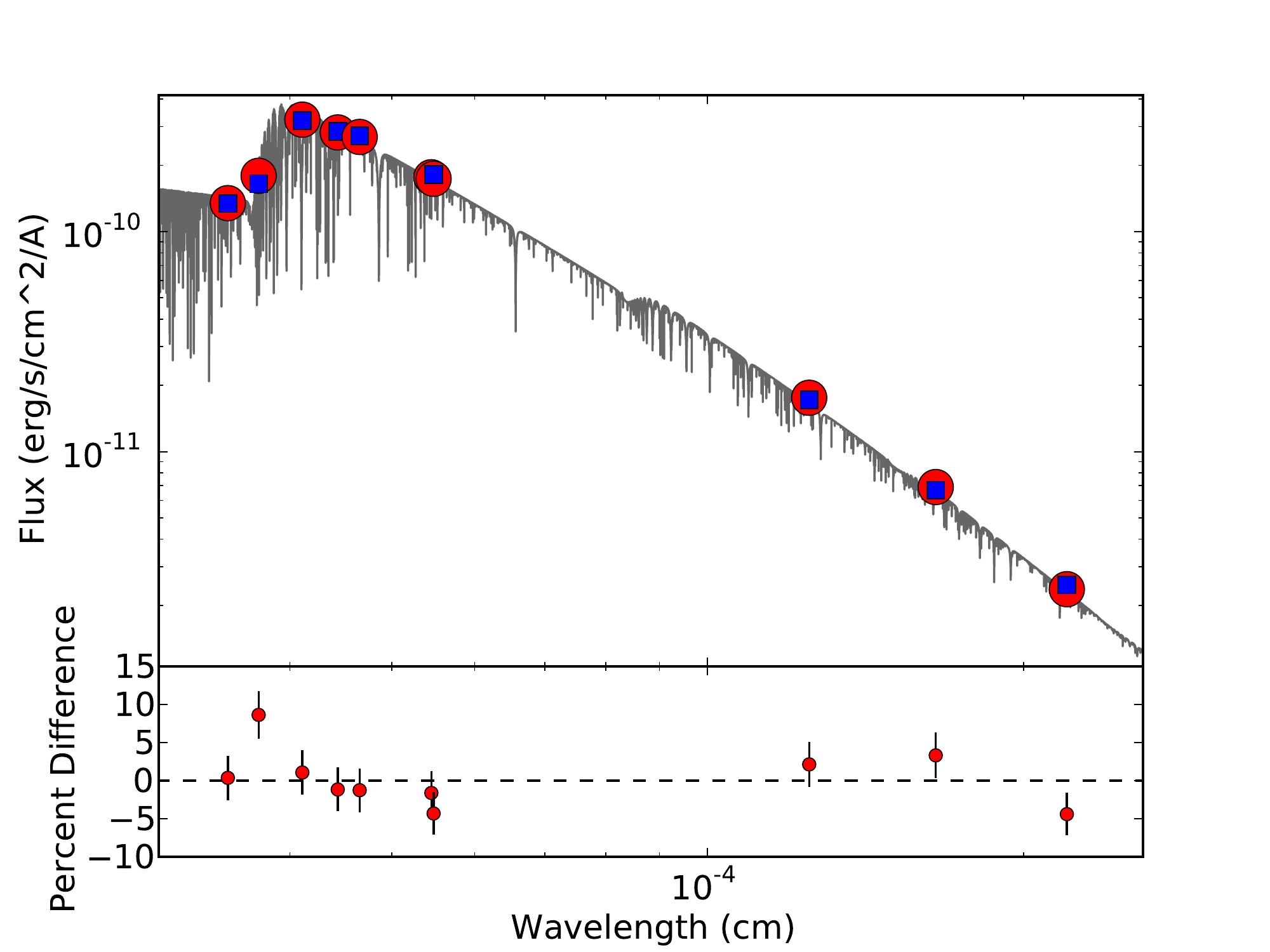}} \\
	\subfloat[\label{fig:HD106591_vZ_vis}]{\includegraphics[height =2.5in]{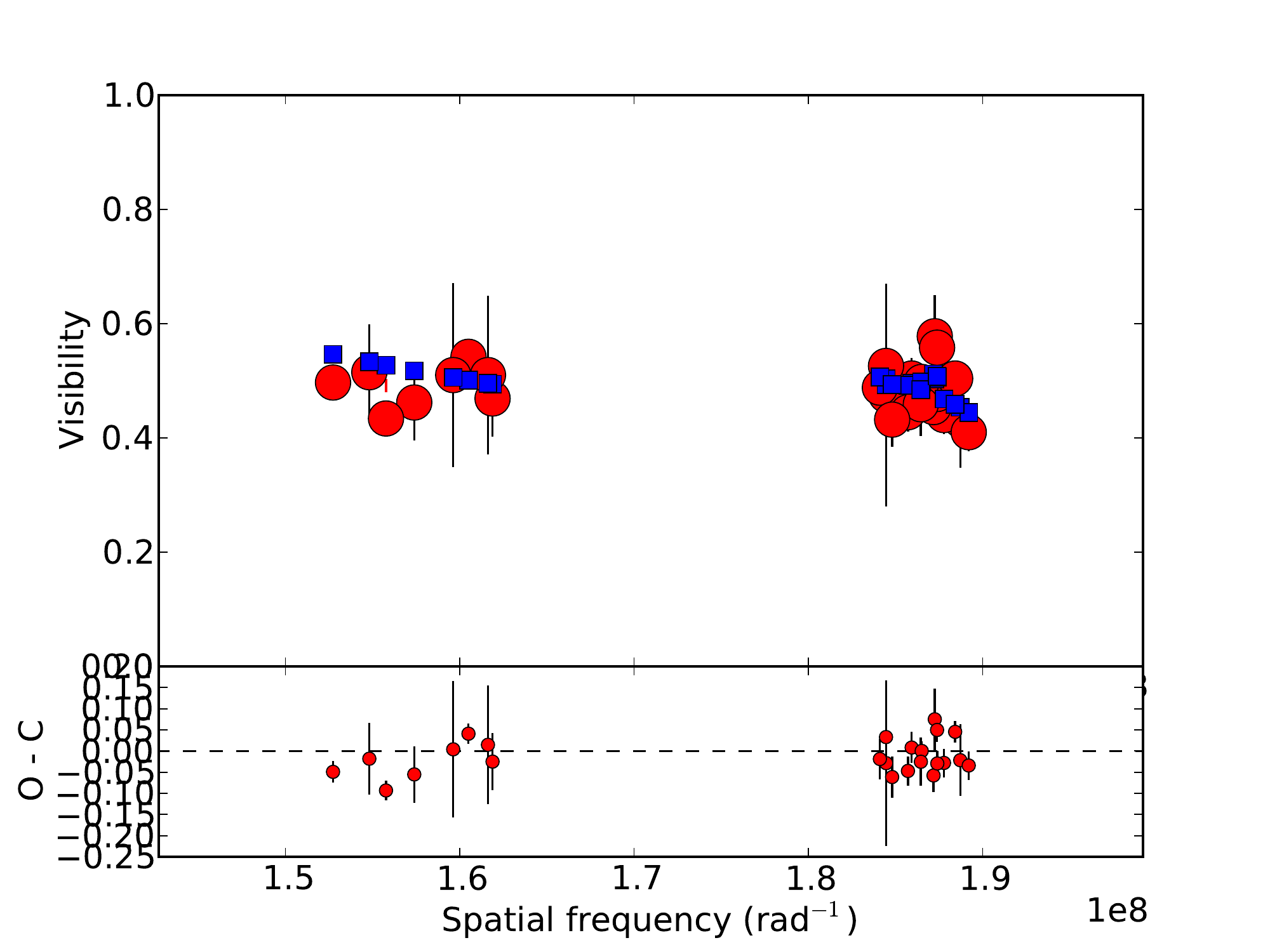}}
	\subfloat[\label{fig:HD106591_vZ_phot}]{\includegraphics[height =2.5in]{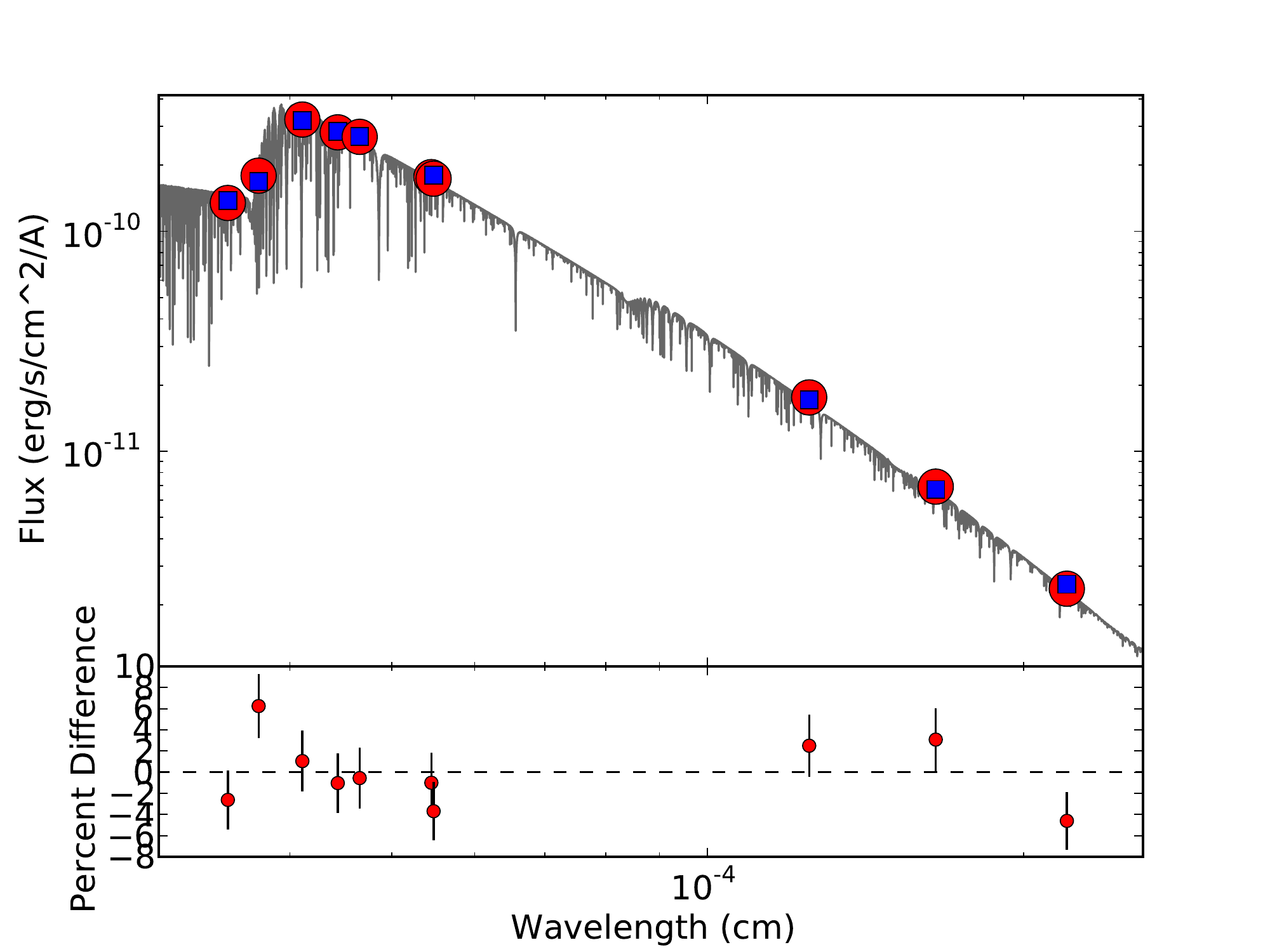}}
	\caption{Same as Figure \ref{fig:HD103287_plots}, but for Megrez (HD 106591).}
	\label{fig:HD106591_plots}
\end{figure*}

\begin{figure*}
	\subfloat[\label{fig:HD116842_ELR_vis}]{\includegraphics[height =2.5in]{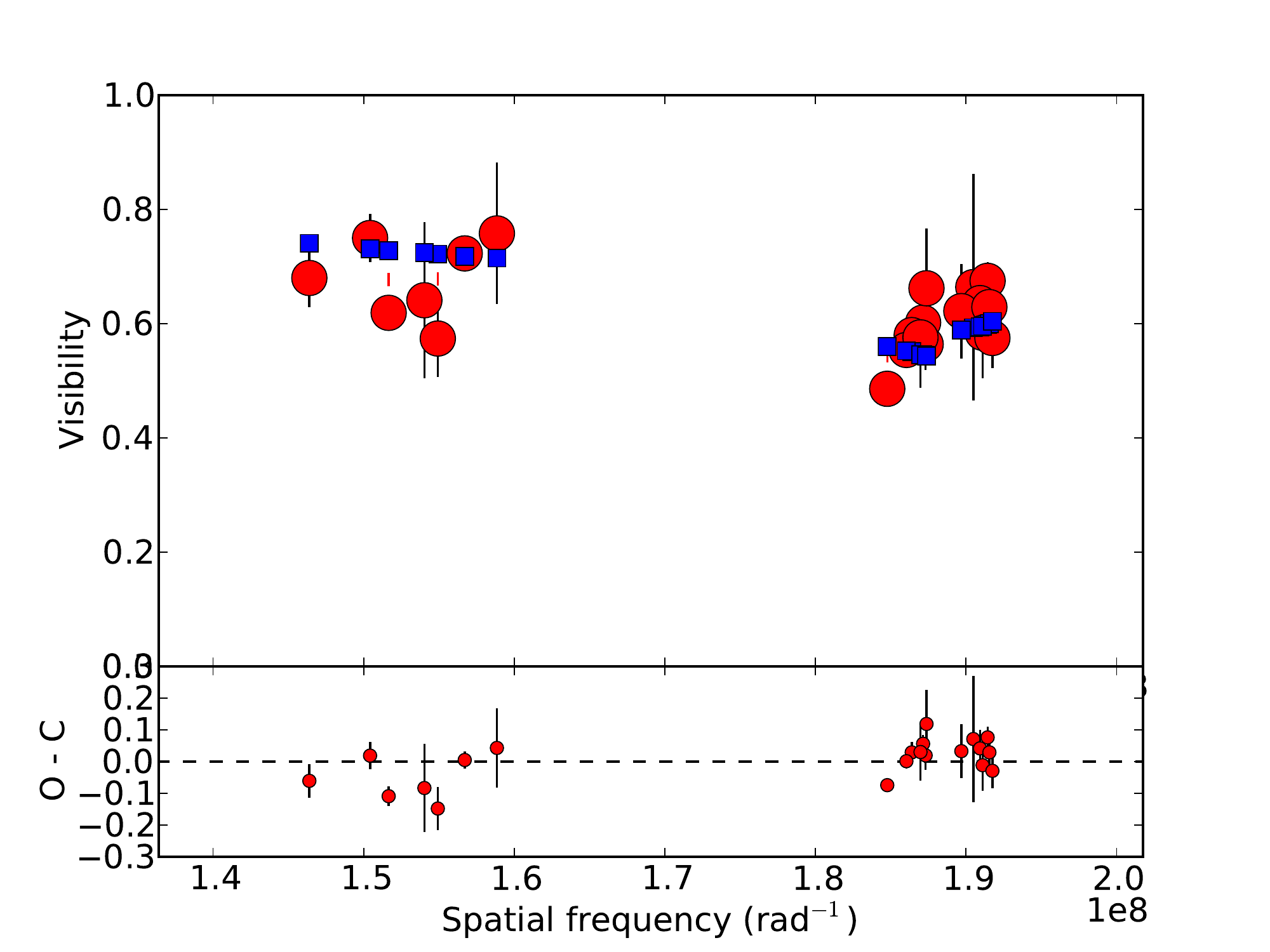}}
	\subfloat[\label{fig:HD116842_ELR_phot}]{\includegraphics[height =2.5in]{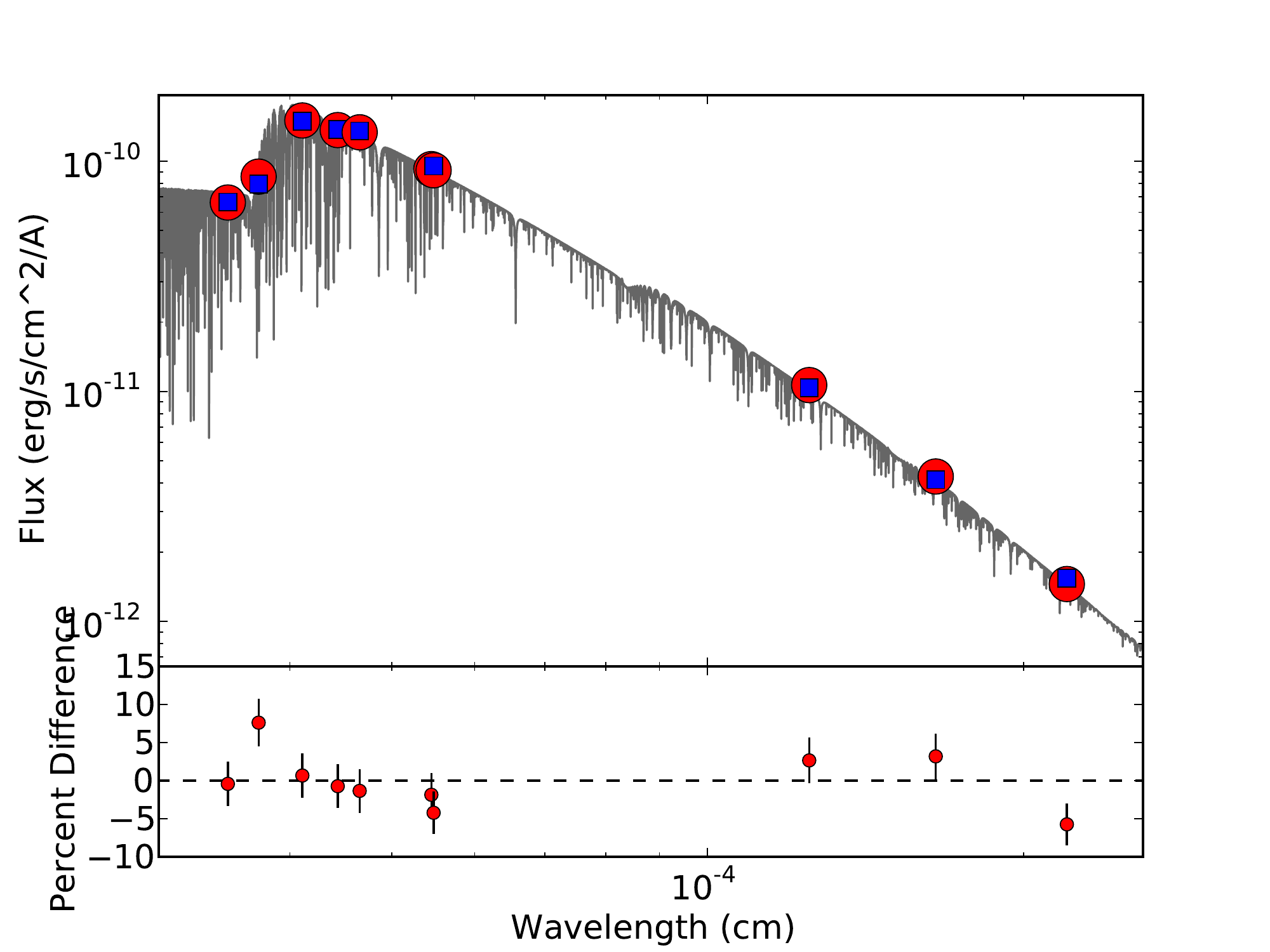}} \\
	\subfloat[\label{fig:HD116842_vZ_vis}]{\includegraphics[height =2.5in]{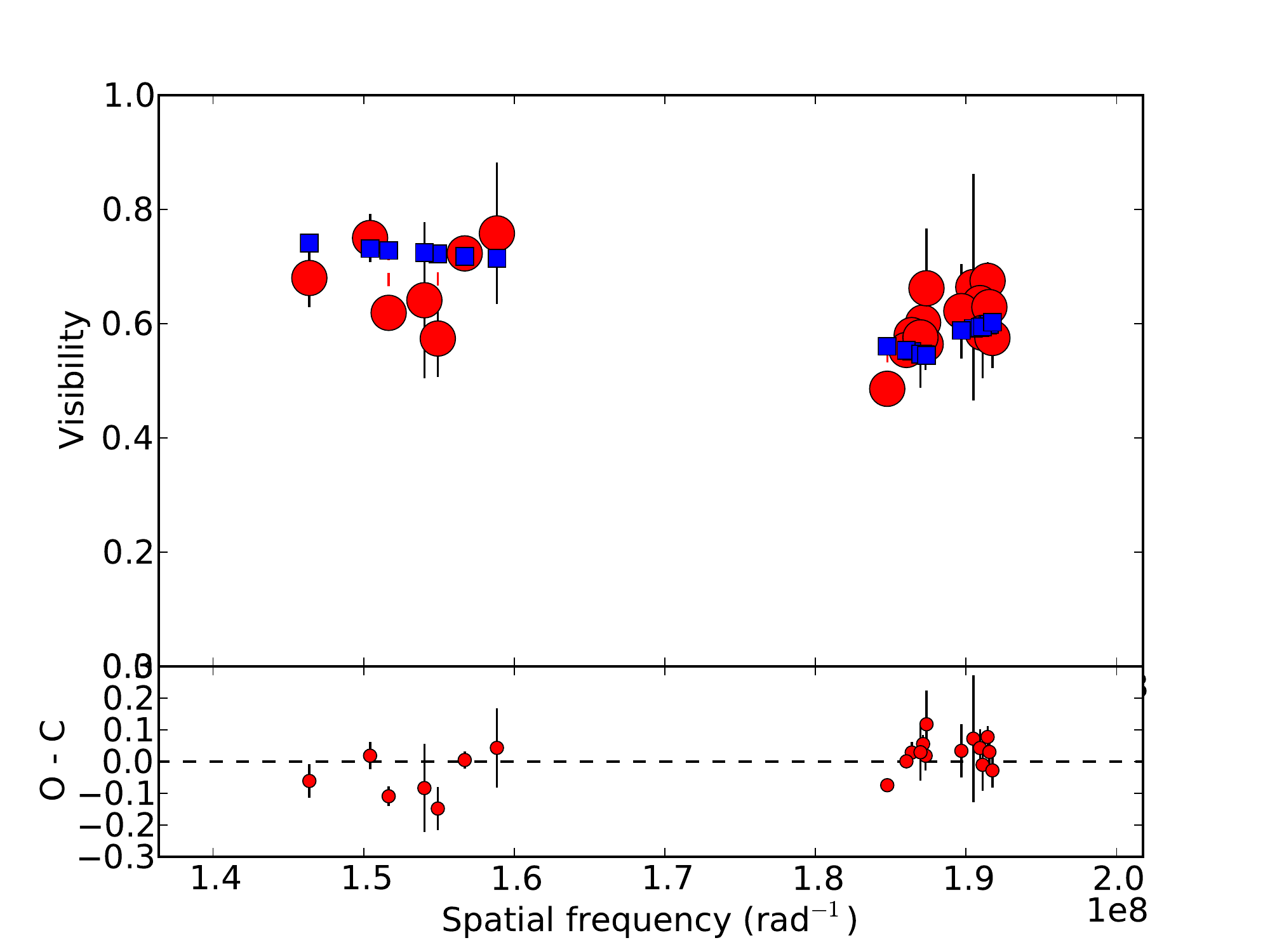}}
	\subfloat[\label{fig:HD116842_vZ_phot}]{\includegraphics[height =2.5in]{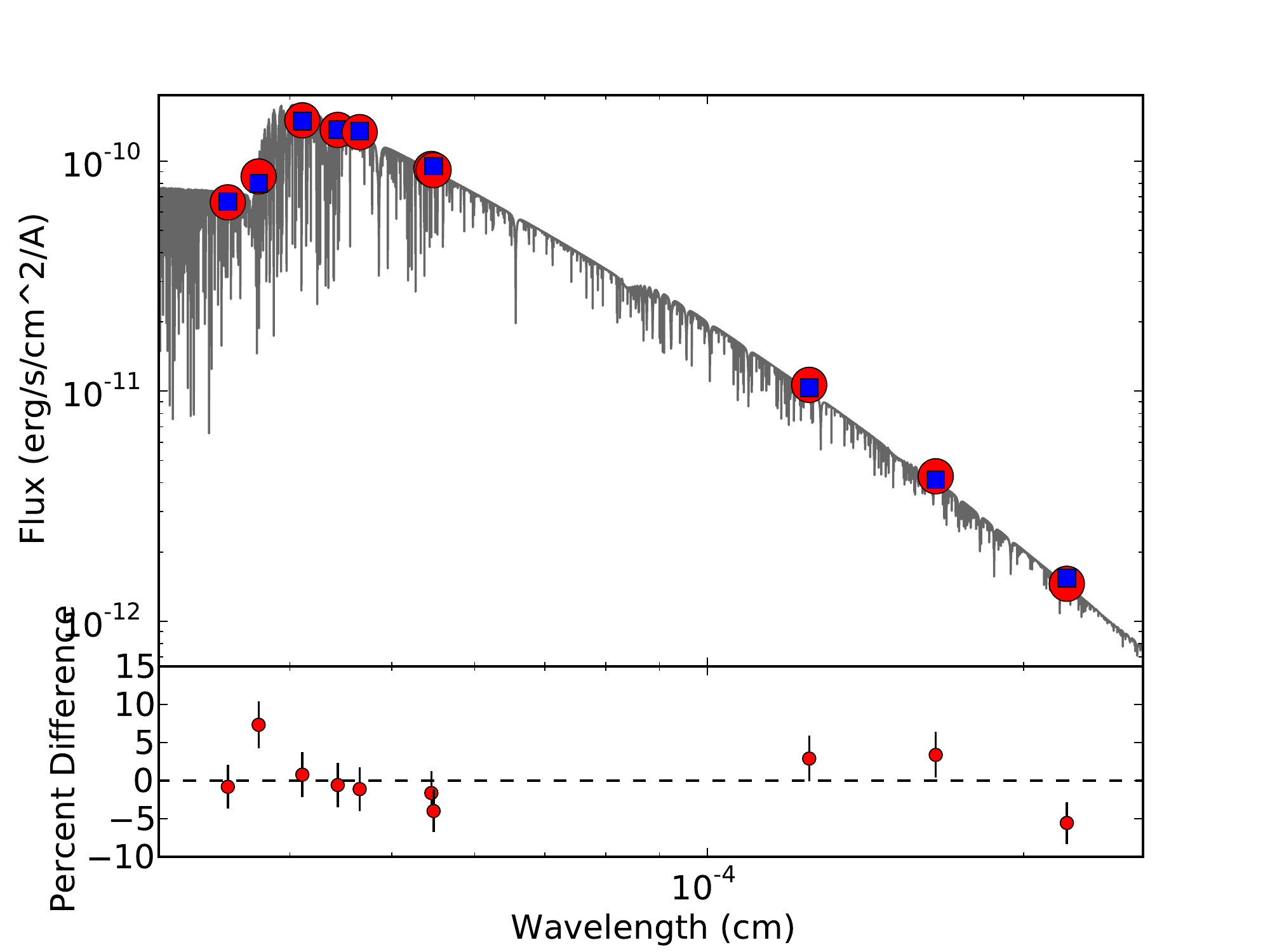}}
	\caption{Same as Figure \ref{fig:HD103287_plots}, but for Alcor (HD 116842).}
	\label{fig:HD116842_plots}
\end{figure*}

\begin{figure*}
	\subfloat[\label{fig:HD141003_ELR_vis}]{\includegraphics[height =2.5in]{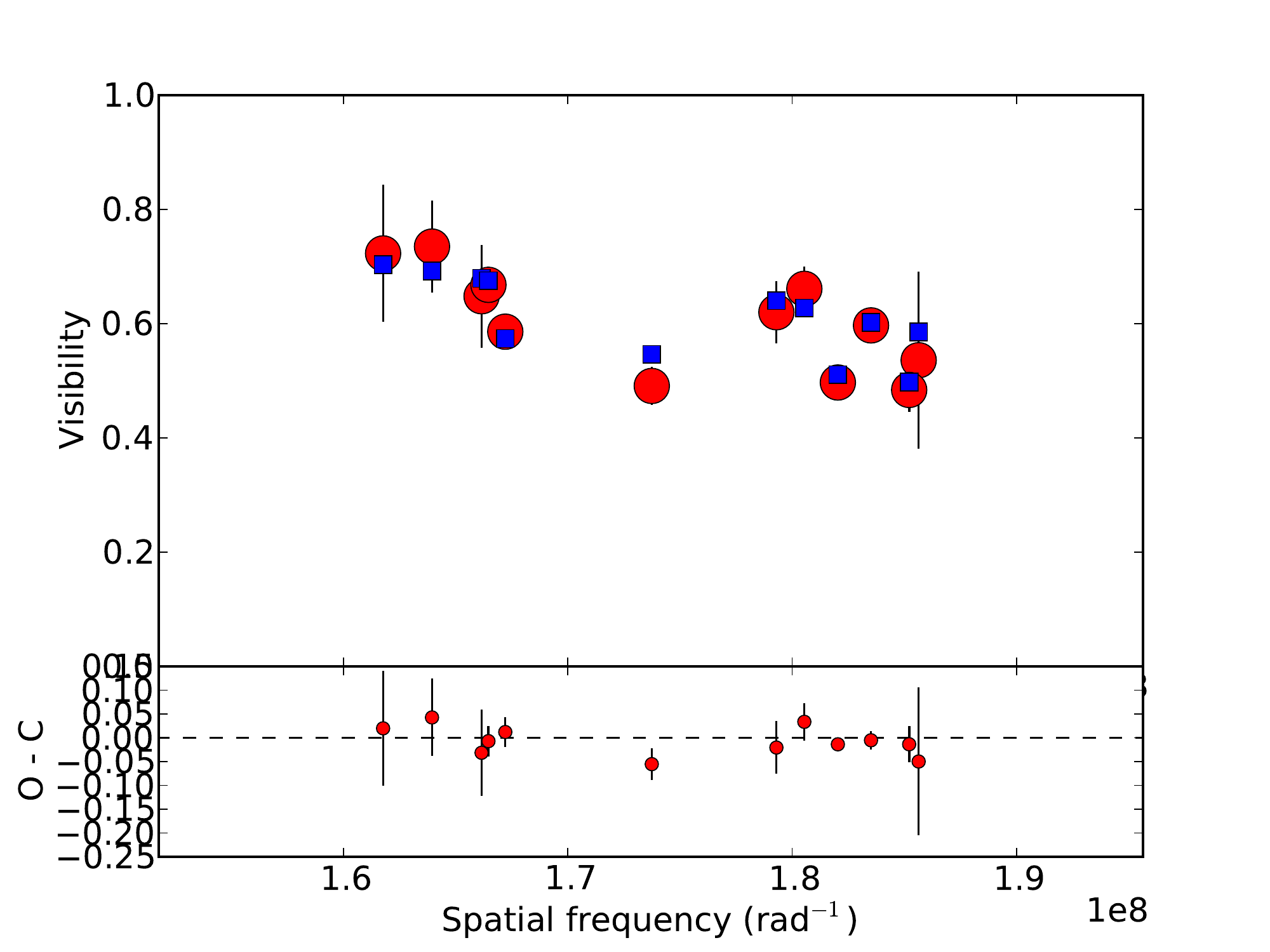}}
	\subfloat[\label{fig:HD141003_ELR_phot}]{\includegraphics[height =2.5in]{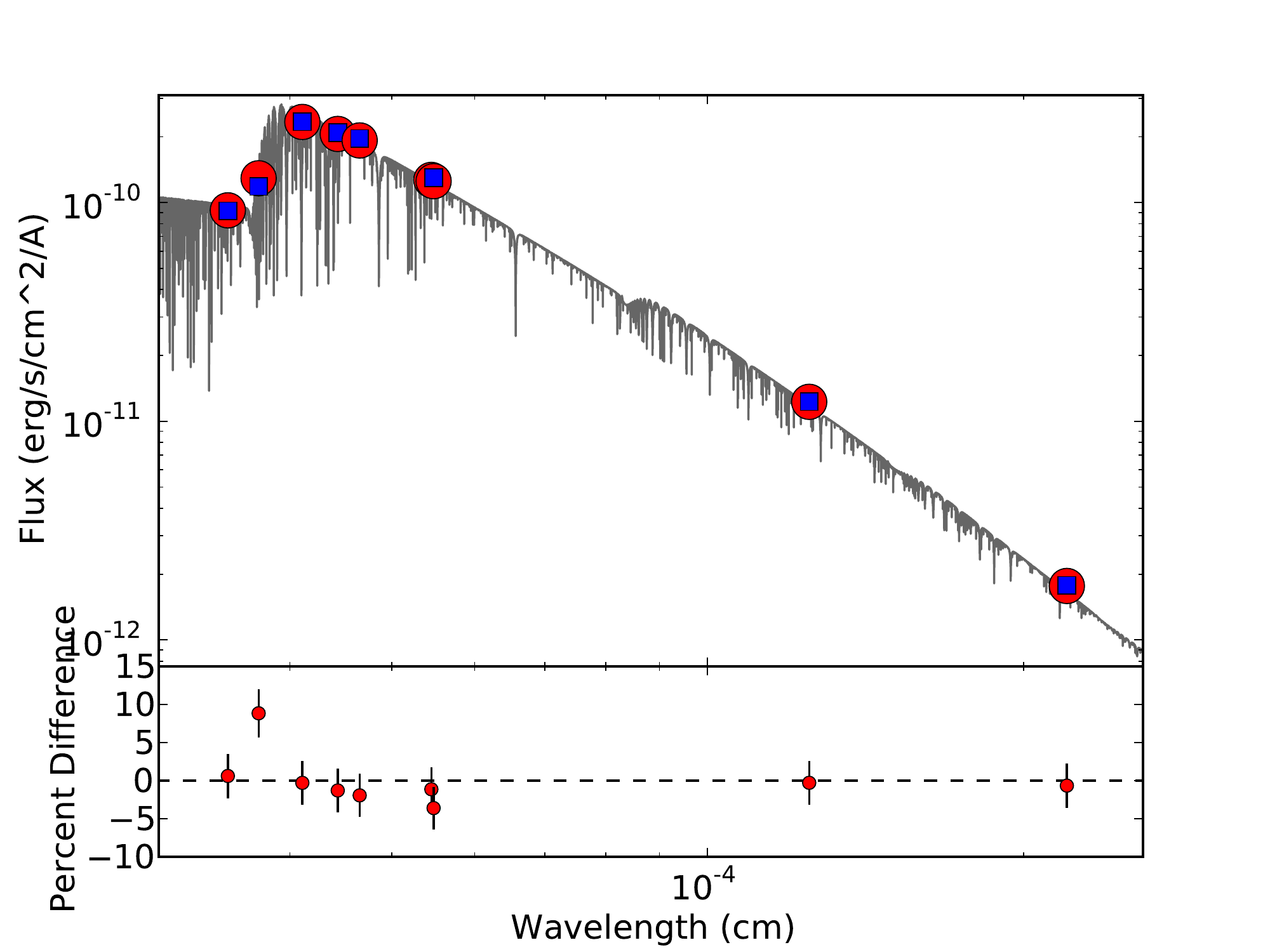}} \\
	\subfloat[\label{fig:HD141003_vZ_vis}]{\includegraphics[height =2.5in]{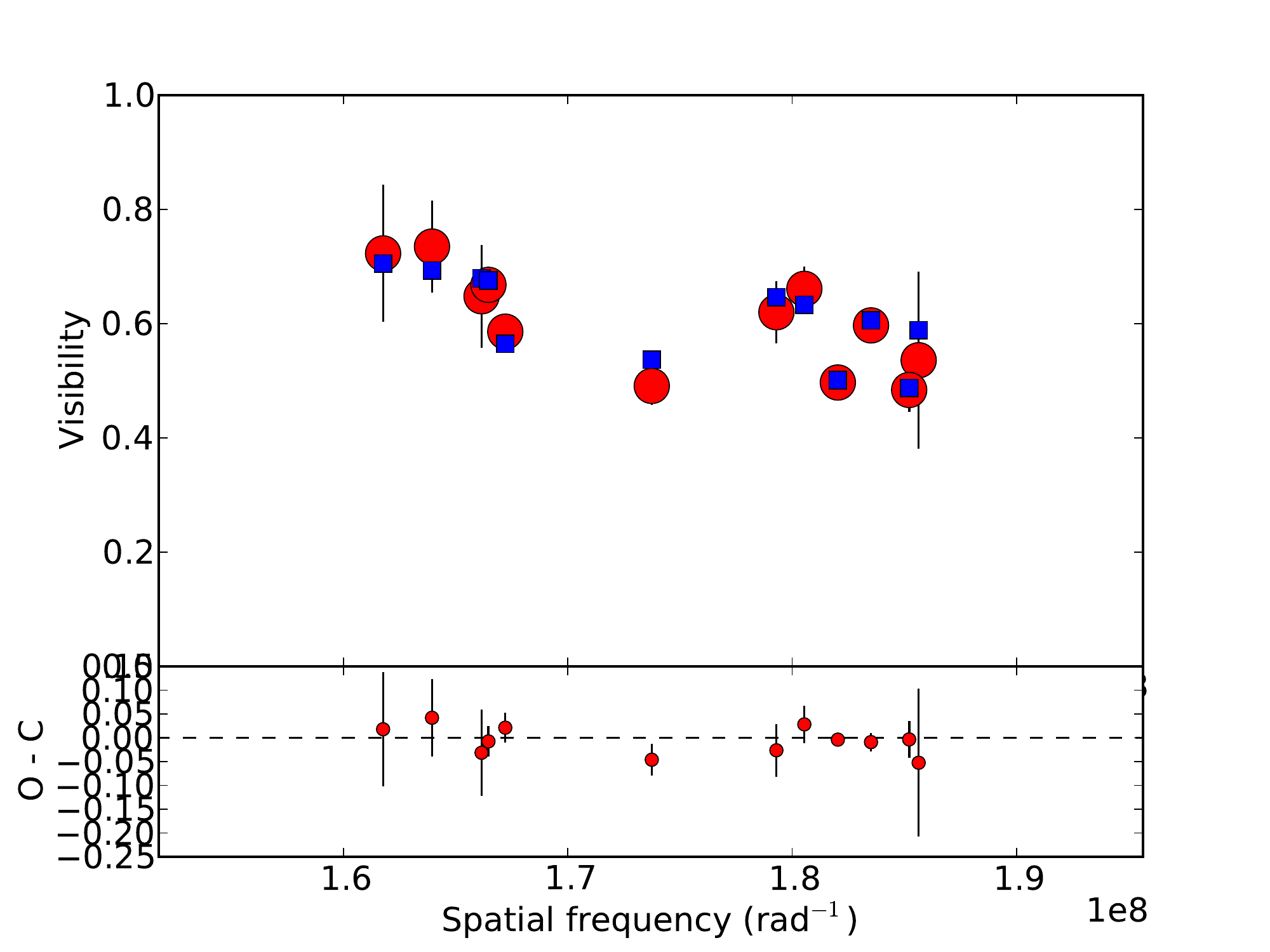}}
	\subfloat[\label{fig:HD141003_vZ_phot}]{\includegraphics[height =2.5in]{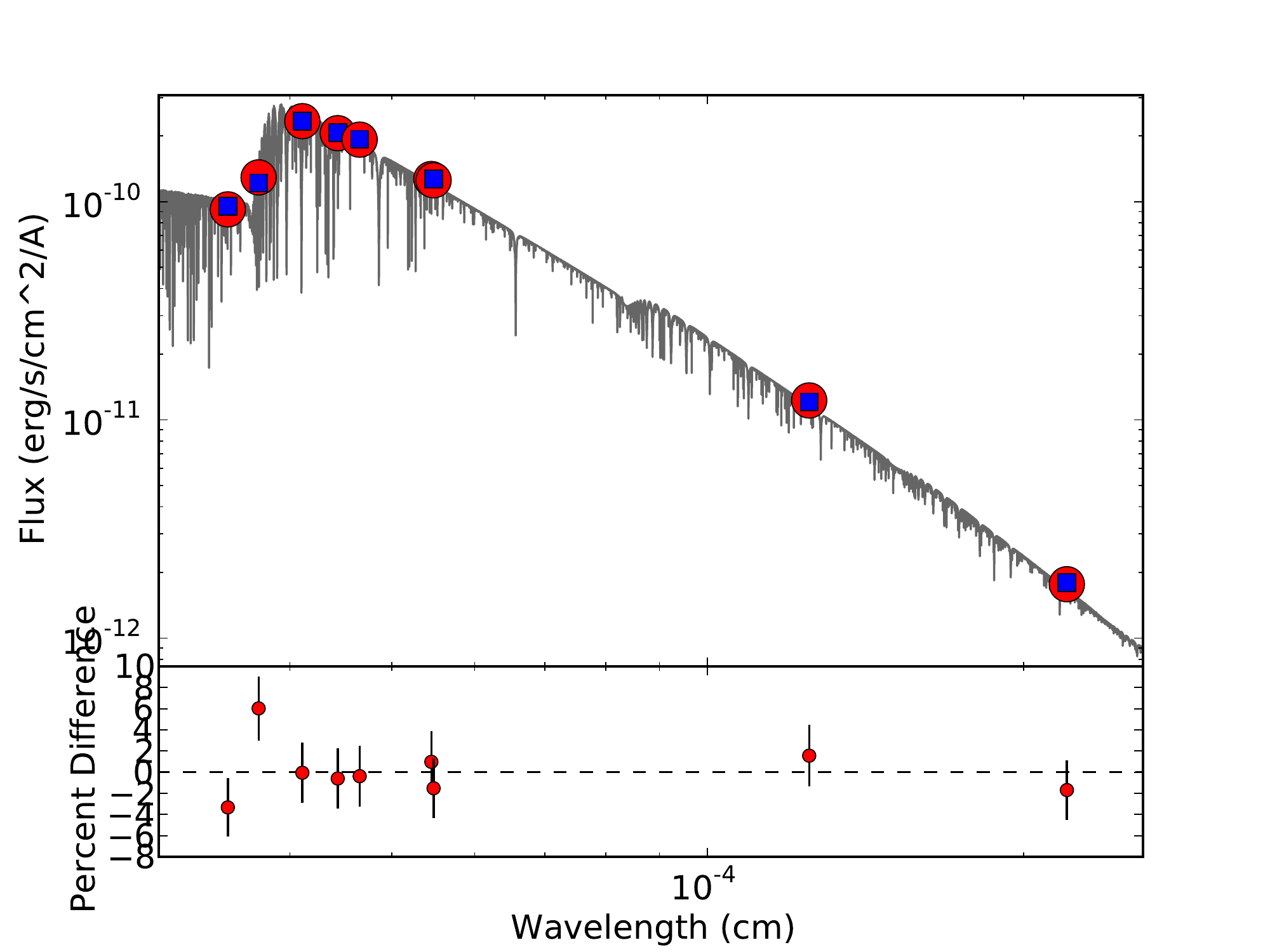}}
	\caption{Same as Figure \ref{fig:HD103287_plots}, but for Chow (HD 141003).}
	\label{fig:HD141003_plots}
\end{figure*}

\begin{figure*}
	\subfloat[\label{fig:HD177196_ELR_i57_vis}]{\includegraphics[height =2.5in]{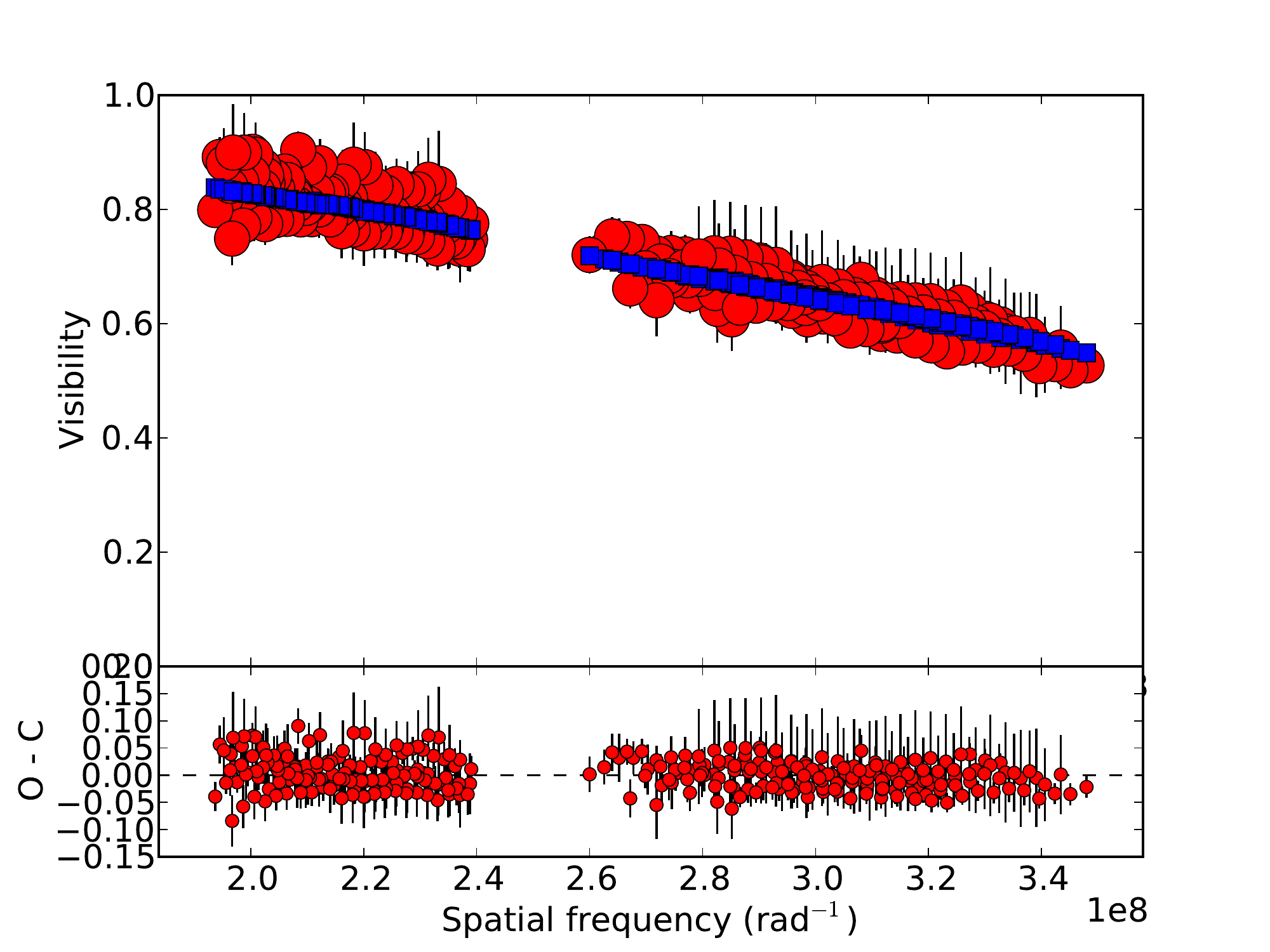}}
	\subfloat[\label{fig:HD177196_ELR_i57_phot}]{\includegraphics[height =2.5in]{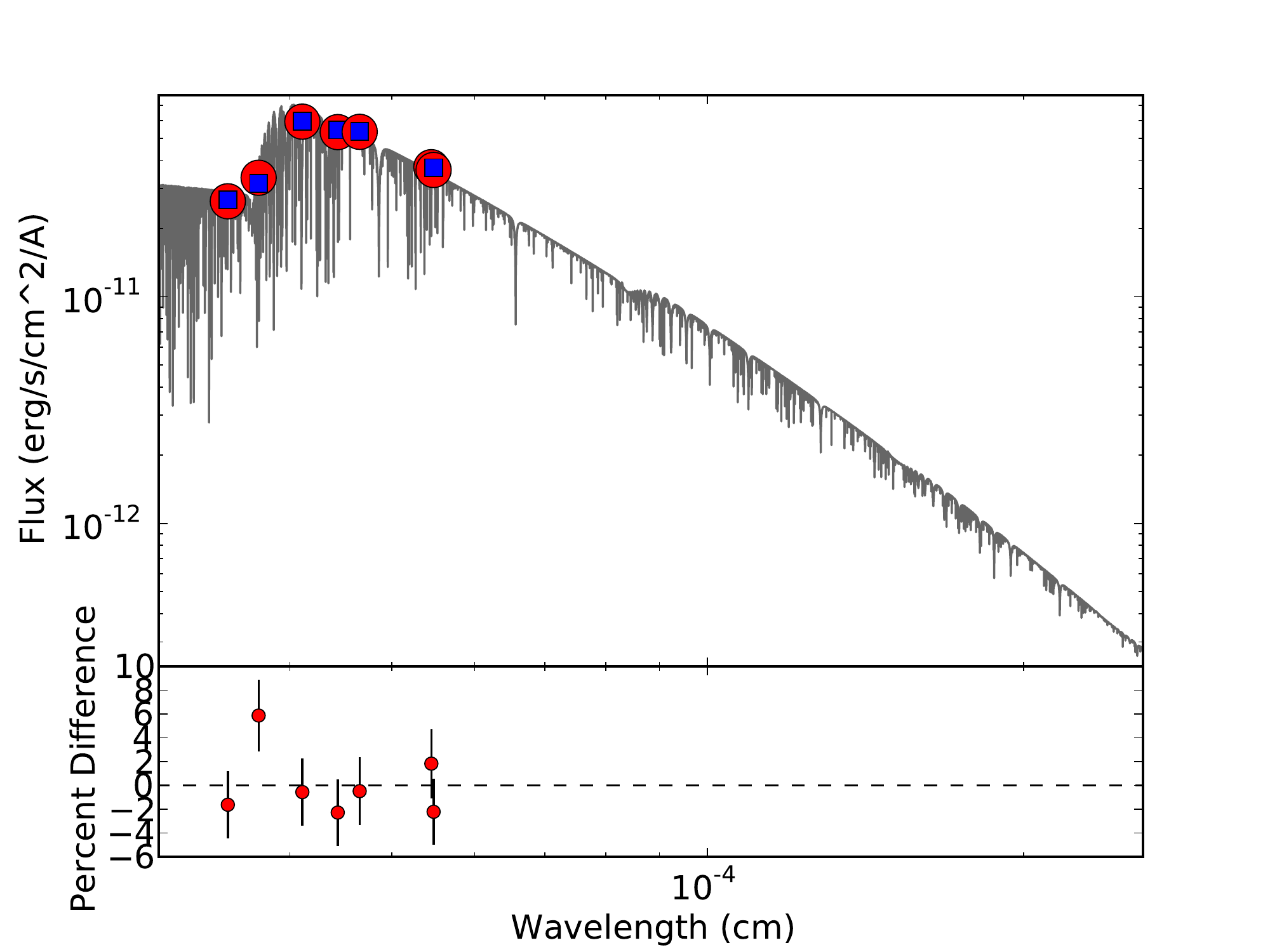}} \\
	\subfloat[\label{fig:HD177196_vZ_i57_vis}]{\includegraphics[height =2.5in]{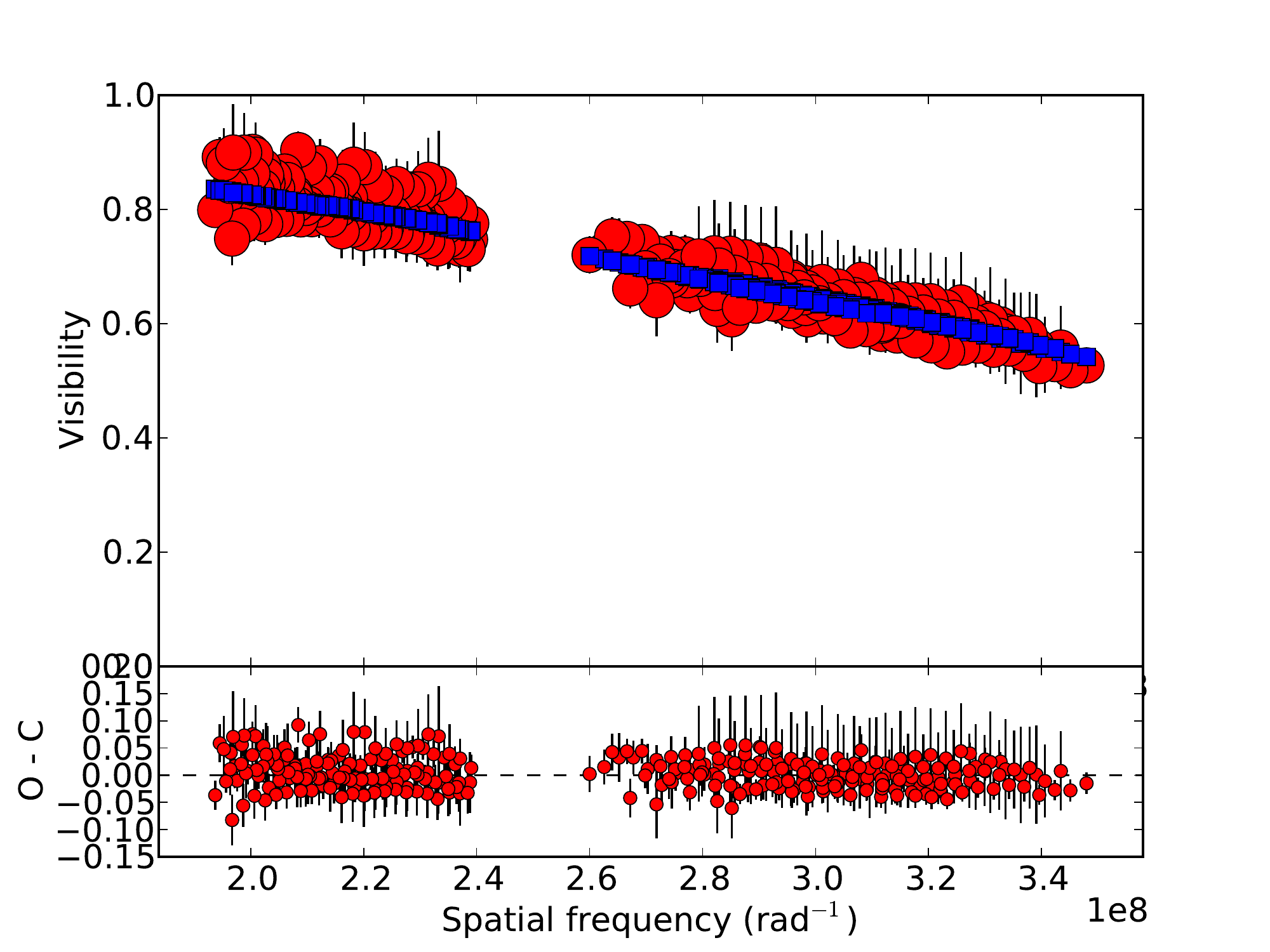}}
	\subfloat[\label{fig:HD177196_vZ_i57_phot}]{\includegraphics[height =2.5in]{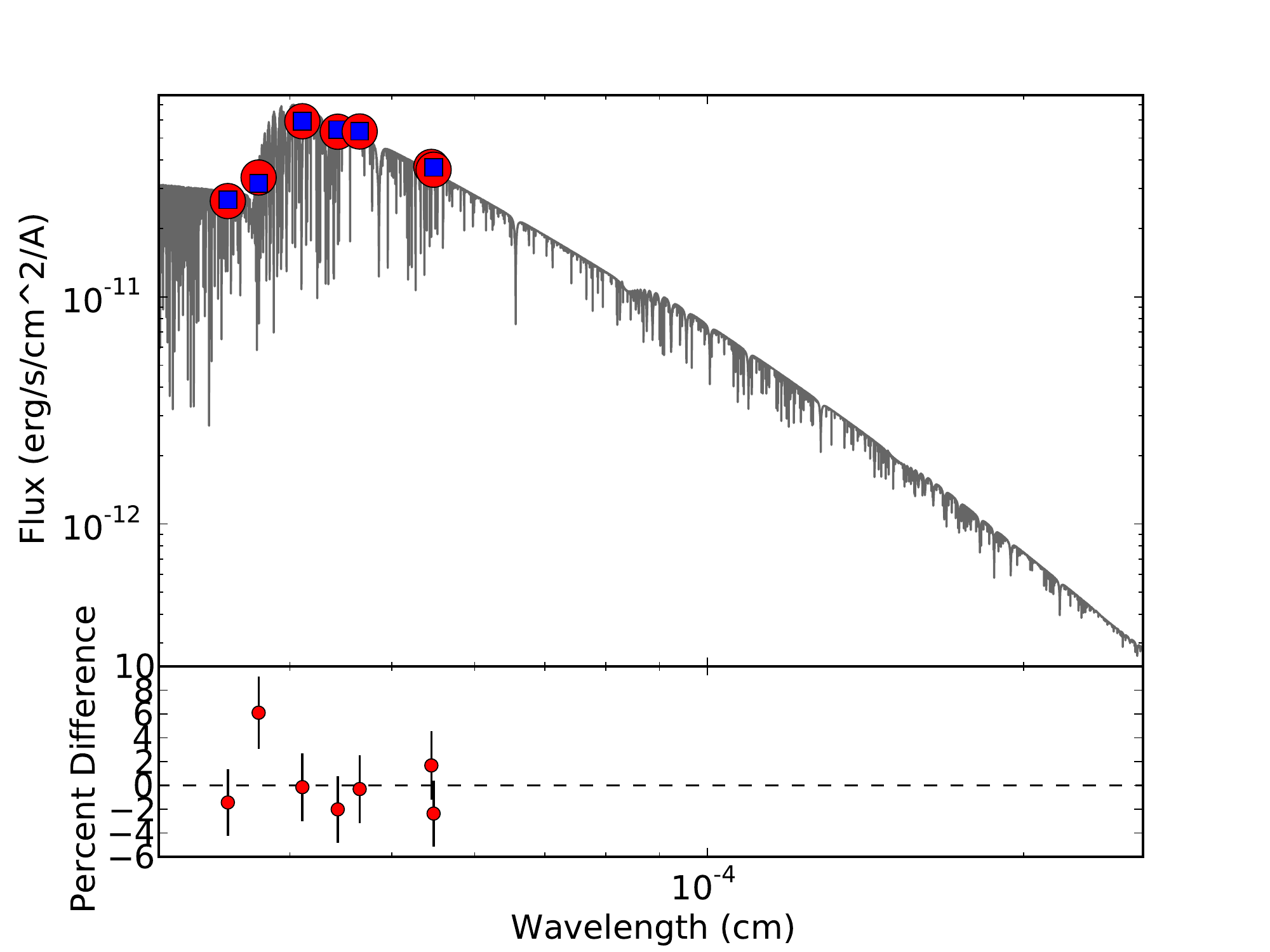}} 
	\caption{Same as Figure \ref{fig:HD103287_plots}, but for 16 Lyr (HD 177196)}
	\label{fig:HD177196_plots}
\end{figure*}

\begin{figure*}
	\subfloat[\label{fig:HD180777_ELR_i28_vis}]{\includegraphics[height =2.5in]{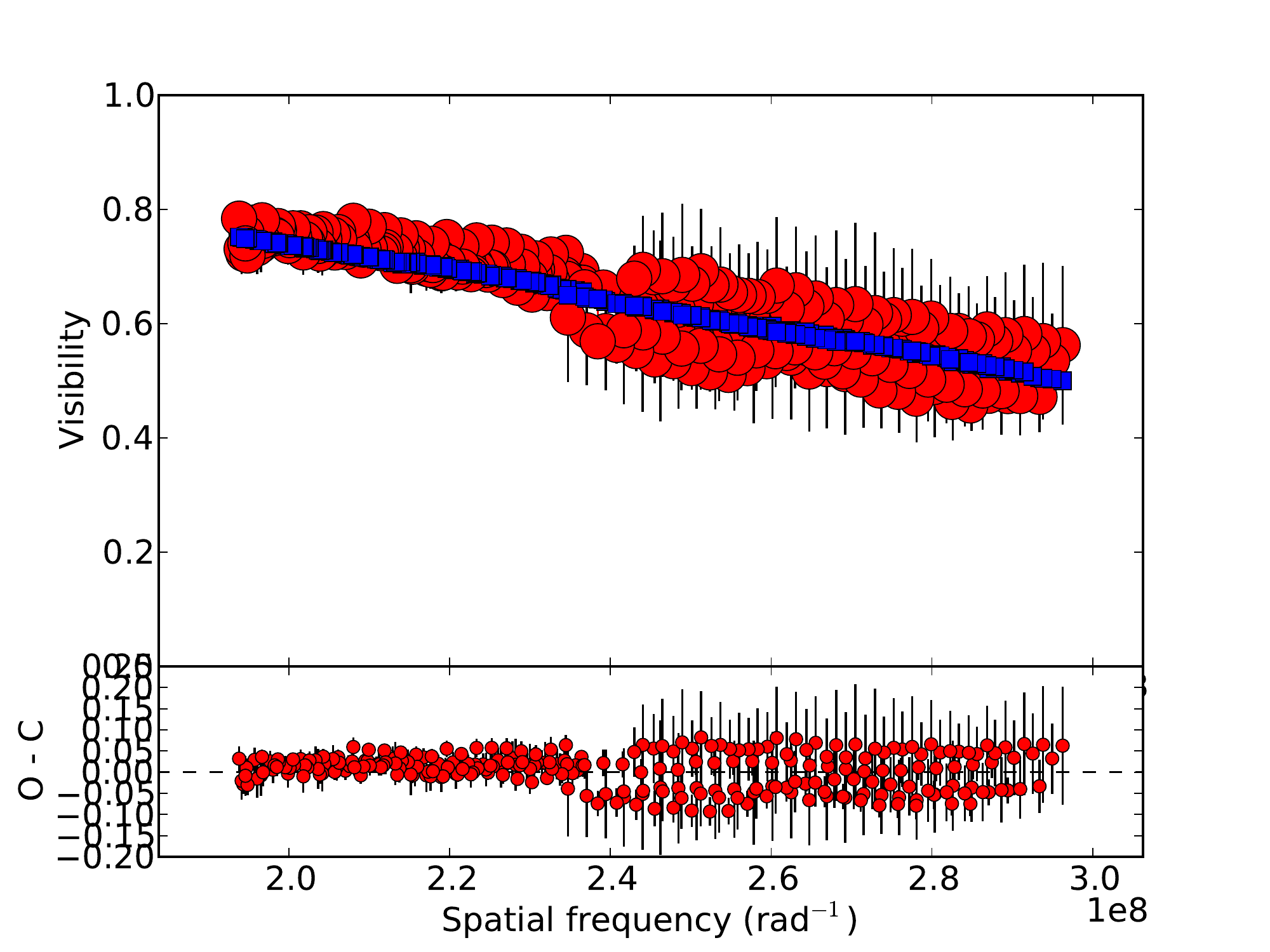}}
	\subfloat[\label{fig:HD180777_ELR_i28_phot}]{\includegraphics[height =2.5in]{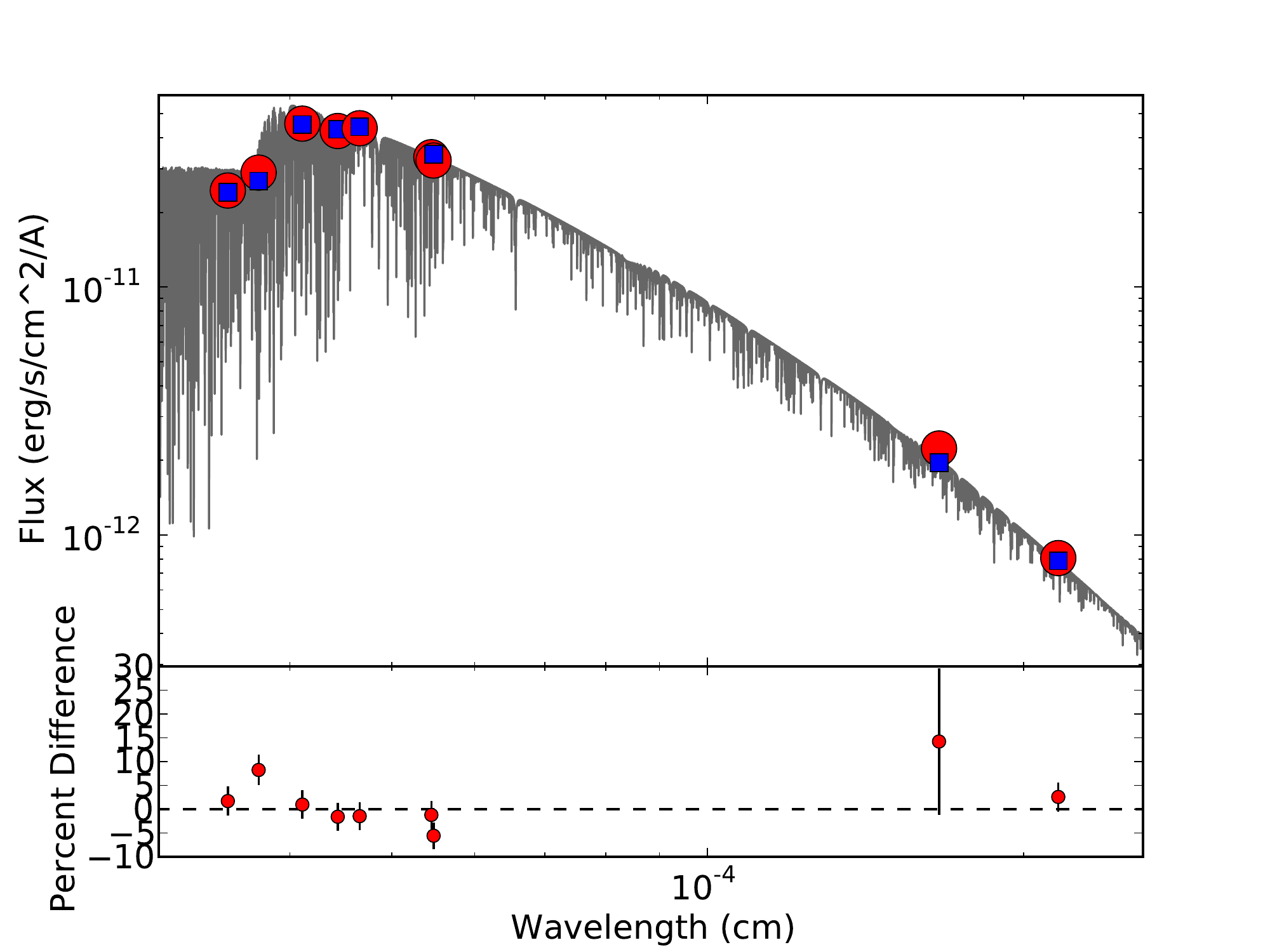}} \\
	\subfloat[\label{fig:HD180777_vZ_i28_vis}]{\includegraphics[height =2.5in]{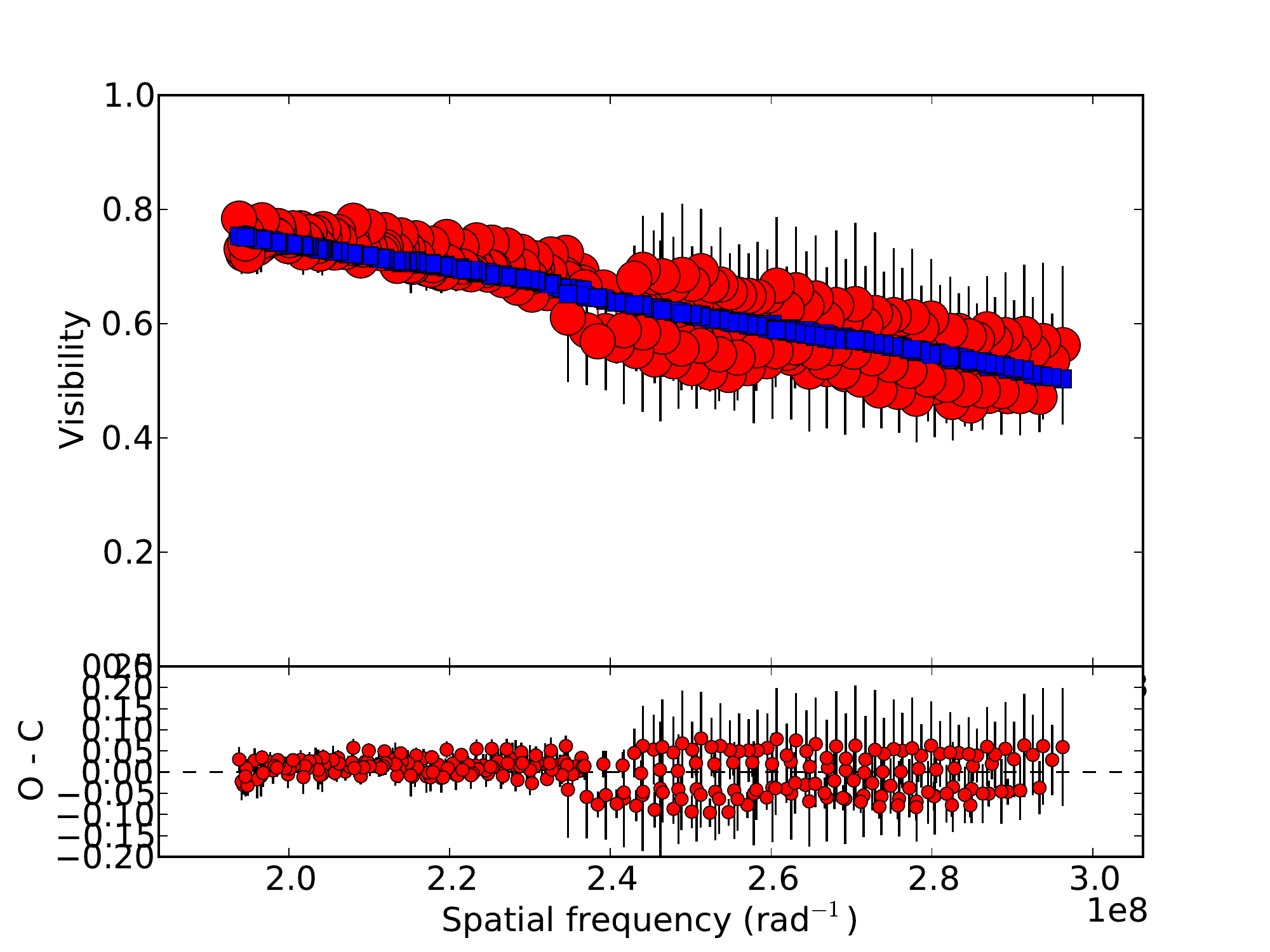}}
	\subfloat[\label{fig:HD180777_vZ_i28_phot}]{\includegraphics[height =2.5in]{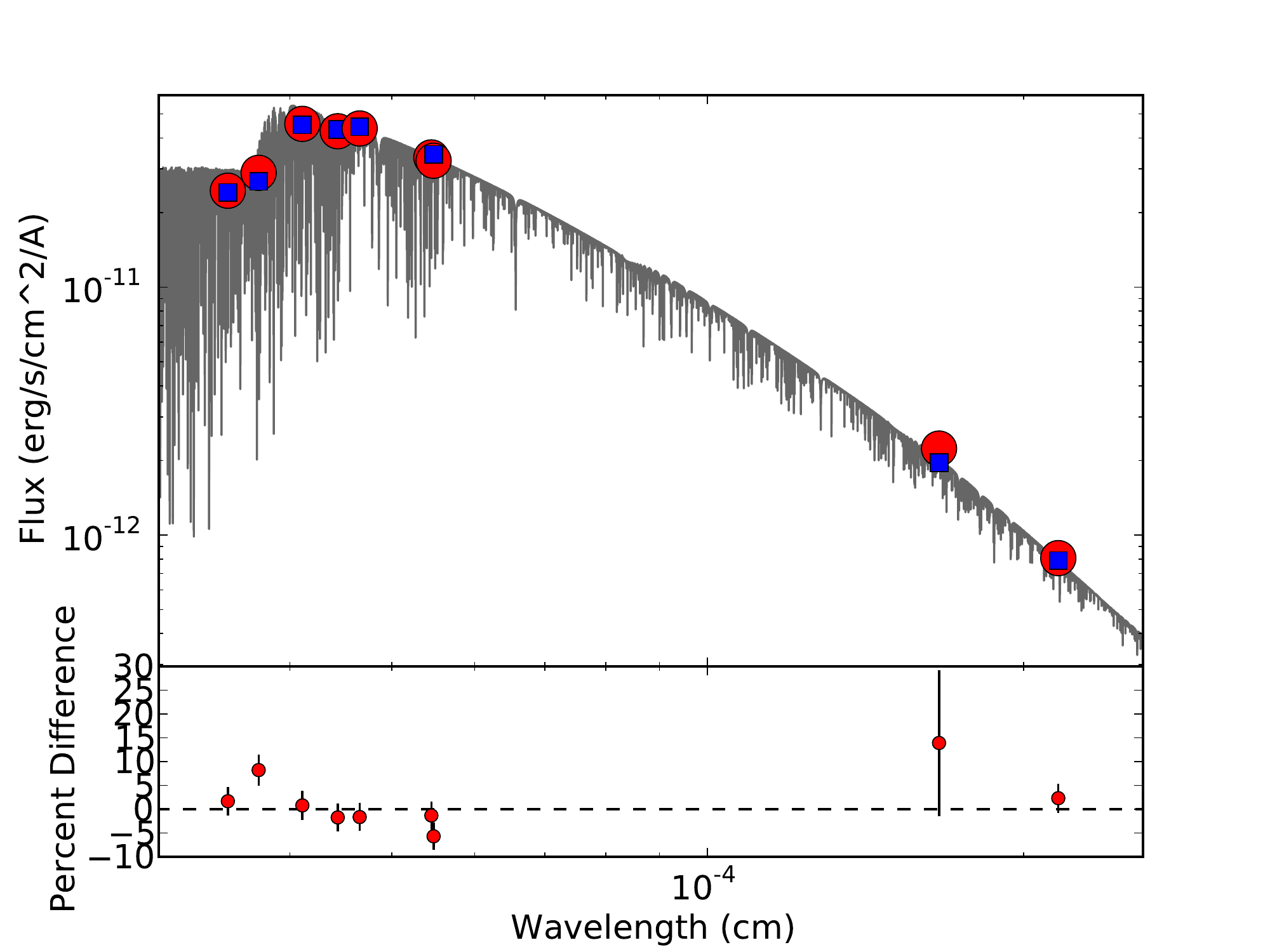}} 
	\caption{Same as Figure \ref{fig:HD103287_plots}, but for 59 Dra (HD 180777)}
	\label{fig:HD180777_plots}
\end{figure*}

		To determine the errors in the age and mass, the age and mass are calculated for the ten points which represent the 1$\sigma$-errors of the five parameters in the oblate star 
			model (i.e., [$R_\mathrm{e} \pm \sigma_{R_\mathrm{e}}$, $V_\mathrm{e}$, $i$, $T_\mathrm{p}$, $\psi$], [$R_\mathrm{e}$, $V_\mathrm{e} \pm \sigma_{V_\mathrm{e}}$, $i$, $T_\mathrm{p}$, $\psi$], etc.).
		The lowest and highest values that come from this procedure represent the lower and upper bounds of the statistical errors presented here.
		We note that this method does not take into account any correlations that may be present between the free parameters. 
		The final best fitting parameters and their errors for the vZ and ELR gravity darkening laws can be found in Tables \ref{tab:fast_res_old_beta} 
			and \ref{tab:fast_res_new_beta}, respectively.
		Figures \ref{fig:HD103287_plots} - \ref{fig:HD180777_plots} illustrate the best fitting model visibilities and photometry, as constrained by the observations, for both gravity darkening prescriptions.
		
	\subsection{Initial Model Parameters}
		\label{sec:imp}
		The $\chi^2$ minimization technique that is used to determine the best-fitting model (see Section \ref{sec:omod}) is especially sensitive to the initial guess given for the star's inclination. 
		To account for this, for each star, the model is run a number of times using various fixed inclinations. 
		The inclinations chosen range from 90$^{\circ}$ (edge-on) down to an inclination that would have the model star rotating at breakup velocity given its $v \sin i$.
		The best-fitting set of parameters of these fixed-inclination models is chosen as the set of input parameters for the process described in Section \ref{sec:omod}. 

		The initial guess value for $M_\mathrm{*}$ that is supplied for the model runs at fixed inclinations is determined based on the star's spectral type and the spectral type-mass relations found in \cite{aaq}.
		The initial guess values for $R_\mathrm{e}$ and $T_\mathrm{p}$ are based on the angular diameters and effective temperatures listed in the JMMC Stellar Diameter Catalog \citep[JSDC,][]{jsdc} for each star. 
		The initial value for $\psi$ is determined by fitting a uniform ellipse to the visibilities in the cases where multiple baseline orientations have been used or is set to 0$^{\circ}$ in the cases where they have not. 
		
	\subsection{Merak}
		\label{sec:merak}
		The apparent slow rotator, Merak (HD 95418), was observed using the Classic beam combiner on the CHARA Array previously by \cite{boyajian_2012}. 
		We have taken the radius and luminosity determined by that study as well as its $v \sin i$ to determine its age and mass using the MESA evolution model using a similar process described in Section \ref{sec:omod}, but 
			without any iteration.
		Because of this, we do not determine the inclination of this star nor its equatorial velocity. 
		We assume an edge-on inclination of 90$^\circ$. 
		The results are compiled in Table \ref{tab:merak}.
		
	\subsection{16 Lyr and 59 Dra}
		\label{sec:slowrot}
		When running the model described above, the two stream stars, 16 Lyr and 59 Dra, both yield best fitting values for $R_\mathrm{avg}$ and $L_\mathrm{tot}$ that correspond to unphysical positions 
			below the zero-age main sequence for their respective best fit values for $V_\mathrm{e}$.
		One way to reconcile this discrepancy would be for the stars to have a metallicity of Z $\lesssim$ 0.013 ($\sim$0.1 dex lower than the moving group). 
		We are cautious against advocating for this interpretation since, as discussed in Section \ref{sec:sampobs}, we have insufficient baseline orientations to fully measure the oblateness and gravity darkening in these cases.
		We note that the best fitting values for $V_\mathrm{e}$ for both 16 Lyr and 59 Dra are sufficiently large that they shift the zero-age main sequence above the best fitting values for $R_\mathrm{avg}$ and $L_\mathrm{tot}$.
		If these $V_\mathrm{e}$ values are too large, this could explain the unphysical $R_\mathrm{avg}$ and $L_\mathrm{tot}$ without changing the metallicity.
		Figure \ref{fig:HRD} illustrates how the zero-age main sequence is raised by rapid rotation.
		With this in mind, we run the model for these two stars constraining the equatorial velocity to be within the more modest range of 94 to 202 $\mathrm{km~s^{-1}}$ for each star.
		This range corresponds to the dispersion about the maximum of the probability distribution of equatorial rotation velocities for late-type A-stars as determined by \cite{zorec_2012}.
		We make this constraint by fixing the stars' inclinations such that $i=\arcsin(\frac{v \sin i}{\mathrm{E}[V_\mathrm{e}]})$ where $\mathrm{E}[V_\mathrm{e}]$ is the maximum of the aforementioned probability distribution.
		This corresponds to inclinations of $\sim$57$^{\circ}$ and $\sim$28$^{\circ}$ for 16 Lyr and 59 Dra, respectively.
\section{Model Results}
	\subsection{Photospheric Properties}
		\label{disc:phot_prop}
		Using the procedure described in Section \ref{sec:omod}, the best fitting models for all six of the stars observed show $\chi^2_\mathrm{tot}$ values ranging from 3.1$-$13.4. 
		The model fitting using the vZ gravity darkening law yield a high inclination ($i$ $>$ 70$^{\circ}$) for one star (Alcor), moderate inclinations (40$^{\circ}$ $>$ $i$ $<$ 70$^{\circ}$) for two stars 
			(Megrez and Chow), and a low inclination ($i$ $<$ 40$^{\circ}$) for one star (Phecda); both 16 Lyr and 59 Dra have fixed inclinations (see Section \ref{sec:slowrot}). 
		These results also show an oblateness, $\rho=(R_\mathrm{e}-R_\mathrm{p})/R_\mathrm{p}$ that ranges from 3\% to 54\% with an average of 26\% and temperature differences across the photosphere, 
			$\Delta T=T_\mathrm{p}-T_\mathrm{e}$ that range from 214 K to 6414 K with an average of 2965 K. 
		The same analysis done using the ELR gravity darkening law also yields a high inclinations for Alcor, moderate inclinations for four Megrez and Chow, and a low inclination for Phecda.
		These results show an oblateness range of 3\% to 55\% with an average of 24\% and temperature differences across the photosphere that range from 192 K to 3769 K with an average of 1696 K. 
		The smaller mean temperature gradient seen with the ELR law is because that law yields a smaller gravity darkening coefficient, $\beta$, which lessens the effect the local surface gravity has on the local temperature.
		Using the vZ law, $\beta$ is 0.25 for all four observed rapid rotators. 
		The ELR law has $\beta$ ranging from 0.138 to 0.242. 

	\subsection{Masses and Ages}
		\label{disc:age_res}
		The masses calculated by the procedures discussed in Section \ref{sec:mod} range from 1.447 to 2.509 M$_{\sun}$ for all seven stars in the sample using the vZ gravity darkening law and 
			1.443 to 2.509 M$_{\sun}$ using the ELR law (Figure \ref{fig:MvA}). 
		The mass estimates for the individual stars are consistent between the two laws within their 1-3\% uncertainties with the exception of Chow, whose mass is $2.333^{+0.015}_{-0.015}$ M$_{\sun}$ using the vZ law
			or $2.388^{+0.036}_{-0.021}$ M$_{\sun}$ using the ELR law. 
		The ages calculated by the procedures described above range from 401 to 659 Myr for all seven stars in the sample using the vZ gravity darkening law and 333 to 610 Myr using the ELR law.
		With the exception of the star Chow, these age estimates are consistent with being coeval using either the vZ and ELR laws, despite their larger uncertainties, that range from 2 to 41\% and with a mean and 
			median uncertainty of 14\% and 12\%, respectively.
		It is worth noting that the uncertainty in the age is partially dependent on the mass because the radius, luminosity, and temperature of more massive stars evolve more rapidly, thus allowing for a more precise 
			determination of the age because fixed uncertainties in these parameters will correspond to a smaller percent error in the age.
		We caution that these uncertainties are only statistical. 
		Systematic uncertainties (such as those in gravity darkening and metallicity) can lead to more substantial errors. 
		Only Chow shows a disparity in its age estimates between the two gravity darkening laws.
		Chow's age is determined to be $659^{+11}_{-10}$ Myr when using the vZ law or $610^{+14}_{-35}$ Myr when using the ELR law. 
		The final ages and masses for the are presented in Table \ref{tab:age_mass}.

	\subsection{Comparison with Other Evolution Models}
		\label{disc:evo_comp}
		In order to test the accuracy of the MESA evolution models and to begin to address some of the systematic errors that may be introduced by them, we compare the results from one of the stars in our sample 
			across four different evolution models: the MESA models; the Geneva models \citep{georgy_2013}, which do take rotation into account; the Padova models \citep{girardi_2002}, which do not account for rotation; 
			and the MESA models again, but without accounting for rotation. 
		We use the total luminosity, average radius, and equatorial rotation velocity determined for Alcor (HD 116842)\footnote{Using the vZ gravity darkening law} as our point of comparison between the four models.
		We chose Alcor for this comparison because it is the only rapidly rotating nucleus member whose rotation speed is less than the maximum predicted by the Geneva models, which are restricted to values of $\omega$ 
			of $\lesssim$ 0.9 for the masses and ages in question.
		The results are listed in Table \ref{tab:evo_comp}.
		
		The absolute ages agree extremely well between the two rotating models, with a percentage difference of only 0.5\% (0.02-$\sigma$).
		The determined stellar masses also show good agreement, with a percentage difference of 3.1\% (1.4-$\sigma$).
		The ages determined by the non-rotating models also agree with each other extremely well with a percentage difference of 0.9\% (0.07-$\sigma$), but as expected, they are systematically older than those 
			determined using the models that account for rotation.
		The masses determined by the non-rotating models also show good agreement with each other with a percentage difference of 2.1\% (1.0-$\sigma$).
			
	\begin{table}
	\begin{center}
		\caption{Comparing Evolution Models. \label{tab:evo_comp}}
		\begin{tabular}{cc}
			\tableline\tableline
			\multicolumn{2}{c}{Fundamental Parameters for Alcor (HD 116842)} \\
			\tableline
			Average Radius (R$_{\sun}$) 			& $1.846^{+0.057}_{-0.057}$ \\
			Total Luminosity, $L_\mathrm{tot}$ (L$_{\sun}$) 	& $13.98^{+0.75}_{-0.75}$ \\
			Equatorial Velocity ($\mathrm{km~s^{-1}}$) 	& $238.6^{+10.0}_{-9.2}$ \\
			\tableline
			\multicolumn{2}{c}{MESA (with rotation)} \\
			\tableline
			Age (Myr) 				& $422^{+67}_{-75}$ \\
			Mass (M$_{\sun}$) 				& $1.842^{+0.027}_{-0.031}$ \\
			\tableline
			\multicolumn{2}{c}{Geneva (with rotation)} \\
			\tableline
			Age (Myr) 				& $424^{+69}_{-75}$ \\
			Mass (M$_{\sun}$) 				& $1.899^{+0.026}_{-0.029}$ \\
			\tableline
			\multicolumn{2}{c}{MESA (without rotation)} \\
			\tableline
			Age (Myr) 				& $575^{+45}_{-41}$ \\
			Mass (M$_{\sun}$) 				& $1.817^{+0.027}_{-0.027}$ \\
			\tableline
			\multicolumn{2}{c}{Padova (without rotation)} \\
			\tableline
			Age (Myr) 				& $580^{+54}_{-56}$ \\
			Mass (M$_{\sun}$) 				& $1.855^{+0.027}_{-0.029}$ \\
			\tableline\tableline
		\end{tabular}
	\end{center}
	\end{table}

	\subsection{A New Age Estimate for the UMa Moving Group}
		\label{disc:fin_age}
		The mean age, uncertainty in the mean, and standard deviation of the 7 Ursa Major moving group A-stars presented here are 451, 32, and 86 Myr when using the vZ gravity darkening law 
			and 451, 37, and 98 Myr when using the ELR law.
		These large standard deviations are due in large part to the relatively old age we estimate for Chow ($659^{+11}_{-10}$ Myr for the vZ law or $610^{+14}_{-35}$ Myr for the ELR law).
		
		The discrepant age for Chow questions its association with the moving group.
		Of the seven stars studied here, Chow is one of two stars considered to be a ``probable member" by \cite{king_2003}; the other five are classified as members.
		As assembled in \cite{king_2003}, its space motion is consistent with that of nucleus members, despite being 23 pc further away (Table \ref{tab:sample}). 
		Since we cannot confidently exclude Chow as a member, we give statistics both with and without it.
		If Chow is excluded, we determine a mean age and standard deviation for the 6 remaining stars to be 416 $\pm$ 11 Myr when using the vZ law and 424 $\pm$ 79 Myr when using the ELR law.
		
\begin{figure*}
	\subfloat[\label{fig:MvA_vZ}]{\includegraphics[height =2.5in]{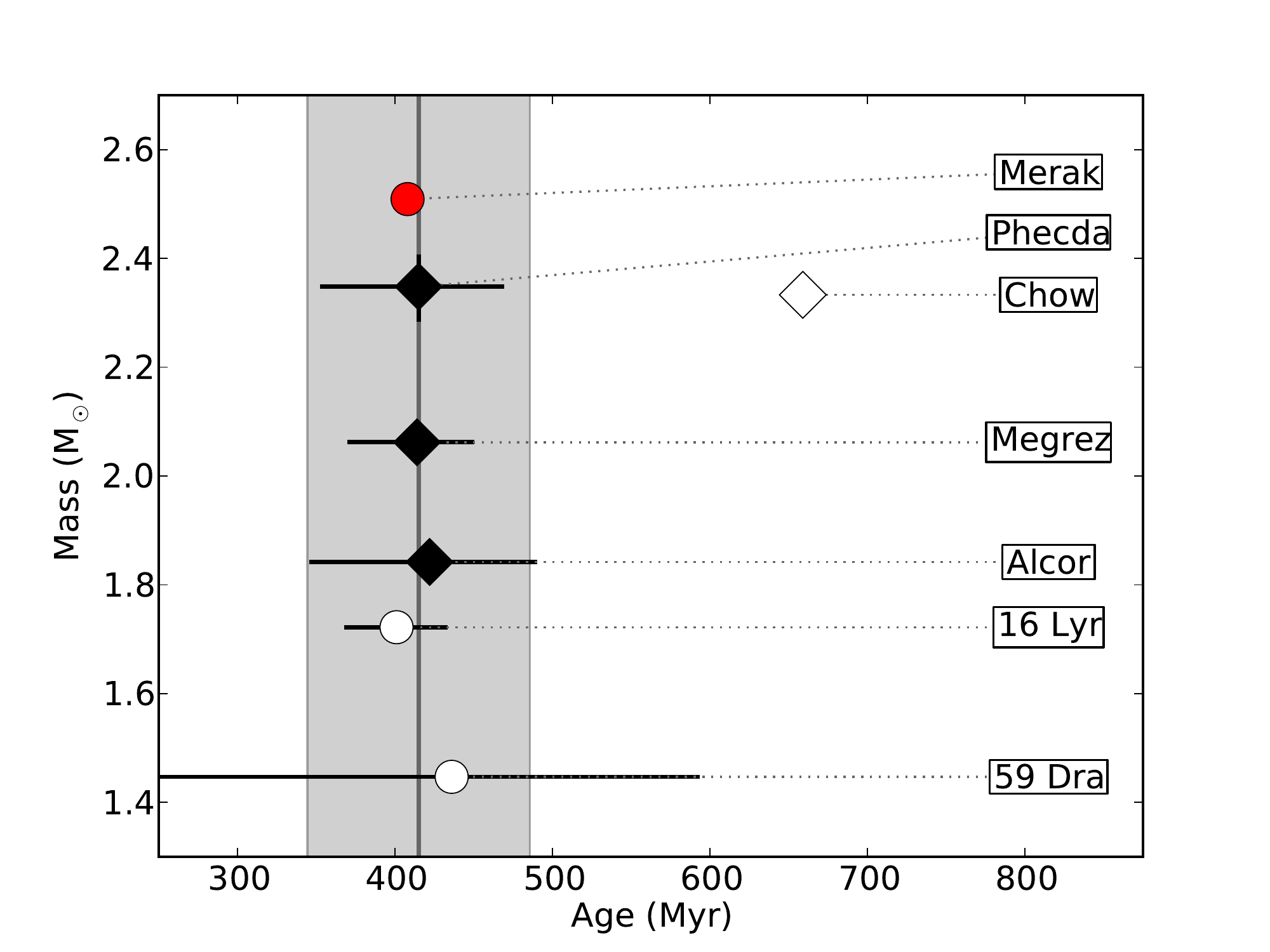}}
	\subfloat[\label{fig:MvA_ELR}]{\includegraphics[height =2.5in]{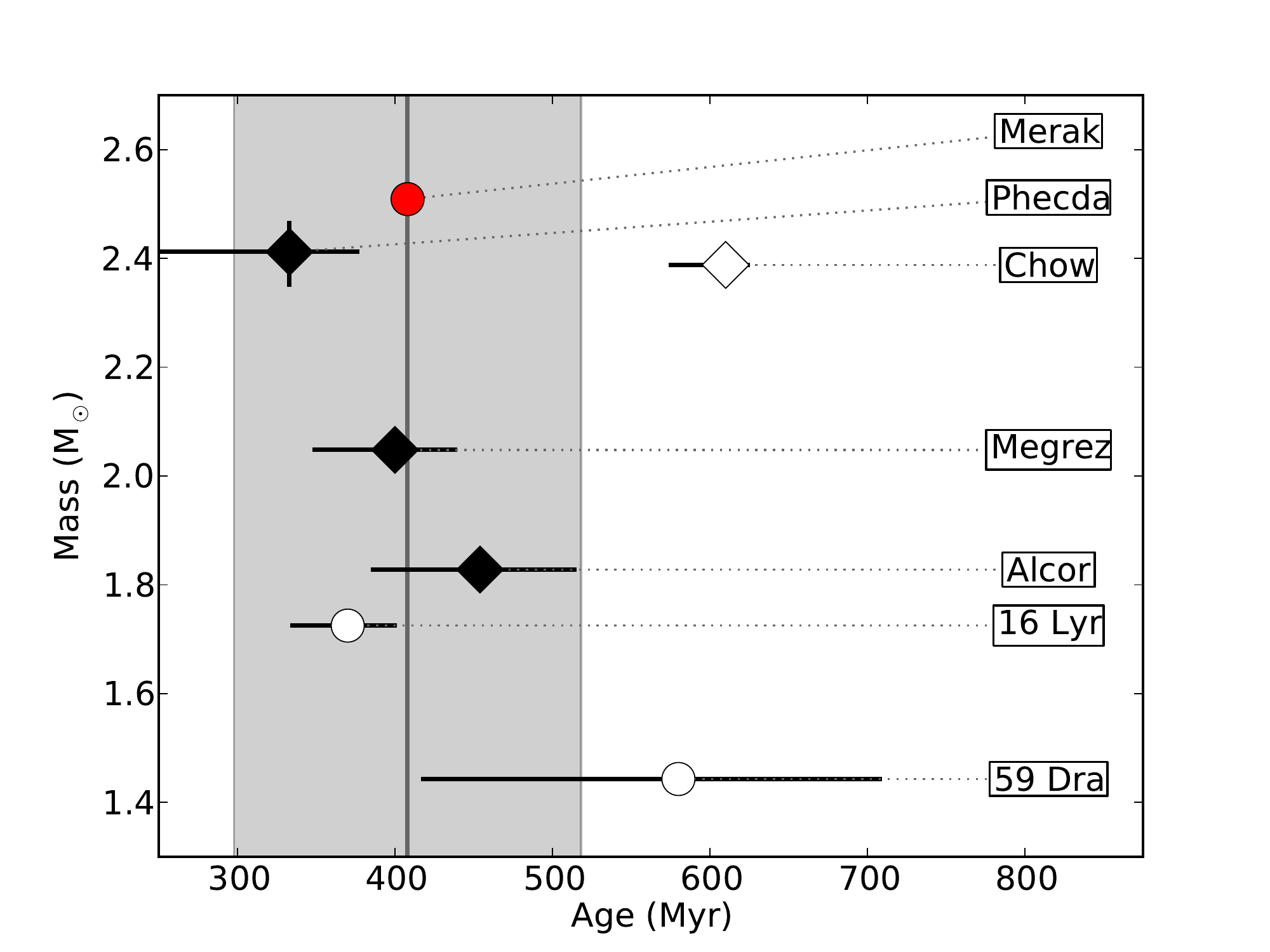}} \\
	\subfloat[\label{fig:MvA_both}]{\includegraphics[height =2.5in]{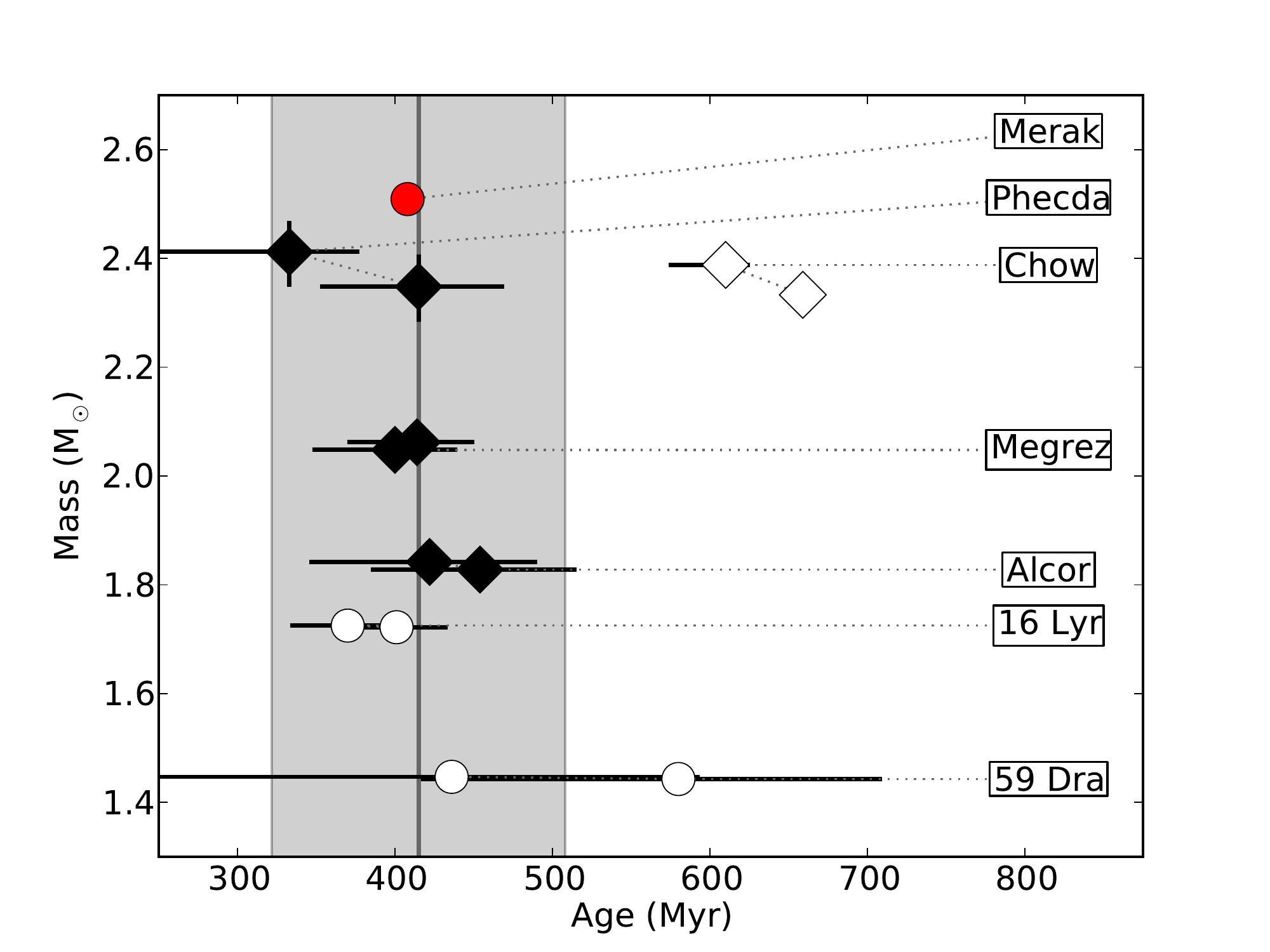}}
	\caption{
	Distribution of stellar masses versus age for 7 stars in the Ursa Major moving group as determined using the vZ gravity darkening law (\ref{fig:MvA_vZ}), ELR law (\ref{fig:MvA_ELR}), and both (\ref{fig:MvA_both}) 
		with the model described in Section \ref{sec:omod}.
	The circles are slowly rotating stars ($V_\mathrm{e} < 170$ $\mathrm{km~s^{-1}}$) and the diamonds are rapidly rotating ($V_\mathrm{e} > 170$ $\mathrm{km~s^{-1}}$).
	The black points are nucleus members and the white points are stream members. 
	The red point shows the mass and age of the nucleus member, Merak, that was previously observed by \cite{boyajian_2012} and is discussed here in Section \ref{sec:merak}. 
	In some cases, the size of the statistical error bar is smaller than the size of the symbol.
	The dark vertical lines represent the median in the ages, the shaded regions represent the gapper scale (the standard deviation equivalent discussed in Section \ref{disc:fin_age}).
	The dotted lines in \ref{fig:MvA_both} connect the age and mass estimates from the two different laws.}
	\label{fig:MvA}
\end{figure*}

		A primary goal of this work is to use the ensemble of stellar ages to provide a new, independent age estimate for the Ursa Major moving group. 
		The distributions of individual ages in Figure \ref{fig:MvA}, however, illustrates the challenge of doing this robustly as the determined ages contain systematic uncertainties (e.g., gravity darkening), a broad range of 
			statistical uncertainties (that can bias weighted values), and possible non-members (e.g., Chow).
		\cite{beers_1990} discuss a variety of statistically robust techniques for computing the central location (``mean") and scale (``dispersion") of small samples that are potentially contaminated with outliers or that have 
			and unknown underlying distribution. 
		Following their recommendations, we choose to compute a median for the central location of the age and use a technique known as the ``gapper" to estimate the dispersion in our sample (see \citealt{wainer_1976}).
		A median is better in this case because it is influenced much less by any individual point than a mean would be.
		A median is also preferred over a weighted mean for this sample because of the broad range of uncertainties that may not account for all systematic uncertainties.
		The gapper method is based on the size of the intervals (or ``gaps") in an ordered set of measurements with the ``gaps" near the median being weighted more heavily. 
		The gapper is normalized such that it is equivalent to a standard deviation.
		The median age and gapper scale ($\sigma_\mathrm{g}$) of the seven A-stars presented here are 415 $\pm$ 71 Myr when using the vZ law and 408 $\pm$ 110 when using the ELR law.
		
		Since the gapper scale is intended to approximate the standard deviation for a Gaussian distribution, we use it to define an uncertainty in the median as $\frac{\sigma_g}{\sqrt{n}}$, following standard convention.
		The median, gapper scale, uncertainty in the median, mean, and standard deviation are presented in Table \ref{tab:age_subsamps} for three distinct subsamples of the seven stars observed.
		The first of these subsamples is the four nucleus stars (Merak, Phecda, Megrez, and Alcor) which are considered bona fide members of the moving group, and so are of greater interest in determining the age of the group.
		We find a median age and gapper scale of 415 $\pm$ 6 Myr and 404 $\pm$ 55 Myr for the vZ and ELR laws, respectively.
		The second of these samples is the full sample of seven stars with an age of 415 $\pm$ 71 Myr (vZ) and 408 $\pm$ 110 Myr (ELR). 
		The final sample is the full sample excluding Chow which, due to its estimated old age, may be an interloper.
		Without Chow, we find a vZ age of 415 $\pm$ 13 Myr and an ELR age of 404 $\pm$ 88 Myr.
		
		As discussed in Section \ref{disc:phot_prop}, the model results using the two gravity darkening laws show no considerable difference for individual stars.  
		The vZ law, as illustrated in Figures \ref{fig:MvA}-\ref{fig:MvA_nochow}, does yield more consistent age estimates ($\sigma_\mathrm{g}$ = 13 Myr) among the observed stars (excluding Chow) than the 
			ELR law does ($\sigma_\mathrm{g}$ = 88 Myr).
		However, given that many of the uncertainties in the individual measurements are as large or larger than the dispersion in the age estimates, we consider that this may be a statistical anomaly.
		Because of this, we hesitate to favor one law over the other.
		
		To estimate the age of the moving group, we combine the following into one set of age estimates: the age of Merak determined using the method described in Section \ref{sec:merak}; the ages of Phecda, 
			Megrez, Alcor, 16 Lyr, and 59 Dra as determined using the vZ law; and the ages of those same five stars as determined using the ELR law.
		This combined set of ages allow us to sample what our technique can achieve by accounting for the full spread in ages we estimate using two gravity darkening laws.
		With this combined set, we find the median age and uncertainty in the median of the moving group to be 414 $\pm$ 23 Myr.
		
	\subsection{Model Precision in the Age Estimate for Isolated A-Stars}
		Under the assumption that these stars are the same age, the resulting coeval ages provide validation of not only the model presented here, but also the MESA evolution model and the physics assumed therein.
		The dispersion of ages can be used to quantify the precision of this technique when applied to isolated adolescent-age A-stars.
		Only three stars (Phecda, Megrez, and Alcor) of the observed seven are both considered bona fide nucleus members of the moving group and were fully modeled by the technique presented in Section \ref{sec:omod}.
		The median and gapper scale of their six age estimates (an age estimate using the vZ law and one using the ELR law for each star) is 415 $\pm$ 40 Myr.
		We use this scale value to determine a precision in our model of $\sim$10\% for stars with masses ranging from $\sim$1.8 - 2.4 M$_\sun$ and at a few hundred Myr age.
		Therefore when using this technique on field A-stars we expect an overall uncertainty of 10\% in the age estimates.
	
\section{Summary}
	We present new interferometric observations for six A-type stars in the Ursa Major moving group and new age and mass estimates for these six plus one that was observed previously. 
	All observations were made using the Classic, CLIMB, and PAVO beam combiners on the CHARA Array. 
	Four of the observed stars are known to be rapidly rotating with $v \sin i$ $\gtrsim$ 170 $\mathrm{km~s^{-1}}$ causing them to be measurably oblate. 
	
	To properly account for this rapid rotation, a model was constructed with a Roche geometry based on eight parameters: $R_\mathrm{e}$, $M_\mathrm{*}$, $V_\mathrm{e}$, $i$, $\beta$, $T_\mathrm{p}$, $\pi_\mathrm{plx}$, and $\psi$. 
	Visibilities and photometry were calculated for each star using model-generated images and PEDs, and then compared to measured visibilities and photometry. 
	Five of the model parameters ($R_\mathrm{e}$, $V_\mathrm{e}$, $i$, $T_\mathrm{p}$, and $\psi$) were allowed to vary, with $V_\mathrm{e}$ constrained by $i$ and the measured $v \sin i$. 
	Age and mass estimates were made for each of the stars in this sample by comparing their modeled average radius, luminosity, and equatorial velocity of to those parameters determined by MESA evolution models. 
	The mass determined by the MESA model was then used in the Roche model and this process was repeated until the models converged.
	
	Two different gravity darkening laws were studied. 
	Neither law was favored by the interferometric and photometric data, nor was either law favored by the final age estimates.
	The dispersion in the age estimates was significantly smaller for the ages estimated using the vZ law than the ELR law.
	However, because this dispersion is of the same order of magnitude as the statistical uncertainties in the ages, we consider that this may be a statistical anomaly. 
	The age estimated for Chow makes it older than the moving group as a whole and is thus excluded as a potential interloper in our final age estimate.
	Because neither gravity darkening law was favored, we combined the ages estimated with the vZ and ELR laws to determine the overall age of the moving group.
	
	By determining the ages of these coeval stars, we validate this technique for use on individual field stars and determine a model uncertainty of approximately 10\% for stars with masses ranging from $\sim$1.8 - 2.5 M$_{\sun}$.
	Using the independent technique described here, we find the age of the Ursa Major moving group to be 414 $\pm$ 23 Myr. 
	This result is consistent with previous age estimates for the Ursa Major moving group.
	
\acknowledgements
	The authors would like to thank John Monnier for his suggestions on how to construct the rapid rotator model and for providing test data, Michel Rieutord for his help in calculating $\beta$ for the ELR gravity darkening law, Doug Gies 
		for his suggestions on how to handle model SEDs, and Brian Kloppenborg for his advice in making our model more computationally efficient.
	J.J. and R.W. acknowledge support from the NSF AAG grant number 1009643. 
	TSB acknowledged partial support from grants 12ADAP120172, 14-XRP14\_2-0147, and ADAP14-0245.
	
\begin{table*}
\begin{center}
	\caption{Model Results using the vZ gravity darkening law. \label{tab:fast_res_old_beta}}
	\begin{tabular}{ccccccc}
		\tableline\tableline
									& Phecda 		& Megrez 		& Alcor 			& Chow 			& 16 Lyr 			& 59 Dra \\
		\tableline
		HD Number 						& 103287 		& 106591 		& 116842 		& 141003 		& 177196 		& 180777 \\
		\tableline
		Equatorial Radius, $R_\mathrm{e}$ (R$_{\sun}$) 			& $3.435^{+0.154}_{-0.148}$ 	& $2.512^{+0.075}_{-0.076}$ 	& $2.002^{+0.068}_{-0.067}$ 	& $4.486^{+0.098}_{-0.082}$ 	& $1.664^{+0.025}_{-0.023}$ 	& $1.524^{+0.033}_{-0.035}$ \\
		Equatorial Velocity, $V_\mathrm{e}$ ($\mathrm{km~s^{-1}}$) 	& $374.7^{+15.0}_{-15.1}$ 	& $310.4^{+9.8}_{-8.7}$ 	& $238.6^{+10.0}_{-9.2}$ 	& $307.8^{+5.3}_{-5.9}$ 	& $101.6^{+14.1}_{-17.4}$ 	& $131.8^{+24.8}_{-27.2}$ \\
		Inclination, $i$ ($^\circ$) 					& $27.1^{+5.4}_{-6.1}$ 	& $52.0^{+3.6}_{-3.2}$ 	& $90.0^{+0.0}_{-19.0}$ 	& $44.8^{+1.5}_{-1.7}$ 	& $^a56.9^{+30.1}_{-25.1}$ 	& $^a28.2^{+20.5}_{-25.7}$ \\
		Polar Temperature, $T_\mathrm{p}$ (K) 			& $11138^{+220}_{-225}$ 	& $10030^{+129}_{-139}$ 	& $8985^{+116}_{-124}$ 	& $10091^{+89}_{-84}$ 	& $8242^{+56}_{-53}$ 	& $7231^{+68}_{-72}$ \\
		Polar Position Angle, $\psi$ ($^{\circ}$) 			& $12.1^{+71.6}_{-54.6}$	& $51.6^{+42.9}_{-43.4}$	& $154.9^{+71.4}_{-74.8}$	& $161.3^{+19.8}_{-20.2}$	& $82.5^{+15.0}_{-15.0}$	& $6.1^{+75.0}_{-56.9}$ \\
		\tableline
		Gravity Darkening, $\beta$ 					& $0.25$ 			& $0.25$ 			& $0.25$ 			& $0.25$ 			& $0.25$		 	& $0.25$ \\
		Angular Rotation Rate, $\omega$ 				& $0.999^{+0.001}_{-0.003}$ 	& $0.964^{+0.008}_{-0.008}$ 	& $0.835^{+0.020}_{-0.019}$ 	& $0.999^{+0.001}_{-0.001}$ 	& $0.404^{+0.051}_{-0.065}$ 	& $0.530^{+0.082}_{-0.098}$ \\
		Polar Radius, $R_\mathrm{p}$ (R$_{\sun}$)  			& $2.233^{+0.064}_{-0.064}$ 	& $1.921^{+0.044}_{-0.044}$ 	& $1.723^{+0.050}_{-0.050}$ 	& $3.037^{+0.045}_{-0.038}$ 	& $1.622^{+0.023}_{-0.022}$ 	& $1.455^{+0.030}_{-0.032}$ \\
		Average Radius, $R_\mathrm{avg}$ (R$_{\sun}$)  			& $2.557^{+0.077}_{-0.079}$ 	& $2.147^{+0.053}_{-0.054}$ 	& $1.846^{+0.057}_{-0.057}$ 	& $3.479^{+0.053}_{-0.045}$ 	& $1.643^{+0.024}_{-0.023}$ 	& $1.488^{+0.031}_{-0.033}$ \\
		Average Diameter, $\theta_\mathrm{avg}$ (mas) 		& $0.932^{+0.028}_{-0.029}$ 	& $0.808^{+0.020}_{-0.020}$ 	& $0.684^{+0.021}_{-0.021}$ 	& $0.680^{+0.010}_{-0.009}$ 	& $0.408^{+0.006}_{-0.006}$ 	& $0.507^{+0.011}_{-0.011}$ \\
		Equatorial Temperature, $T_\mathrm{e}$ (K) 			& $4724^{+914}_{-1953}$ 	& $6909^{+195}_{-234}$ 	& $7556^{+109}_{-123}$ 	& $3825^{+634}_{-1116}$ 	& $8028^{+67}_{-63}$ 	& $6887^{+126}_{-140}$ \\
		$v \sin i$ ($\mathrm{km~s^{-1}}$) 				& $171.0^{+30.8}_{-36.2}$ 	& $244.6^{+11.6}_{-11.1}$ 	& $238.6^{+10.0}_{-13.0}$ 	& $217.0^{+5.6}_{-6.5}$ 	& $85.1^{+16.3}_{-31.6}$ 	& $62.3^{+36.8}_{-56.6}$ \\
		Total Luminosity, $L_\mathrm{tot}$ (L$_{\sun}$) 			& $42.37^{+3.47}_{-3.34}$ 	& $23.00^{+1.21}_{-1.24}$ 	& $13.98^{+0.75}_{-0.75}$ 	& $52.87^{+1.88}_{-1.73}$ 	& $10.45^{+0.30}_{-0.28}$ 	& $4.861^{+0.285}_{-0.290}$ \\
		Apparent Luminosity, $L_\mathrm{app}$ (L$_{\sun}$) 		& $66.94^{+5.55}_{-5.34}$ 	& $23.88^{+1.26}_{-1.28}$ 	& $11.84^{+0.68}_{-0.64}$ 	& $64.00^{+2.28}_{-2.11}$ 	& $10.42^{+0.30}_{-0.28}$ 	& $5.126^{+0.213}_{-0.222}$ \\
		Age (Myr) 						& $415^{+53}_{-61}$ 	& $414^{+35}_{-43}$ 	& $422^{+67}_{-75}$ 	& $659^{+11}_{-10}$ 	& $401^{+31}_{-32}$ 	& $436^{+156}_{-203}$ \\
		Mass (M$_{\sun}$) 						& $2.348^{+0.055}_{-0.060}$ 	& $2.062^{+0.030}_{-0.033}$ 	& $1.842^{+0.027}_{-0.031}$ 	& $2.333^{+0.015}_{-0.015}$ 	& $1.722^{+0.013}_{-0.013}$ 	& $1.447^{+0.014}_{-0.015}$ \\
		\tableline
		Visibility $\chi^2$ 						& 7.646 			& 2.719 			& 4.498 			& 0.763 			& 1.083			& 1.488 \\
		Photometry $\chi^2$ 					& 5.798 			& 3.214 			& 4.021 			& 2.329 			& 6.313			& 5.100 \\
		Total $\chi^2$ 						& 13.45 			& 5.933 			& 8.519 			& 3.092 			& 7.396			& 6.588 \\
		\tableline
	\end{tabular}
\end{center}
\small
$^a$As discussed in Section \ref{sec:slowrot}, the inclinations of 16 Lyr and 59 Dra were fixed.
\end{table*}

\begin{table*}
\begin{center}
	\caption{Model Results using the ELR gravity darkening law. \label{tab:fast_res_new_beta}}
	\begin{tabular}{cccccccc}
		\tableline\tableline
									& Phecda 		& Megrez 		& Alcor 			& Chow 			& 16 Lyr 			& 59 Dra \\
		\tableline
		HD Number 						& 103287 		& 106591 		& 116842 		& 141003 		& 177196 		& 180777 \\
		\tableline
		Equatorial Radius, $R_\mathrm{e}$ (R$_{\sun}$) 			& $3.385^{+0.204}_{-0.257}$ 	& $2.511^{+0.074}_{-0.068}$ 	& $2.001^{+0.062}_{-0.065}$ 	& $4.195^{+0.092}_{-0.084}$ 	& $1.651^{+0.023}_{-0.024}$ 	& $1.518^{+0.033}_{-0.033}$ \\
		Equatorial Velocity, $V_\mathrm{e}$ ($\mathrm{km~s^{-1}}$) 	& $386.3^{+10.5}_{-8.4}$ 	& $318.9^{+15.5}_{-15.6}$ 	& $234.1^{+12.9}_{-11.8}$ 	& $282.2^{+10.1}_{-9.7}$ 	& $101.3^{+16.1}_{-17.6}$ 	& $100.9^{+32.0}_{-42.3}$ \\
		Inclination, $i$ ($^\circ$) 					& $28.5^{+5.7}_{-5.9}$ 	& $50.0^{+4.0}_{-4.1}$ 	& $86.8^{+2.9}_{-17.3}$ 	& $50.1^{+2.7}_{-2.7}$ 	& $^a56.9^{+30.1}_{-24.7}$ 	& $^a28.2^{+34.2}_{-25.9}$ \\
		Polar Temperature, $T_\mathrm{p}$ (K) 			& $10520^{+194}_{-220}$ 	& $9550^{+143}_{-126}$ 	& $8762^{+112}_{-119}$ 	& $9539^{+104}_{-93}$ 	& $8270^{+53}_{-57}$ 	& $7164^{+68}_{-68}$ \\
		Polar Position Angle, $\psi$ ($^{\circ}$) 			& $18.4^{+72.6}_{-54.3}$	& $50.9^{+44.4}_{-42.6}$ 	& $154.0^{+71.7}_{-74.5}$	& $159.8^{+27.0}_{-25.0}$	& $13.0^{+25.9}_{-25.6}$	& $161.2^{+1.0}_{-37.2}$ \\
		\tableline
		Gravity Darkening, $\beta$				 	& $0.138^{+0.008}_{-0.019}$ 	& $0.170^{+0.007}_{-0.007}$ 	& $0.207^{+0.004}_{-0.004}$ 	& $0.161^{+0.005}_{-0.006}$ 	& $0.242^{+0.003}_{-0.003}$ 	& $0.241^{+0.006}_{-0.006}$ \\
		Angular Rotation Rate, $\omega$ 				& $0.999^{+0.001}_{-0.002}$ 	& $0.972^{+0.011}_{-0.014}$ 	& $0.827^{+0.026}_{-0.026}$ 	& $0.985^{+0.006}_{-0.007}$ 	& $0.401^{+0.058}_{-0.066}$ 	& $0.417^{+0.116}_{-0.168}$ \\
		Polar Radius, $R_\mathrm{p}$ (R$_{\sun}$)  			& $2.186^{+0.083}_{-0.110}$ 	& $1.893^{+0.046}_{-0.045}$ 	& $1.729^{+0.046}_{-0.048}$ 	& $3.070^{+0.057}_{-0.059}$ 	& $1.609^{+0.022}_{-0.023}$ 	& $1.477^{+0.031}_{-0.031}$ \\
		Average Radius, $R_\mathrm{avg}$ (R$_{\sun}$)  			& $2.500^{+0.088}_{-0.121}$ 	& $2.124^{+0.051}_{-0.048}$ 	& $1.849^{+0.053}_{-0.055}$ 	& $3.472^{+0.061}_{-0.056}$ 	& $1.630^{+0.023}_{-0.024}$ 	& $1.497^{+0.032}_{-0.032}$ \\
		Average Diameter, $\theta_\mathrm{avg}$ (mas) 		& $0.912^{+0.032}_{-0.044}$ 	& $0.800^{+0.019}_{-0.018}$ 	& $0.685^{+0.020}_{-0.020}$ 	& $0.678^{+0.012}_{-0.011}$ 	& $0.405^{+0.006}_{-0.006}$ 	& $0.510^{+0.011}_{-0.011}$ \\
		Equatorial Temperature, $T_\mathrm{e}$ (K) 			& $6751^{+304}_{-1025}$ 	& $7244^{+199}_{-218}$ 	& $7630^{+97}_{-108}$ 	& $6967^{+170}_{-199}$ 	& $8066^{+63}_{-67}$ 	& $6972^{+126}_{-132}$ \\
		$v \sin i$ ($\mathrm{km~s^{-1}}$) 				& $184.5^{+32.6}_{-35.6}$ 	& $244.2^{+13.6}_{-15.4}$ 	& $233.7^{+12.9}_{-14.4}$ 	& $216.6^{+8.1}_{-8.7}$ 	& $84.9^{+16.3}_{-30.9}$ 	& $47.7^{+41.7}_{-43.6}$ \\
		Total Luminosity, $L_\mathrm{tot}$ (L$_{\sun}$) 			& $44.57^{+3.39}_{-3.61}$ 	& $22.04^{+1.34}_{-1.14}$ 	& $13.67^{+0.72}_{-0.74}$ 	& $58.17^{+2.57}_{-2.25}$ 	& $10.45^{+0.29}_{-0.30}$ 	& $4.966^{+0.302}_{-0.292}$ \\
		Apparent Luminosity, $L_\mathrm{app}$ (L$_{\sun}$) 		& $64.74^{+4.99}_{-5.32}$ 	& $23.33^{+1.43}_{-1.20}$ 	& $11.85^{+0.66}_{-0.66}$ 	& $61.72^{+2.73}_{-2.39}$ 	& $10.42^{+0.29}_{-0.31}$ 	& $5.118^{+0.219}_{-0.216}$ \\
		Age (Myr) 						& $333^{+43}_{-83}$ 	& $400^{+38}_{-51}$ 	& $454^{+60}_{-68}$ 	& $610^{+14}_{-35}$ 	& $370^{+30}_{-35}$ 	& $580^{+128}_{-162}$ \\
		Mass (M$_{\sun}$) 						& $2.412^{+0.053}_{-0.060}$ 	& $2.048^{+0.035}_{-0.030}$ 	& $1.828^{+0.027}_{-0.030}$ 	& $2.388^{+0.036}_{-0.021}$ 	& $1.725^{+0.013}_{-0.014}$ 	& $1.443^{+0.015}_{-0.015}$ \\
		\tableline
		Visibility $\chi^2$ 						& 6.897 			& 2.664 			& 4.481 			& 1.080 			& 1.141			& 1.542 \\
		Photometry $\chi^2$ 					& 6.045 			& 4.133 			& 4.235 			& 3.835 			& 6.265			& 5.060 \\
		Total $\chi^2$ 						& 12.94 			& 6.797 			& 8.716 			& 4.915 			& 7.406			& 6.602 \\
		\tableline
	\end{tabular}
\end{center}
\small
$^a$As discussed in Section \ref{sec:slowrot}, the inclinations of 16 Lyr and 59 Dra were fixed.
\end{table*}

\begin{table}
\begin{center}
	\caption{Fundamental properties of Merak (HD 95418). \label{tab:merak}}
	\begin{tabular}{ccc}
		\tableline\tableline
							& Value 			& Source\\
		\tableline
		Radius (R$_{\sun}$) 				& $3.0210 \pm 0.0383$ 	& \cite{boyajian_2012} \\
		Temperature (K) 				& $9193 \pm 56 $ 		& \cite{boyajian_2012} \\
		Luminosity, $L_\mathrm{tot}$ (L$_{\sun}$) 	& $58.46 \pm 0.47$ 	& \cite{boyajian_2012} \\
		$v \sin i$ ($\mathrm{km~s^{-1}}$) 		& $46 \pm 2.3$ 		& \cite{royer_3} \\
		Inclination, $i$ ($^\circ$) 			& 90 			& Assumed \\
		Age (Myr) 				& $408 \pm 6$ 		& This work \\
		Mass (M$_{\sun}$) 				& $2.509\pm 0.005$ 	& This work \\
		\tableline\tableline
	\end{tabular}
\end{center}
\end{table}

\begin{table*}
\begin{center}
	\caption{Ages and Masses for Individual Stars. \label{tab:age_mass}}
	\begin{tabular}{ccccc}
		\tableline\tableline
		Star & \multicolumn{2}{c}{Mass (M$_{\sun}$)} & \multicolumn{2}{c}{Age (Myr)}  \\
		Name & vZ law & ELR Law & vZ law & ELR Law \\
		\tableline
		Merak 	& \multicolumn{2}{c}{$2.509 \pm 0.005$} 		& \multicolumn{2}{c}{$408 \pm 6$}			\\
		Phecda 	& $2.348^{+0.055}_{-0.060}$ 	& $2.412^{+0.053}_{-0.060}$ 	& $415^{+53}_{-61}$ 	& $333^{+43}_{-83}$ 	\\
		Megrez 	& $2.062^{+0.030}_{-0.033}$ 	& $2.048^{+0.035}_{-0.030}$ 	& $414^{+35}_{-43}$ 	& $400^{+38}_{-51}$ 	\\
		Alcor 	& $1.842^{+0.027}_{-0.031}$ 	& $1.828^{+0.027}_{-0.030}$ 	& $422^{+67}_{-75}$ 	& $454^{+60}_{-68}$ 	\\
		Chow 	& $2.333^{+0.015}_{-0.015}$ 	& $2.388^{+0.036}_{-0.021}$ 	& $659^{+11}_{-10}$ 	& $610^{+14}_{-35}$ 	\\
		16 Lyr 	& $1.722^{+0.013}_{-0.013}$ 	& $1.725^{+0.013}_{-0.014}$ 	& $401^{+31}_{-32}$ 	& $370^{+30}_{-35}$ 	\\
		59 Dra 	& $1.447^{+0.014}_{-0.015}$ 	& $1.443^{+0.015}_{-0.015}$ 	& $436^{+156}_{-203}$ 	& $580^{+128}_{-162}$ 	\\
		\tableline\tableline
	\end{tabular}
\end{center}
\end{table*}


\begin{figure*}
	\subfloat[\label{fig:MvA_vZ_nochow}]{\includegraphics[height =2.5in]{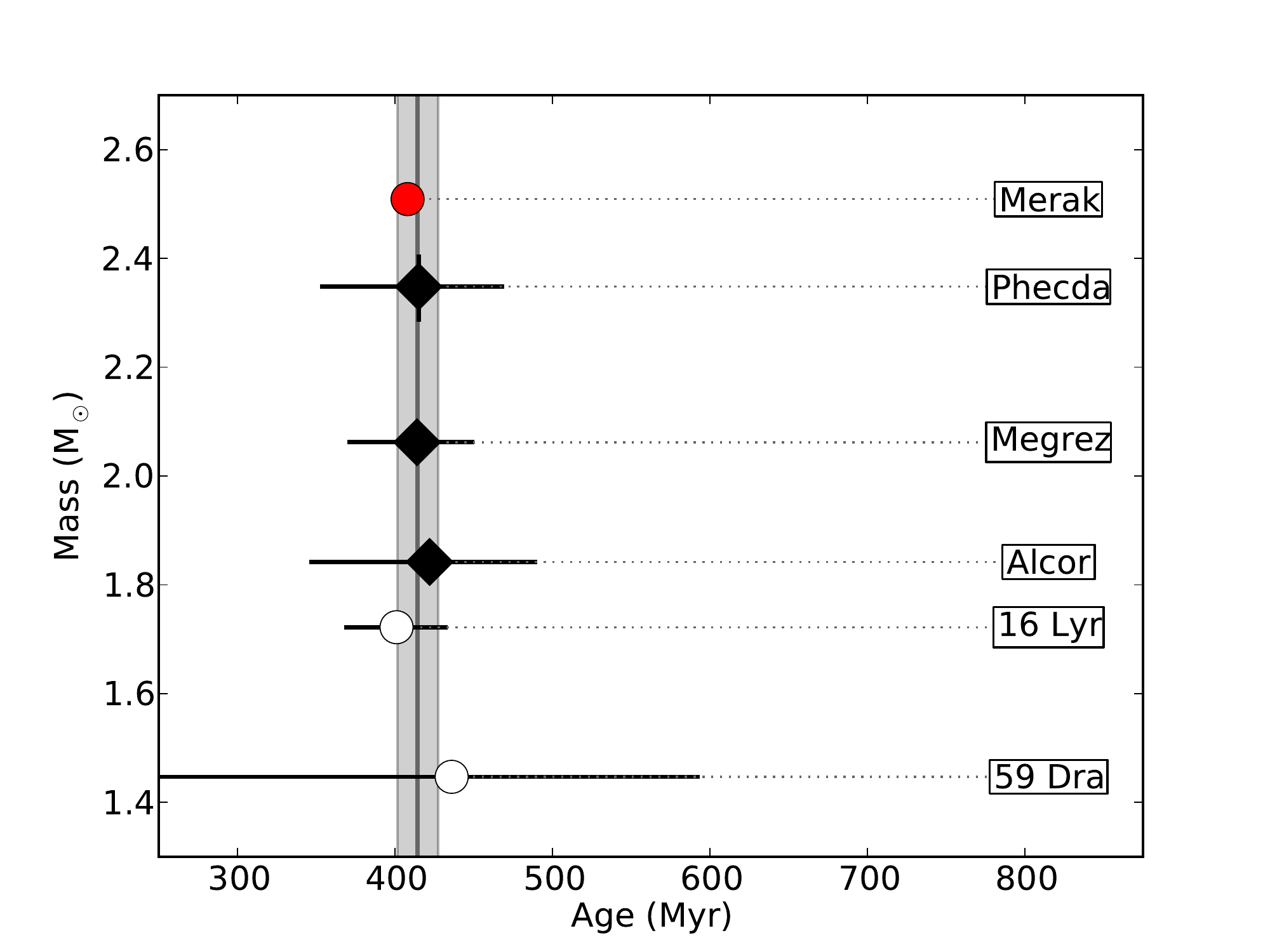}}
	\subfloat[\label{fig:MvA_ELR_nochow}]{\includegraphics[height =2.5in]{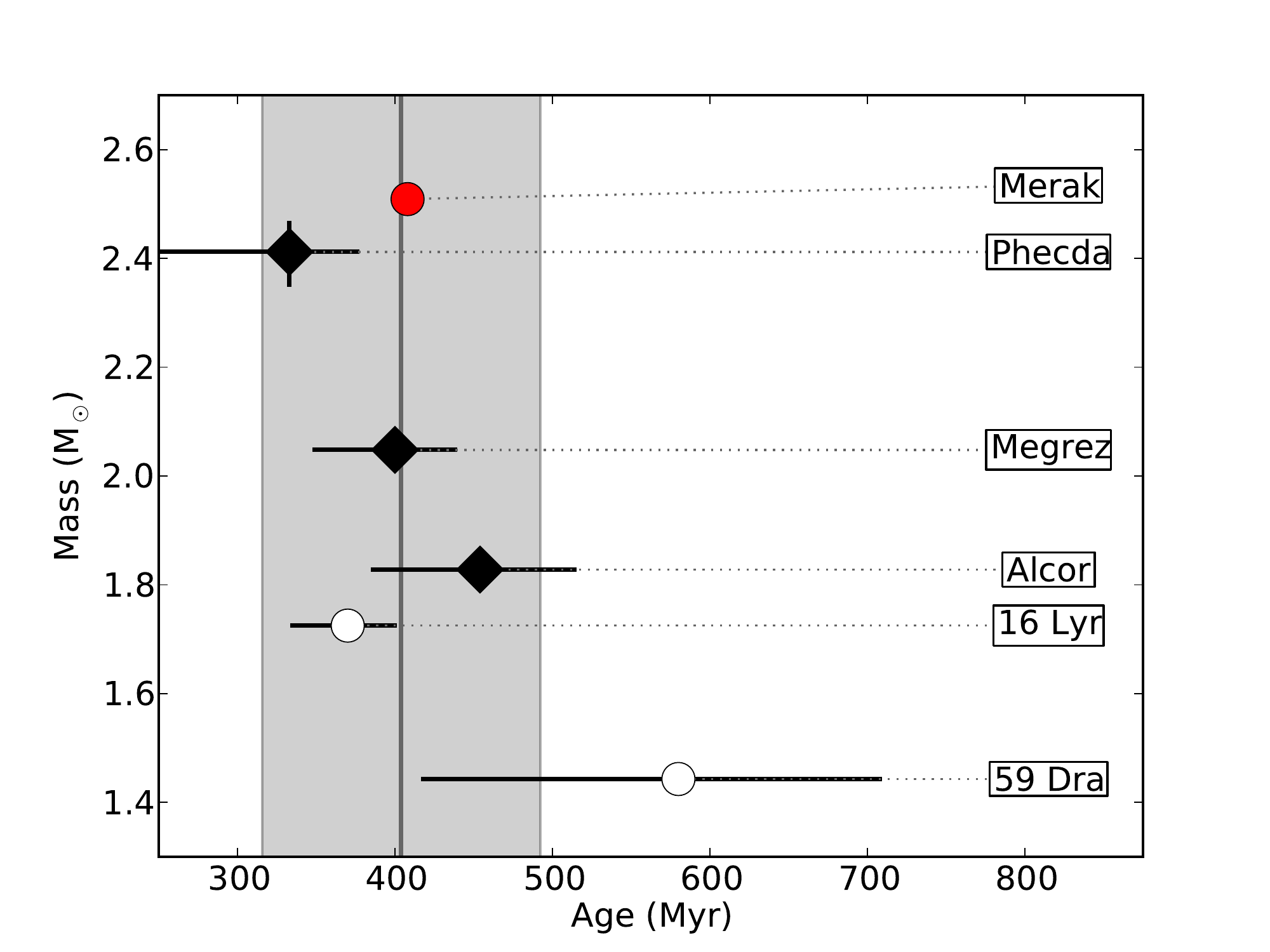}} \\
	\subfloat[\label{fig:MvA_nochow}]{\includegraphics[height =2.5in]{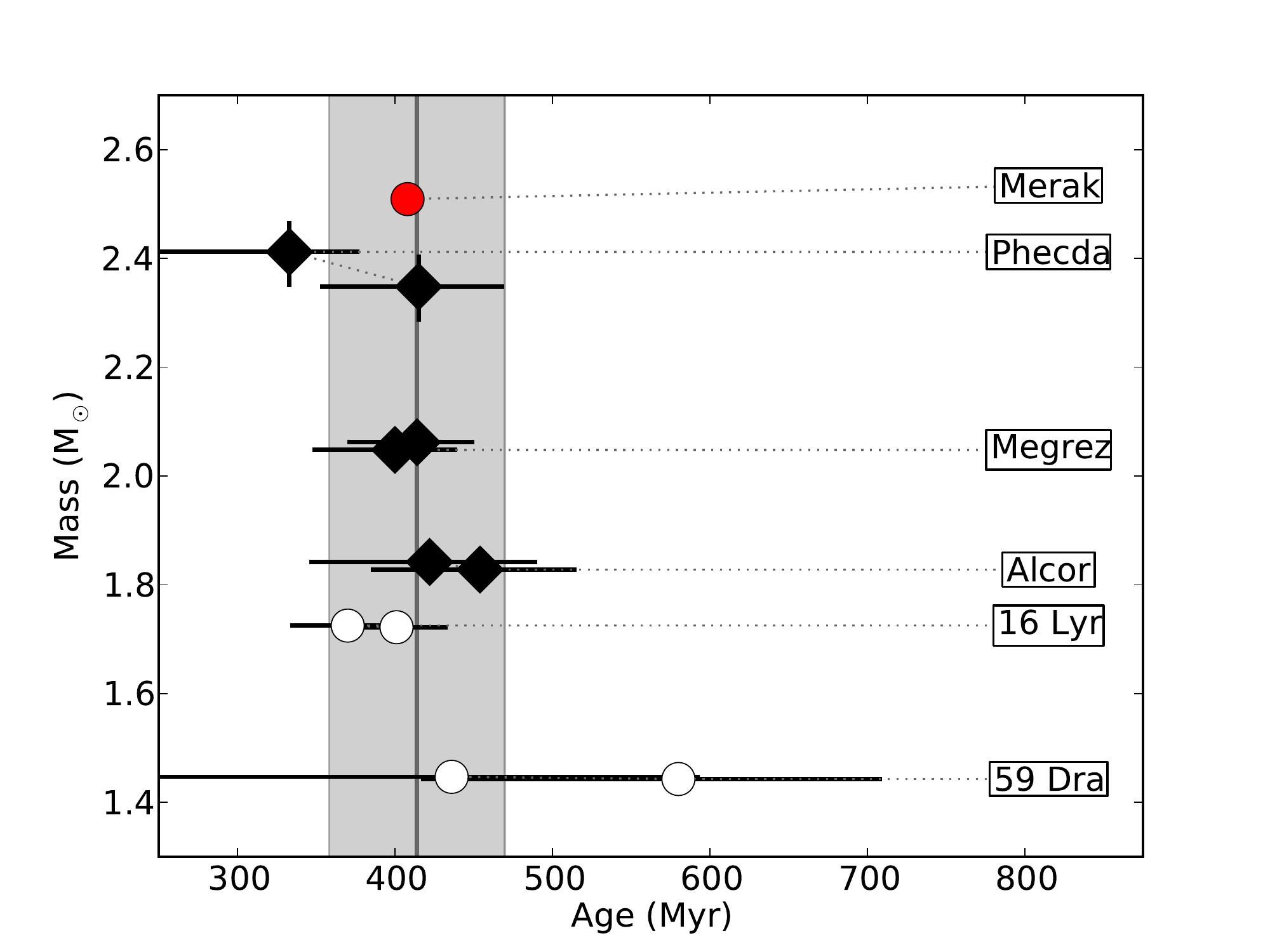}}
	\caption{
	Same as Figure \ref{fig:MvA}, but excluding Chow.}
	\label{fig:MvA_nochow}
\end{figure*}

\begin{table*}
\begin{center}
	\caption{
	Age Estimates and Uncertainties (in Myr) for Various Subsets
	\label{tab:age_subsamps}}
	\begin{tabular}{cccccccccc}
		\tableline\tableline
						&  	&	& \multicolumn{2}{c}{vZ law}						& \multicolumn{2}{c}{ELR law}						& \multicolumn{3}{c}{Combined} 				\\ 
						& n	& n$^*$	& Mean $\pm$ $\sigma$ 	& Median $\pm$ $\sigma_\mathrm{g}$		& Mean $\pm$ $\sigma$ 	& Median $\pm$ $\sigma_\mathrm{g}$		& Mean $\pm$ $\sigma$ 	& Median $\pm$ $\sigma_\mathrm{g}$	& $\frac{\sigma_\mathrm{g}}{\sqrt{n}}$	\\
		\tableline
		Nucleus Members 			& 4	& 7	& 415 $\pm$ 5		& 415 $\pm$ 6 				& 399 $\pm$ 43		& 404 $\pm$ 55 				& 407 $\pm$ 34		& 414 $\pm$ 35 			& 17				\\
		All Members 			& 7 	& 13 	& 451 $\pm$ 86		& 415 $\pm$ 71 				& 451 $\pm$ 98		& 408 $\pm$ 110 				& 454 $\pm$ 95		& 415 $\pm$ 93 			& 35				\\
		All excluding Chow			& 6 	& 11	& 416 $\pm$ 11		& 415 $\pm$ 13				& 424 $\pm$ 79		& 404 $\pm$ 88				& 421 $\pm$ 59		& 414 $\pm$ 56			& 23				\\
		\tableline
	\end{tabular}
\end{center}
\small
n is the number of stars in each subset and also corresponds to the number of age estimates in the vZ and ELR subsets. n$^*$ is the number of age estimates in the combined subsets and corresponds to $\mathrm{2n-1}$.
\end{table*}
\newpage
\appendix
\section{Photometry}
\label{app:phot}

The photometry used to construct photometric energy distributions are presented in Table \ref{tab:phot}.
The optical photometry is taken from \citet[$UBV$]{mermilliod_1991} and \citet[$uvby$]{hauck_1997}.
Values in these compilations are adopted over others because of the large number of observations that they average to compute final values, and because all eight stars are included in these surveys.
\citet{hauck_1997} do not report errors for Str{\"o}mgren y band measurements of Phecda and Megrez. 
For these stars, an error of 0.01 mag is assumed, which is consistent with the photometric uncertainties of stars of similar brightness in their survey. 
In addition, \citet{hauck_1997} do not report any uncertainties in Str{\"o}mgren photometry for 16 Lyr or 59 Dra.  
Near-infrared ($JHK$) photometry are assembled from various sources and is either already in the Johnson photometric system or converted to it. 
2MASS $JHK$ photometry are listed in Table \ref{tab:phot}, but are not adopted for many of the sample stars because they are saturated. 

\textbf{Merak} (HD 95418) - $J$- and $K$-band photometry is adopted from \cite{morel_1978}, and errors of 0.05 mag are assumed. No $H$-band photometry is available.

\textbf{Phecda} (HD 103287) - $J$-band photometry is adopted from \cite{morel_1978}, and errors of 0.05 mag are assumed. $K$-band photometry is adopted from \cite{ducati_2002}. No $H$-band photometry is available. 

\textbf{Megrez} (HD 106591) - $J$-, $H$-, and $K$-band photometry is adopted from \cite{aumann_1991}, and the adopted uncertainties are the average uncertainties of that survey.

\textbf{Alcor} (HD 116842) - $J$-, $H$-, and $K$-band photometry is adopted from \cite{kidger_2003} after converting the $JHK$ measurements found there to the Johnson system with the method found in \cite{alonso_1994}. 
	The adopted uncertainties are the average uncertainties of that survey.

\textbf{Chow} (HD 141003) - $J$-band photometry adopted from \cite{selby_1988} after converting the $J_n$ measurement found there to the Johnson system with the method found there. 
	The adopted uncertainty is the reported uncertainty in that conversion. $K$-band photometry is adopted from \cite{ducati_2002}. No $H$-band photometry is available.

\textbf{16 Lyr} (HD 177196) -While 2MASS $K$-band photometry is unsaturated for this star, it is not adopted because including it causes the model to find a best fit with a $\chi^2$ in the visibility approximately double what it is without the 
		$K$-band value.
	No $J$- or $H$-band photometry is available. 

\textbf{59 Dra} (HD 177196) - 2MASS $H$- and $K$-band photometry is unsaturated for this star, so it is adopted. No $J$-band photometry is available.

\setcounter{table}{0}
\renewcommand{\thetable}{A\arabic{table}}

\begin{table}
	\footnotesize
	\begin{center}
	\caption{Adopted Photometry.\label{tab:phot}}
	\begin{tabular}{cccccccc}
		\tableline\tableline
		Bandpass 	& Merak 			& Phecda 		& Megrez 		& Alcor 			& Chow 			& 16 Lyr 			& 59 Dra \\
		\tableline
				& HD 95418 		& HD 103287 		& HD 106591  		& HD 116842  		& HD 141003 		& HD 177196 		& HD 180777  \\
		\tableline\tableline
		\multicolumn{8}{c}{Adopted Optical Photometry} \\
		\tableline\tableline
		\multicolumn{8}{c}{Mermilliod 1991} \\
		\tableline
		Johnson $U$ 	& 2.349 $\pm$ 0.014 	& 2.451 $\pm$ 0.010 	& 3.460 $\pm$ 0.009 	& 4.260 $\pm$ 0.008 	& 3.816 $\pm$ 0.023 	& 5.281 $\pm$ 0.018 	& 5.440 $\pm$ 0.016 \\
		Johnson $B$ 	& 2.346 $\pm$ 0.011 	& 2.437 $\pm$ 0.006 	& 3.389 $\pm$ 0.009 	& 4.176 $\pm$ 0.007 	& 3.731 $\pm$ 0.013 	& 5.199 $\pm$ 0.015 	& 5.442 $\pm$ 0.016 \\
		Johnson $V$ 	& 2.366 $\pm$ 0.009 	& 2.437 $\pm$ 0.005 	& 3.312 $\pm$ 0.007 	& 4.009 $\pm$ 0.006 	& 3.667 $\pm$ 0.010 	& 5.013 $\pm$ 0.014 	& 5.136 $\pm$ 0.014 \\
		\tableline
		\multicolumn{8}{c}{Hauck 1997} \\
		\tableline
		Str{\"o}mgren $u$ 	& 3.741 $\pm$ 0.022 	& 3.860 $\pm$ 0.016 	& 4.849 $\pm$ 0.011 	& 5.620 $\pm$ 0.015 	& 5.261 $\pm$ 0.029 	& 6.619 			& 6.699 \\
		Str{\"o}mgren $v$ 	& 2.501 $\pm$ 0.012 	& 2.587 $\pm$ 0.011 	& 3.572 $\pm$ 0.010 	& 4.400 $\pm$ 0.012 	& 3.922 $\pm$ 0.019 	& 5.412 			& 5.696 \\
		Str{\"o}mgren $b$ 	& 2.349 $\pm$ 0.006 	& 2.426 $\pm$ 0.010 	& 3.350 $\pm$ 0.010 	& 4.110 $\pm$ 0.011 	& 3.715 $\pm$ 0.012 	& 5.106 			& 5.324 \\
		Str{\"o}mgren $y$ 	& 2.355 $\pm$ 0.005 	& 2.420 $\pm$ 0.010 	& 3.312 $\pm$ 0.010 	& 4.014 $\pm$ 0.011 	& 3.670 $\pm$ 0.010 	& 5.000 			& 5.120 \\
		\tableline\tableline
		\multicolumn{8}{c}{Infrared Photometry from Literature} \\
		\tableline\tableline
		\multicolumn{8}{c}{Cutri 2003} \\
		\tableline
		2MASS $J$ 	& 2.269 $\pm$ 0.244 	& 2.381 $\pm$ 0.290 	& 3.316 $\pm$ 0.248 	& 3.291 $\pm$ 0.226 	& 3.440 $\pm$ 0.290 	& 4.776 $\pm$ 0.282 	& 4.338 $\pm$ 0.222 \\
		2MASS $H$ 	& 2.359 $\pm$ 0.164 	& 2.487 $\pm$ 0.174 	& 3.306 $\pm$ 0.252 	& 3.295 $\pm$ 0.228 	& 3.539 $\pm$ 0.276 	& 4.578 $\pm$ 0.036 	& 4.264 $\pm$ 0.144 \\
		2MASS $K$ 	& 2.285 $\pm$ 0.244 	& 2.429 $\pm$ 0.288 	& 3.104 $\pm$ 0.338 	& 3.145 $\pm$ 0.244 	& 3.546 $\pm$ 0.318 	& 4.505 $\pm$ 0.016 	& 4.313 $\pm$ 0.018 \\
		\tableline
		\multicolumn{8}{c}{Ducati 2002} \\
		\tableline
		Johnson $K$ 	& \nodata 		& 2.33 $\pm$ 0.02 		& \nodata 		& 1.76 $\pm$ 0.01 		& 3.42 $\pm$ 0.01 		& \nodata 		& \nodata \\
		\tableline
		\multicolumn{8}{c}{Morel 1978} \\
		\tableline
		Johnson $J$ 	& 2.350 			& 2.400 			& 3.110 			& \nodata 		& \nodata 		& \nodata 		& \nodata \\
		Johnson $K$ 	& 2.350 			& 2.370 			& 3.090 			& \nodata 		& \nodata 		& \nodata 		& \nodata \\
		\tableline
		\multicolumn{8}{c}{Kidger 2003} \\
		\tableline
		Johnson $J$ 	& \nodata 		& \nodata 		& \nodata 		& 3.674 $\pm$ 0.004 	& \nodata 		& \nodata 		& \nodata \\
		Johnson $H$ 	& \nodata 		& \nodata 		& \nodata 		& 3.623 $\pm$ 0.004 	& \nodata 		& \nodata 		& \nodata \\
		Johnson $K$ 	& \nodata 		& \nodata 		& \nodata 		& 3.631 $\pm$ 0.004 	& \nodata 		& \nodata 		& \nodata \\
		\tableline
		\multicolumn{8}{c}{Neugebauer 1969} \\
		\tableline
		Johnson $K$ 	& 2.38 $\pm$ 0.06 		& 2.34 $\pm$ 0.10 		& \nodata 		& \nodata 		& \nodata 		& \nodata 		& \nodata \\
		\tableline
		\multicolumn{8}{c}{Aumann 1991} \\
		\tableline
		Johnson $J$ 	& \nodata 		& \nodata 		& 3.13 $\pm$ 0.02 		& \nodata 		& \nodata 		& \nodata 		& \nodata \\
		Johnson $H$ 	& \nodata 		& \nodata 		& 3.10 $\pm$ 0.02 		& \nodata 		& \nodata 		& \nodata 		& \nodata \\
		Johnson $K$ 	& \nodata 		& \nodata 		& 3.10 $\pm$ 0.02 		& \nodata 		& \nodata 		& \nodata 		& \nodata \\
		\tableline
		\multicolumn{8}{c}{Selby 1988} \\
		\tableline
		Johnson $J$ 	& \nodata 		& \nodata 		& \nodata 		& 3.77 $\pm$ 0.03 		& 3.52 $\pm$ 0.03 		& \nodata 		& \nodata \\
		Johnson $K$ 	& \nodata 		& \nodata 		& \nodata 		& 3.63 $\pm$ 0.03 		& 3.43 $\pm$ 0.03 		& \nodata 		& \nodata \\
		\tableline\tableline
		\multicolumn{8}{c}{Adopted Infrared Photometry} \\
		\tableline\tableline
		Johnson $J$ 	& 2.35 $\pm$ 0.05 		& 2.40 $\pm$ 0.05 		& 3.13 $\pm$ 0.02 		& 3.674 $\pm$ 0.004 	& 3.52 $\pm$ 0.03 		& \nodata 		& \nodata \\
		Johnson $H$ 	& \nodata 		& \nodata 		& 3.10 $\pm$ 0.02 		& 3.623 $\pm$ 0.004 	& \nodata 		& \nodata 		& \nodata \\
		Johnson $K$ 	& 2.38 $\pm$ 0.06 		& 2.33 $\pm$ 0.02 		& 3.10 $\pm$ 0.02 		& 3.631 $\pm$ 0.004 	& 3.42 $\pm$ 0.01 		& \nodata 		& \nodata \\
		2MASS $H$ 	& \nodata 		& \nodata 		& \nodata 		& \nodata 		& \nodata 		& \nodata 		& 4.264 $\pm$ 0.144 \\
		2MASS $K$ 	& \nodata 		& \nodata 		& \nodata 		& \nodata 		& \nodata 		& \nodata 	& 4.313 $\pm$ 0.03 \\
		\tableline
	\end{tabular}
	\end{center}
\small
Note - The photometric errors listed here are those reported by the respective surveys. We adopt errors of 0.03 mag for all photometric points with reported errors $<$ 0.03 mag (see Section \ref{sec:omod}).
\end{table}

\bibliography{Astars_UMa}{}
\bibliographystyle{asp2010}

\end{document}